\documentstyle[emulateapj,psfig]{article}

\makeatletter

\newenvironment{inlinefigure}{%
\def\@captype{figure}%
\noindent\begin{minipage}{0.999\linewidth}\begin{center}}
{\end{center}\end{minipage}\smallskip}
\makeatother

\lefthead{Barger et al.}

\begin{document}
\title{X-ray, Optical, and Infrared Imaging and Spectral Properties 
of the 1~Ms {\it Chandra} Deep Field North Sources}
\author{
A.~J.~Barger,$\!$\altaffilmark{1,2,3}
L.~L.~Cowie,$\!$\altaffilmark{3}
W.~N.~Brandt,$\!$\altaffilmark{4}
P.~Capak,$\!$\altaffilmark{3}
G.~P.~Garmire,$\!$\altaffilmark{4}
A.~E.~Hornschemeier,$\!$\altaffilmark{4}
A.~T.~Steffen,$\!$\altaffilmark{1}
E.~H.~Wehner$\!$\altaffilmark{1}
}

\altaffiltext{1}{Department of Astronomy, University of Wisconsin-Madison,
475 North Charter Street, Madison, WI 53706}
\altaffiltext{2}{Department of Physics and Astronomy,
University of Hawaii, 2505 Correa Road, Honolulu, HI 96822}
\altaffiltext{3}{Institute for Astronomy, University of Hawaii,
2680 Woodlawn Drive, Honolulu, HI 96822}
\altaffiltext{4}{Department of Astronomy \& Astrophysics,
525 Davey Laboratory, The Pennsylvania State University,
University Park, PA 16802}

\slugcomment{Submitted to the Astronomical Journal}

\begin{abstract}
We present the optical, near-infrared, submillimeter, and radio
follow-up catalog of the X-ray selected sources from an $\approx 1$~Ms
{\it Chandra} observation of the Hubble Deep Field North region.
We have $B$, $V$, $R$, $I$, and $z'$ magnitudes for the 370 X-ray 
point sources, $HK^\prime$ magnitudes for 276, and 
spectroscopic redshifts for 182. We present high-quality spectra 
for 175 of these. The redshift distribution shows indications of 
structures at $z=0.843$ and $z=1.0175$ (also detected in optical surveys) 
which could account for a part of the field-to-field variation
seen in the X-ray number counts; however, these structures 
do not dominate the
number of X-ray sources in the sample and hence should not strongly 
affect the redshift distribution.
All of the X-ray sources with $z>1.6$ are either broad-line AGN or
have narrow Ly$\alpha$ and/or CIII]~1909~\AA\ emission; none of
the known $z>1.6$ absorption-line galaxies in the field are detected
individually in X-rays.
We estimate photometric redshifts for the sources with $B-I>1.5$ 
(bluer than this it is hard to distinguish between low
redshift irregulars and luminous high-redshift AGN) 
and find agreement (most are within 25\%) with the 
available spectroscopic redshifts. The majority of the
galaxies in both the $2-8$~keV (hard) and $0.5-2$~keV (soft) 
samples have absolute magnitudes comparable to or more luminous 
than $M_I^\ast=-22$.
The flux contributions separated into unit bins of redshift show 
that the $z<1$ spectroscopically identified sources already contribute 
about one-third of the total flux in both the hard and soft bands. 
Thus, major accretion onto supermassive black holes has occurred 
since the Universe was half its present age. We find from
ratios of the X-ray counts that the X-ray spectra are 
well-described by absorption of an 
intrinsic $\Gamma=1.8$ power-law, with $N_H$ values
ranging from about $10^{21}$~cm$^{-2}$ 
to $5\times 10^{23}$~cm$^{-2}$. 
We find very little evolution in the maximum rest-frame opacity-corrected 
and $K$-corrected $2-8$~keV X-ray luminosities with decreasing redshift
until $z\lesssim 0.5$, where 
the volume becomes too small to probe effectively very high 
luminosity sources. We estimate that the {\it Chandra} sources that
produce 87\% of the {\it HEAO-A} X-ray background (XRB)
at 3~keV produce 57\% at 20~keV, provided that at high energies 
the spectral shape of the sources continues to be well-described 
by a $\Gamma=1.8$ power-law. However,
when the {\it Chandra} contributions are renormalized to the
{\it BeppoSAX} XRB at 3~keV, the shape matches fairly well the
observed XRB at both energies.  Thus, whether a substantial 
population of as-yet undetected Compton-thick sources, or some 
change in the spectral shape of the current sources from the 
simple power-law dependence, is required to completely resolve 
the XRB above $10$~keV depends critically on how the currently 
discrepant XRB measurements in the $1-10$~keV energy range tie 
together with the higher energy XRB.

\end{abstract}


\section{Introduction}
\label{secintro}

X-ray surveys most directly trace accretion onto supermassive
black holes and hence provide our best window on
black hole evolution. Some primary 
observational goals of X-ray studies include the measurement 
of supermassive black hole properties (such as number density 
and accretion rates) versus redshift and the determination 
of the nature of the host galaxies. Since much of the accretion 
power in the Universe may be absorbed by substantial neutral
hydrogen column 
densities (e.g., \markcite{fabian99}Fabian \& Iwasawa 1999),
the challenge is to construct a {\it complete census} of 
supermassive black holes to the earliest epoch,
including sources which are heavily obscured from soft X-ray 
energies to the near-infrared (NIR). The census is well 
underway for unobscured sources (i.e., quasars with broad
emission lines and big blue bumps), but the obscured
population is not well-characterized, 
except for the relatively rare radio galaxies.

A fundamental goal of the {\it Chandra X-ray Observatory} was 
to resolve the hard ($2-8$~keV) X-ray background (XRB) into 
discrete sources. At these energies the photons can penetrate all
but the highest column densities ($>10^{24}$~cm$^{-2}$)
of gas and dust, so many obscured active galactic nuclei (AGN) 
can now be detected.
The two recent $\approx 1$~Ms exposures of
the {\it Chandra} Deep Field North 
(CDF-N; \markcite{brandt01}Brandt et al.\ 2001c, hereafter B01)
and the {\it Chandra} Deep Field South 
(CDF-S; \markcite{giacconi02}Giacconi et al.\ 2002) have resolved 
$>80-90$\% of the $2-8$~keV XRB into discrete sources
(\markcite{campana01}Campana et al.\ 2001;
\markcite{cowie02}Cowie et al.\ 2002; 
\markcite{rosati02}Rosati et al.\ 2002) 
with the main uncertainty being the normalization of the 
XRB itself. The CDF-N exposure has recently been extended 
to a second megasecond. 

Although with the {\it Chandra} and {\it XMM-Newton
Observatories} there has been rapid improvement 
in our ability to detect hard X-ray sources at the faintest 
fluxes, our efforts to understand in detail the nature 
and evolution of the sources creating the hard XRB are still 
in the early stages. Optical and NIR follow-up studies of
$100-300$~ks {\it Chandra} 
(\markcite{mushotzky00}Mushotzky et al.\ 2000;
\markcite{horn01}Hornschemeier et al.\ 2001, hereafter H01;
\markcite{barger01a}Barger et al.\ 2001a, c; 
\markcite{tozzi02}Tozzi et al.\ 2002;
\markcite{stern02b}Stern et al.\ 2002b)
and {\it XMM-Newton} (\markcite{hasinger01}Hasinger et al.\ 2001;
\markcite{baldi02}Baldi et al.\ 2002;
\markcite{fiore01}Fiore et al.\ 2001)
imaging surveys have found that almost half of the hard X-ray 
light arises in optically bright ($I<23.5$) galaxies in the 
$z<1.5$ redshift range. Almost all of these sources can be 
spectroscopically identified, and most are bulge-dominated 
galaxies with near-$L^\ast$ luminosities. 
The other half of the hard X-ray light arises in a mixture of
$z>1.5$ AGN and optically faint galaxies ($I>23.5$) that
cannot be spectroscopically identified.

Contrary to the situation for the faint {\it ROSAT} soft X-ray 
sources (\markcite{hasinger98}Hasinger et al.\ 1998;
\markcite{schmidt98}Schmidt et al.\ 1998), the vast 
majority ($>80$\%) of the spectroscopically identified hard X-ray 
sources do not have broad optical or ultraviolet lines, and 
almost half show no obvious high-ionization signatures of AGN 
activity in their spectra. The hard to soft X-ray flux ratios 
of the latter sources suggest that the sources are highly 
absorbed systems whose high column densities could effectively 
extinguish the optical, ultraviolet, and NIR continua from the 
AGN and render traditional identification techniques ineffective.

Based on the colors, many of the spectroscopically unidentified
$I>23.5$ galaxies may lie at redshifts in the range 
$z=1.5$ to 3 (\markcite{crawford01}Crawford et al.\ 2001;
\markcite{cowie01}Cowie et al.\ 2001;
\markcite{barger01a}Barger et al.\ 2001a, b;
\markcite{alexander01}Alexander et al.\ 2001).
However, a very intriguing possibility suggested by
\markcite{haiman99}Haiman \& Loeb (1999)
is that some of the optically faint sources may instead
be low luminosity quasars at very high redshifts ($z>5$),
since cold dark matter models can allow a large number of
such high-redshift AGN.

The $\approx 1$~Ms {\it Chandra} exposures have also resolved 
$>90$\% of the soft ($0.5-2$) XRB into discrete sources (B01; 
\markcite{rosati02}Rosati et al.\ 2002).
At the faintest X-ray fluxes probed by these exposures,
galaxies whose X-ray light is dominated by processes related 
to star formation rather than AGN activity begin to be significant 
contributors to the number counts, even though they contribute only 
$2-3$\% to the X-ray light 
(e.g., \markcite{brandt01a}Brandt et al.\ 2001a;
\markcite{horn02}Hornschemeier et al.\ 2002).
The large numbers of detections are understandable 
since we now have the sensitivity to detect the most luminous 
($>2\times 10^{39}$~erg~s$^{-1}$) X-ray 
binaries and supernova remnants at $z<0.15$. 
AGN may cease to dominate the X-ray number counts at extremely faint 
fluxes (e.g., \markcite{moran99}Moran, Lehnert, \& Helfand 1999)
in a manner analogous to the dominance of star forming galaxies 
rather than AGN at very faint radio fluxes
(e.g., \markcite{richards00}Richards\ 2000).

In this paper we present follow-up optical, NIR, and 
submillimeter imaging and optical spectroscopy of the X-ray sources 
detected in the $\approx 1$~Ms CDF-N exposure to characterize 
the properties of the faint X-ray source populations.
We then use our data to describe the evolution of the X-ray 
sources and their host galaxies. 

\section{X-ray Sample}
\label{secxray}

B01 presented the 975.3~ks CDF-N X-ray images, along with details
of the observations, data reduction, and technical analysis.
B01 merged their source lists in four standard X-ray 
bands, $0.5-8$~keV (full band), $0.5-2.0$~keV (soft band), $2-8$~keV
(hard band), and $4-8$~keV (ultrahard band), into a catalog (Table~3 
in B01) of 370 significantly detected point sources over an area of 
about 450~arcmin$^2$. Near the aim point the data reach limiting
fluxes of $\approx 3\times 10^{-17}$ ($0.5-2$~keV) and
$\approx 2\times 10^{-16}$~erg~cm$^{-2}$~s$^{-1}$ ($2-8$~keV),
which are, respectively, $\approx 40$ and $\approx 400$ times
fainter than achieved by pre-{\it Chandra} missions.
The number counts and completeness levels are described in B01
and \markcite{cowie02}Cowie et al.\ (2002).
The six extended X-ray sources detected in the CDF-N are discussed 
in great detail in \markcite{bauer02}Bauer et al.\ (2002) and will 
not be considered in the present work.

In this paper we present a follow-up data table to match the full 
B01 X-ray point source catalog.
However, outside a $10'$ radius from the approximate {\it Chandra}
image center J2000 RA $12^h~36^m~51.20^s$ and Dec $62^\circ~13'~43.0''$
(this choice of center maximizes the average exposure time within
a $6.5'$ radius circle), the X-ray point spread function degrades 
rapidly; thus, in order to have a more uniform flux sample with
accurate positions for our analysis, in \S~\ref{secz} we 
introduce a restricted $10'$ radius sample that we will use (in
addition to some even more restricted subsamples) subsequently.

Table~1 lists the full 370 B01 X-ray sources by number (column~1) 
and by RA and Dec coordinates in decimal degrees (columns~2 and 3).
Columns~$4-15$, to be discussed in subsequent subsections,
give submillimeter fluxes, radio fluxes, isophotal $R$ magnitudes, 
corrected aperture $HK'$, $z'$, $I$, $R$, $V$, and $B$ magnitudes, 
optical $\Delta x$ and $\Delta y$ offsets in arcseconds
from the X-ray positions, 
and redshifts. The final column contains a `B' if the spectrum has broad 
emission lines and a `C' if the source is optically compact.

Figure~\ref{fig1} shows a schematic of the 370 X-ray point sources
in the full B01 sample (solid squares denote sources with 
spectroscopic identifications, as discussed in \S~\ref{secz}). 
Radii of $6.5'$ and $10'$ from the approximate X-ray image center 
are indicated with large circles and will be discussed in 
subsequent sections. The dashed contour shows the area covered 
by the $HK'$ imaging data (see \S~\ref{secnir}) and the solid 
contour shows the area covered by the submillimeter data
(see \S~\ref{secsmm}). The optical data cover a $24'\times 24'$
area centered on the Hubble Deep Field North (HDF-N).

\section{Optical Observations and Reduction}
\label{secopt}

The optical imaging data consist of Johnson $B$, Johnson $V$, 
Cousins $R$, Cousins $I$, and Sloan $z'$ observations obtained 
with the Subaru Prime Focus Camera (Suprime-Cam; 
\markcite{miya98}Miyazaki et al.\ 1998) 
on the Subaru\footnote{The Subaru telescope is operated by the 
National Astronomical Observatory of Japan.}
8.2~m telescope between 2001 February and April. 
The images and catalogs of the entire 
sample of galaxies and stars in the field are presented in 
P.~Capak et al., in preparation, where a more extensive discussion 
may be found. Here we briefly summarize the 
Suprime-Cam observations and reductions.

A five step dither pattern with $1'$ steps was used to fill in the 
chip gaps and provide adequate offsets for flat-fielding 
each dither pattern separately. Half the dither patterns were taken 
at a 90 degree rotation to provide better chip-to-chip photometric 
calibration and remove bleeding from saturated stars. The total 
exposure times were 6000~s in $B$ and $V$, 19440~s in $R$, 
4080~s in $I$, and 11450~s in $z'$. Seeing was good
throughout the observations, ranging from $0.5''$ to $1.5''$. 
The median seeing in the final images is $0.75''$ in $B$, 
$1''$ in $V$, $1''$ in $R$, $0.8''$ in $I$, and $0.8''$ in $z'$.  
The inverse second moment of the light in the highest resolution
$B$ image was used to determine the compactness (Kron 1980).
$B<26.5$ sources consistent with being point-like and
unsaturated are marked `C' for compact in the final column 
of Table~1.

The data were reduced using Nick Kaiser's Imcat software package.
The data were overscan corrected and bias subtracted. Bad pixels
were masked. No dark subtraction was performed since Suprime-Cam 
has a negligible dark current. A median sky flat was generated for 
each band to correct for the vignetting pattern.  
For the $I$ and $z'$ bands a surface was tessellated over a grid spaced 
at 128~pixel intervals. This surface was used instead of a median 
sky flat to remove high frequency fringing, which
is additive, from the flat. To generate a fringe frame, a median sky flat
was created for each night in $I$ and $z'$ and divided by the tessellated
surface. The fringe frame was scaled to the background level in each chip 
and subtracted to remove the fringing. Next, a separate median flat was
generated for each dither pattern to remove flattening variations caused
by flexure in the instrument. A second order polynomial was then fit to
the sky background and subtracted to remove the sky. After this reduction
the images were flat to better than 1\% r.m.s.

The astrometric solution was obtained in two stages.  
Stars brighter than 19th magnitude are saturated in most of the
science images making comparisons to the USNOA-2 difficult.  Thus, a 
standard star frame was fit to the USNOA-2 to find a rough initial 
solution for each chip in the mosaic. This initial solution was 
applied to the science frames, and stars appearing within $2''$ of 
the same position in at least four images were identified.  
A cubic polynomial distortion was then fit to each chip of the mosaic
in each exposure to minimize the r.m.s. scatter in
the position of the stars from exposure to exposure.  
The solution was iterated several times to eliminate mismatched 
stars. The absolute positioning was then chosen to match the VLA 
20~cm catalog of \markcite{richards00}Richards (2000; see
\S~\ref{secradio}), which was also used in B01 to determine the 
absolute positions of the {\it Chandra} sources.

We show $z'$-band thumbnail images of the CDF-N sources in
Fig.~2 ordered by B01 catalog number. We use the $z'$-band 
since many of the sources are red and hence more easily seen in 
this band. Each thumbnail is $18''$ 
on a side. The number of the source from Table~1 is printed in the 
upper left corner. If the source is also a submillimeter source
(see \S~\ref{secsmm}), the letter `S' is printed in the upper
right corner. If the source is also a radio source 
(see \S~\ref{secradio}), the letter `R' is printed in the upper right 
corner. If the source is in the B01 $2-8$~keV sample, the letter 
`H' is printed in the bottom right corner. If there is a spectroscopic
identification for a source that lies within $2''$ of the
X-ray position (see \S~\ref{secz}), then the redshift 
(or the word `STAR' for a stellar spectrum) is printed in the bottom 
left corner. If there is a spectroscopic identification for a 
source that lies outside of our chosen cross-identification 
radius of $2''$
but still within $5''$ of the X-ray position, then the 
redshift of that source is printed in fainter text to indicate 
that the redshift has been given for interest. However, apart
from two special cases discussed in \S~\ref{secz}, such sources
have not been cross-identified in Table~1 with the X-ray sources,
nor have the redshifts been used in any subsequent analysis.

Of the 370 sources in the CDF-N catalog, 210 have measured
$R$ magnitudes brighter than 24. Since the surface density of
galaxies and stars in this region is roughly
7.3~arcmin$^{-2}$ to this magnitude limit, the number of 
random overlaps should be small: 5 chance projections within a 
$1.5''$ radius and 22 within a $3''$ radius. For the $R<24$
sources the median offset between the optical and
X-ray positions is $-0.09''$ in RA and $-0.05''$ in Dec,
illustrating that the absolute astrometric solutions are indeed
consistent.

The positional uncertainties are expected primarily to reflect the
centroiding accuracies and the overall distortions in the X-ray data,
both of which increase as we move to larger angles away from the
optical axis of the {\it Chandra} telescope and also
depend on the flux of the source (B01). In order to determine
the relative positional accuracy of the X-ray
and optical data, we measured the spread in offsets
between the X-ray and optical positions for the $R\le 24$
sample. Within a $6.5'$ radius from the approximate X-ray image 
center the r.m.s. dispersion is $0.28''$ in RA and $0.25''$ in Dec,
while between $6.5'$ and $10'$ this rises to $0.58''$ in RA and
$0.7''$ in Dec. The corresponding two dimensional dispersions
are $0.38''$ and $0.9''$. The circles in Fig.~\ref{fig2} show 
$2''$ ($5\sigma$) radii for sources within $6.5'$ of the center
and $3.6''$ radii ($4\sigma$) for sources beyond this radius.
Nearly all the true counterparts should lie within these
error circles.

The Suprime-Cam data were photometrically calibrated
using sources in the HDF-N proper, as given in the
catalog of \markcite{fernandez99} Fern{\'a}ndez-Soto, 
Lanzetta, \& Yahil (1999). To facilitate comparison
with previous work, we give the $B$ and $V$ magnitudes
in the Johnson system and the $R$ and $I$ magnitudes in 
the Kron-Cousins system. The offset between AB magnitude and 
Vega-based magnitude is 0.41, 0.20, 0.0, and $-0.11$
for $I$, $R$, $V$, and $B$, respectively, where the
AB magnitude is defined by $-48.60-2.5\log f_\nu$. Here 
$f_\nu$ is the flux of the source in units of 
erg~cm$^{-2}$~s$^{-1}$~Hz$^{-1}$.
The $z'$ magnitudes are given in the AB system.

We measured the magnitude of each source in a $3''$ diameter 
aperture. We then corrected to an approximate total magnitude 
using an offset calculated as the median difference between 
$3''$ and $6''$ aperture magnitudes measured for a complete 
$I<25$ optical sample in the field. Higher resolution
images were smoothed to match the seeing in the poorer
resolution images in order to give the most precise color
measurements. For sources with a counterpart
brighter than $z'=24$ within $2''$ of the X-ray position,
we centered on the brightest $z'$-band peak
within the $2''$ error circle. The optical $\Delta x$ and 
$\Delta y$ offsets in arcseconds 
from the X-ray position (positive values are 
to the East and North) are given in columns~13 and 14 of Table~1. 
For optically fainter sources we centered on the X-ray position. 

For sources brighter than $R=23$ we
also measured the isophotal magnitude in the $R$-band
to an isophotal level which is 1\% of the
peak surface brightness. At magnitudes fainter than
$R=20$ the isophotal and corrected aperture magnitudes
generally agree well, except for the small number of
X-ray sources which lie on the edges of bright galaxies
(sources 30, 62, 125, 192, 218, 225, 271, and 290). Here the 
isophotal magnitude measures the magnitude of the bright galaxy.
At $R<20$ the sizes of many of the galaxies become
larger than the $3''$ diameter aperture, and the isophotal
magnitudes become brighter than the corrected aperture
magnitudes. For these objects the isophotal magnitude
may be preferred.

The isophotal $R$ magnitudes are given in column~6 of Table~1
and the corrected aperture magnitudes for $z'$, $I$, $R$, $V$, 
and $B$ are given in columns $8-12$. Sources which are saturated 
in a given band are listed at a nominal magnitude of 18 
(given in brackets), which is about a magnitude brighter
than where saturation occurs in the images, while sources which
are contaminated by bright neighbors or by bleeding columns
from bright stars are listed at a nominal magnitude of 30.
Sources with a negative flux in the aperture are listed
with a negative magnitude; in these cases the absolute value of
the magnitude was computed from the absolute value of the 
flux. The $1\sigma$ limits of the images are 27.0 ($z'$), 
27.6 ($I$), 29.2 ($R$), 28.5 ($V$), and 29.0 ($B$).

\section{Near-Infrared Observations and Reduction}
\label{secnir}

We have expanded the wide-field $HK^\prime$ coverage of the
HDF-N region presented in \markcite{barger99}Barger et al.\ (1999)
with new $HK^\prime$ observations obtained with the
University of Hawaii 2.2~m telescope.
The new data were obtained UT 2001 March 1--6
with the University of Hawaii QUick Infrared
Camera (QUIRC; \markcite{hodapp96}Hodapp et al.\ 1996). QUIRC utilizes
a $1024\times 1024$ pixel HgCdTe Astronomical Wide Area Infrared
Imaging (HAWAII) array produced by the Rockwell
International Science Center. At the f/10 focal ratio the
field-of-view of QUIRC is $193''\times 193''$
with a scale of $0.1886$~arcsec~pixel$^{-1}$.
The observations were made using a notched $HK^\prime$ filter with a
central wavelength of $1.8\micron$.
The filter covers the longer wavelength region of the $H$-band and the
shorter wavelength region of the $K$-band (roughly the $K'$ filter).
Because of its broad bandpass and low sky background, this filter is
extremely fast and is roughly twice as sensitive as the $H$, $K'$,
or $K_s$ filters. \markcite{barger99}Barger et al.\ (1999) found
the empirical relation $HK^\prime - K = 0.13 + 0.05 (I-K)$ to convert
between the $HK^\prime$ and $K$ bands; for galaxies at most
redshifts, $HK^\prime - K \approx 0.3$.

We used a dither pattern of thirteen spatially shifted short exposures
of step-size $5''-20''$ in all directions. This observing
strategy enabled us to use the data images for the subsequent sky
subtraction and flat-fielding procedures. Each object is
sampled by a different array pixel on each exposure, and
the signal-to-noise ratio should improve as the square root of the number
of frames. The dark current is insignificant in the QUIRC data.
The data were processed using median sky flats generated from the
dithered images and calibrated from observations of UKIRT faint
standards (\markcite{hawarden01}Hawarden et al.\ 2001)
taken on the first three nights when the sky was photometric.
The FWHM of the total composite image is $0.8''$.
The magnitudes were measured in $3.0''$ diameter apertures and
corrected to approximate total magnitudes
using an offset calculated as the median difference between
$3''$ and $6''$ aperture magnitudes measured for a complete $HK'<20$
sample in the field.

Of the 370 sources in the B01 sample, 276 have $HK^\prime$ 
measurements. The corrected aperture $HK'$ magnitudes 
are given in column~7 of Table~1.

%
%
\begin{inlinefigure}
\psfig{figure=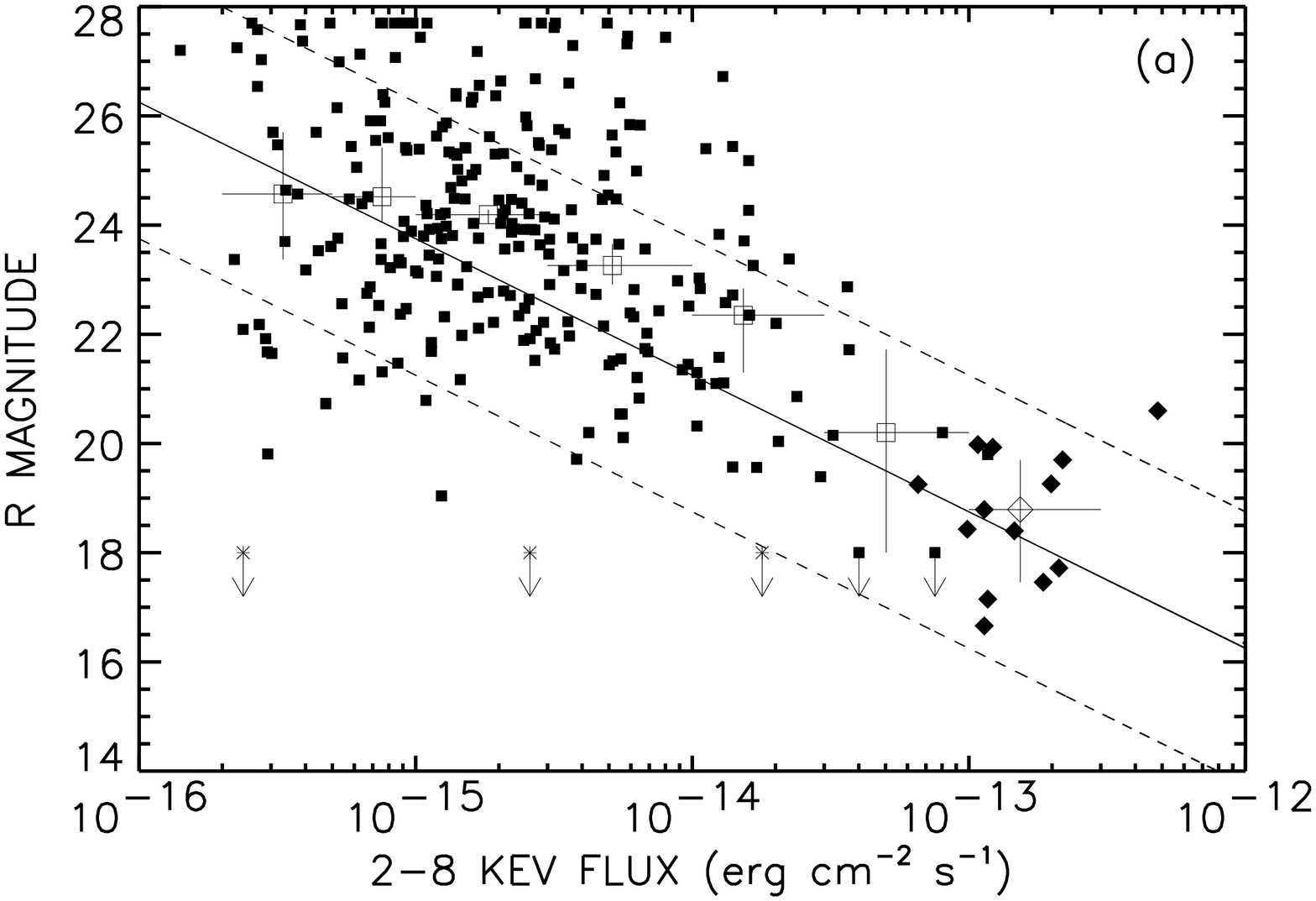,angle=90,width=3.5in}
\psfig{figure=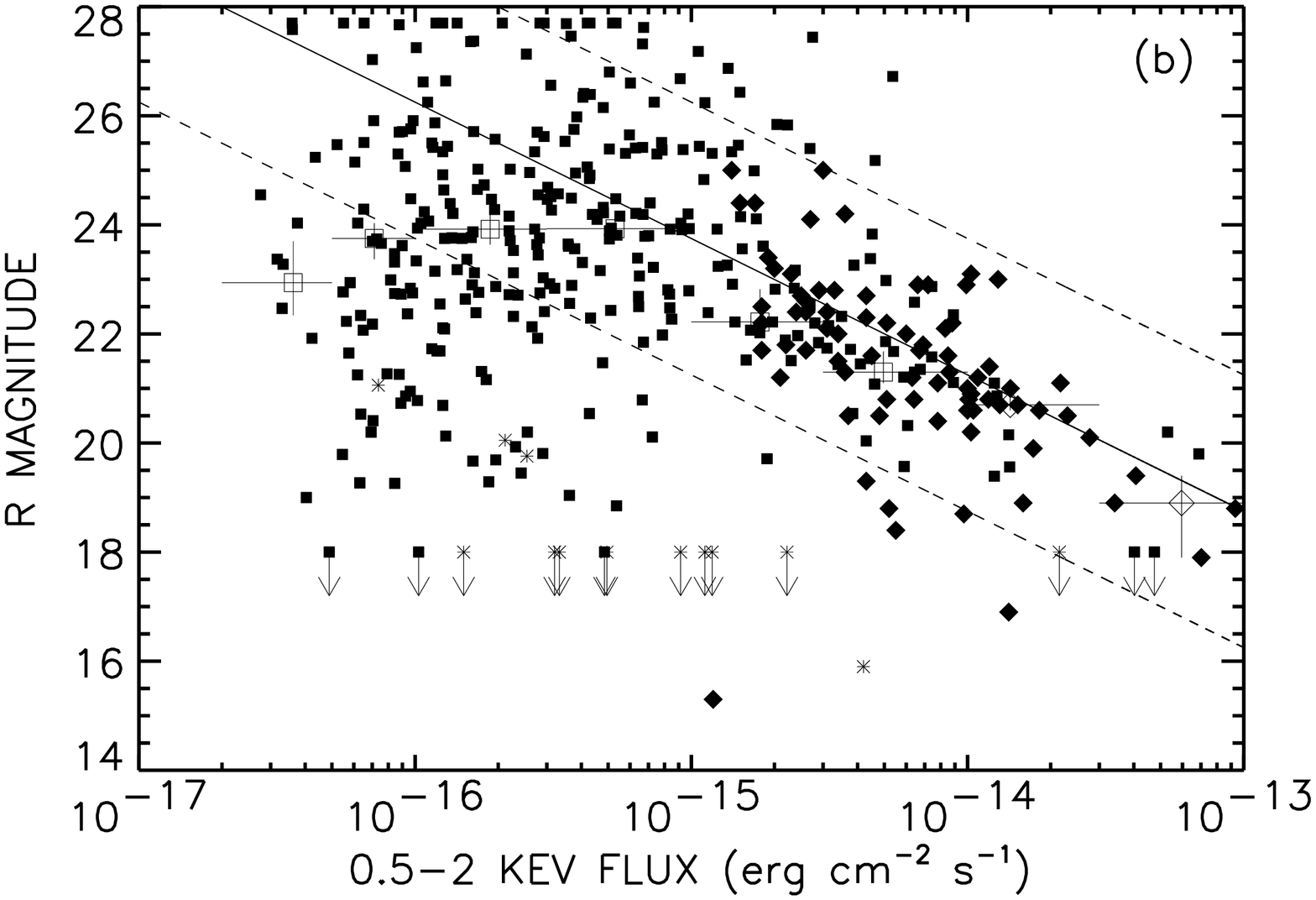,angle=90,width=3.5in}
\vspace{6pt}
\figurenum{3}
\caption{
(a) $R$ magnitude versus $2-8$~keV flux for the CDF-N hard X-ray
sample (solid squares; asterisks denote spectroscopically identified
stars) and for the Akiyama et al.\ (2000) data (solid diamonds).
The large open squares show the median values of the CDF-N
optical magnitudes tabulated in Table~\ref{tab2a}. The large open
diamond is the median value of the Akiyama et al.\ magnitudes.
(b) $R$ magnitude versus $0.5-2$~keV flux for the CDF-N soft X-ray
sample (solid squares; asterisks denote spectroscopically identified stars)
and for the Lehmann et al.\ (2001) data
(solid diamonds). The large open squares show the median values
of the CDF-N optical magnitudes tabulated in Table~\ref{tab2b}.
The two large open diamonds are the
median values of the Lehmann et al.\ magnitudes.
In both (a) and (b) the $\log(f_X/f_R)=0.0$ (solid) and 
$\log(f_X/f_R)=\pm1$ (dashed) lines are loci of constant optical to 
X-ray flux that cover the range
of most of the X-ray selected sources at the brighter X-ray fluxes.
Sources fainter than $R=27.7$ (the 
$4\sigma$ limit) are plotted at this magnitude. Sources saturated
in $R$ are shown at a nominal magnitude of $R=18$
with downward pointing arrows.
\label{fig3}
}
\addtolength{\baselineskip}{10pt}
\end{inlinefigure}

\section{Optical Properties of the X-ray Sources}

In Fig.~\ref{fig3}a we plot $R$ magnitude versus $2-8$~keV flux
for the full CDF-N hard X-ray sample (solid squares).
Spectroscopically identified stars are denoted by asterisks.
At bright X-ray fluxes we include the
\markcite{akiyama00}Akiyama et al.\ (2000) {\it ASCA} Large Sky
Survey data (solid diamonds; the two clusters and one source
without an optical identification have been excluded, and
the star is off-scale), after converting their $2-10$~keV fluxes
to $2-8$~keV assuming $\Gamma=1.7$ (the value Akiyama
et al.\ assumed in their paper for the intrinsic photon index).
The flux in the $R$-band is related to the $R$ magnitude by
$\log f_R=-5.5-0.4 R$ (H01).
The $\log (f_X/f_R)=0$ (solid) and $\log (f_X/f_R)=\pm1$ (dashed)
lines are loci of constant optical to X-ray flux
that cover the range of most of the hard X-ray selected sources
at the brighter X-ray fluxes.

Median optical magnitudes for the CDF-N hard X-ray sample
are summarized in Table~\ref{tab2a} and
are shown in Fig.~\ref{fig3}a as large open squares.
The large open diamond
shows the median optical magnitude for the Akiyama et al.\ sample.
The horizontal bars show the widths of the flux
bins while the vertical bars show the 68\% confidence
range in the median computed using the number of sources
in each bin (\markcite{gehrels86}Gehrels 1986).
At bright X-ray fluxes $\log(f_X/f_R)=0$ roughly matches the
median optical magnitudes, presumably because
an unobscured AGN dominates the total light output from each source.
At fainter X-ray fluxes the medians deviate above $\log(f_X/f_R)=0$
as the sources become obscured in the optical due to the presence
of dust and gas. At the faintest X-ray fluxes the medians
flatten as the optical light from the host galaxy begins to
dominate the total light output from each source.

%
%
\begin{inlinefigure}
\psfig{figure=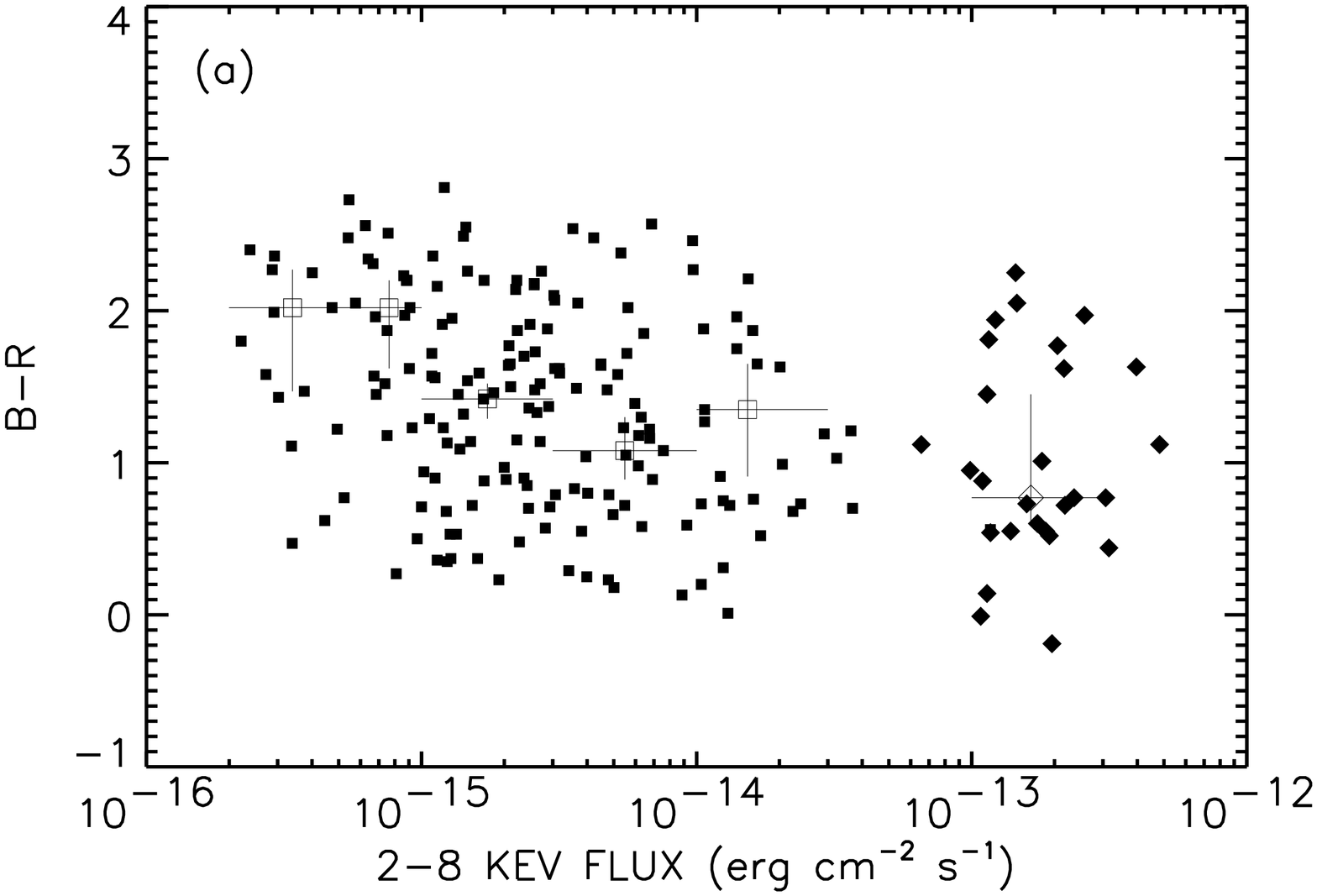,angle=90,width=3.5in}
\psfig{figure=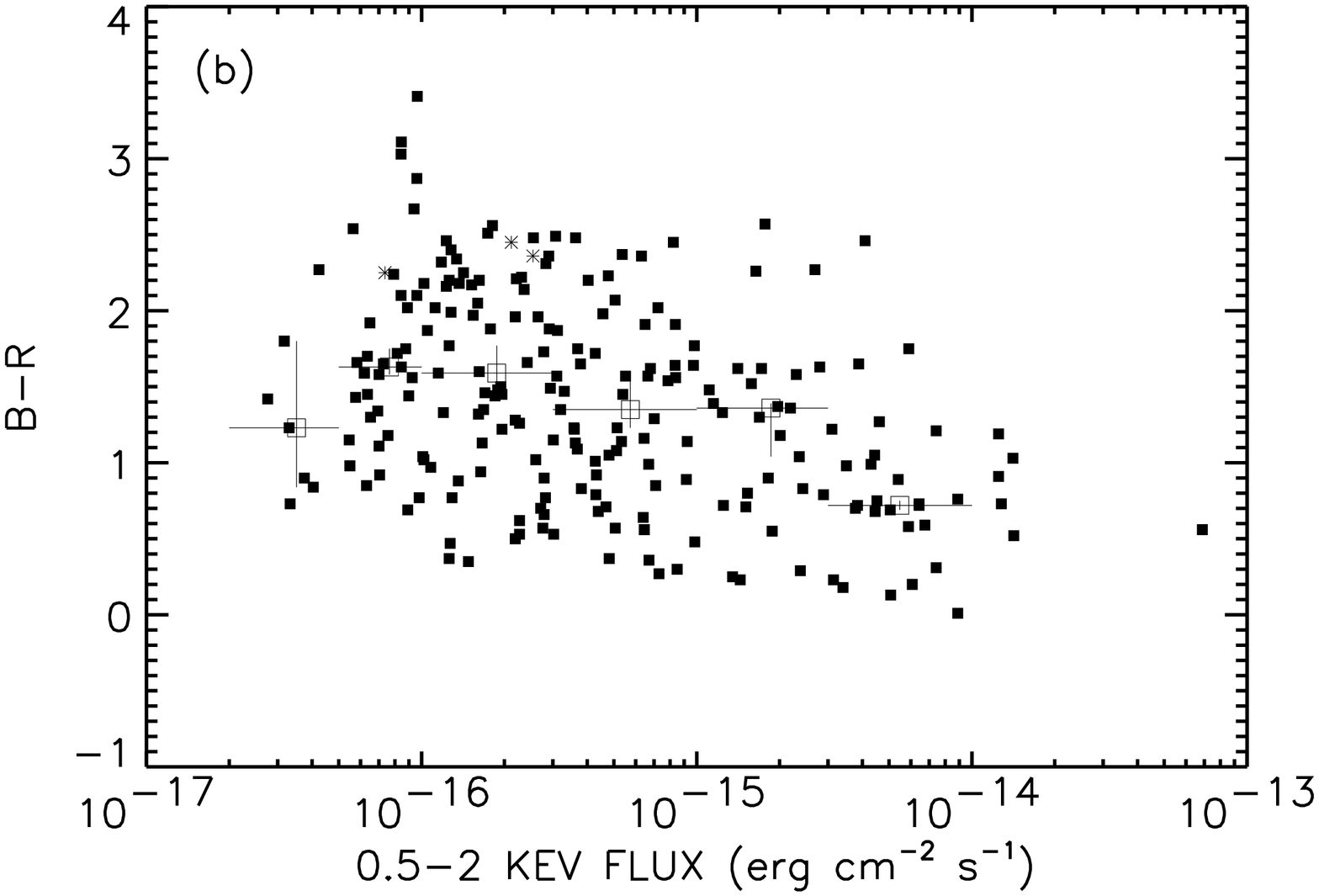,angle=90,width=3.5in}
\vspace{6pt}
\figurenum{4}
\caption{
$B-R$ color versus (a) $2-8$~keV flux for the hard X-ray sample
with $R<25$ and versus (b) $0.5-2$~keV flux for the soft X-ray 
sample with $R<25$. Both exclude any sources that are saturated.
Symbols are as in Fig.~\ref{fig3}. (The Lehmann et al.\ 2001 sample
did not include $B-R$ colors so cannot be plotted in b.)
\label{fig4}
}
\addtolength{\baselineskip}{10pt}
\end{inlinefigure}

In Fig.~\ref{fig3}b we plot $R$ versus $0.5-2$~keV flux for the
full CDF-N soft X-ray sample (solid squares). At bright X-ray fluxes
we include the \markcite{lehmann01}Lehmann et al.\ (2001)
{\it ROSAT} Ultra Deep Survey data (solid diamonds; groups and
clusters and one optically unidentified source have been excluded).
Spectroscopically identified stars are denoted by asterisks.
Again, the $\log (f_X/f_R)=0$ (solid) and $\log (f_X/f_R)=\pm1$
(dashed) lines are loci of constant optical to X-ray flux
that cover the range of most of the soft X-ray selected sources
at the brighter X-ray fluxes.

Median optical magnitudes for the CDF-N soft X-ray 
sample are summarized in Table~\ref{tab2b} and 
are shown as large open squares in Fig.~\ref{fig3}b.
The two large open diamonds
show the median optical magnitudes for the Lehmann et al.\ sample.
Figure~\ref{fig3}b and Table~\ref{tab2b} show that at fluxes above
$10^{-15}$~erg~cm$^{-2}$~s$^{-1}$ the $\log(f_X/f_R)=0$ line matches 
the median optical magnitudes fairly well, again presumably 
because an unobscured AGN dominates the total light output from each
source. Soft X-ray selected AGN samples at
$f_X>8\times 10^{-14}$~erg~s$^{-1}$~cm$^{-2}$ ($0.3-3.5$~keV)
from the {\it Einstein Observatory} Extended Medium Sensitivity
Survey (\markcite{stocke91}Stocke et al.\ 1991) also have values
of $\log (f_X/f_R)$ in a similar range. Unlike 
the hard sample, the medians do not begin to deviate 
above $\log (f_X/f_R)$ at fainter fluxes, presumably because
soft-band sources are mainly unobscured.
Below $10^{-15}$~erg~cm$^{-2}$~s$^{-1}$ the median optical 
magnitudes become constant, within the uncertainties,
as the optical light from the host galaxy begins to dominate
the total light output from each source.

%
%
\begin{deluxetable}{ccrrrcccrc}
\renewcommand\baselinestretch{1.0}
\tablenum{2a}
\tablewidth{0pt}
\tablecaption{Median Optical Magnitudes for the $2-8$~keV CDF-N Sample}
\small
\tablehead{$2-8$~keV flux range & mean $2-8$~keV flux & $B$ & $V$ & $R$ & $I$ & $z'$  \\ ($10^{-16}$~erg~cm$^{-2}$~s$^{-1}$) & ($10^{-16}$~erg~cm$^{-2}$~s$^{-1}$)}
\startdata
  2 --   5  &    3.35  &  25.4  &  24.9  &  24.5  &  23.8  &  23.8  \cr
  5 --  10  &    7.55  &  26.2  &  25.7  &  24.5  &  23.7  &  23.9  \cr
 10 --  30  &    18.2  &  25.5  &  24.8  &  24.2  &  23.3  &  23.4  \cr
 30 -- 100  &    51.4  &  24.7  &  23.8  &  23.2  &  22.2  &  22.2  \cr
100 -- 300  &    152.  &  23.8  &  23.0  &  22.5  &  21.7  &  21.7  \cr
300 -- 1000  &    502.  &  21.1  &  21.4  &  20.2  &  19.9  &  20.1  \cr

\enddata
\label{tab2a}
\end{deluxetable}

\begin{deluxetable}{ccrrrcccrc}
\renewcommand\baselinestretch{1.0}
\tablenum{2b}
\tablewidth{0pt}
\tablecaption{Median Optical Magnitudes for the $0.5-2$~keV CDF-N Sample}
\small
\tablehead{$0.5-2$~keV flux range & mean $0.5-2$~keV flux & $B$ & $V$ & $R$ & $I
$ & $z'$  \\ ($10^{-17}$~erg~cm$^{-2}$~s$^{-1}$) & ($10^{-17}$~erg~cm$^{-2}$~s$^{-1}$)}
\startdata
  2 --   5  &    4.04  &  24.9  &  24.3  &  23.3  &  22.4  &  22.7  \cr
  5 --  10  &    7.40  &  25.0  &  24.1  &  22.9  &  22.0  &  22.4  \cr
 10 --  30  &    17.6  &  25.2  &  24.3  &  23.7  &  23.0  &  22.8  \cr
 30 -- 100  &    55.7  &  25.5  &  24.8  &  24.2  &  23.5  &  23.7  \cr
100 -- 300  &    174.  &  24.5  &  23.9  &  23.5  &  22.9  &  22.9  \cr
300 -- 1000  &    493.  &  22.5  &  22.1  &  21.7  &  21.1  &  21.2  \cr
\enddata
\label{tab2b}
\end{deluxetable}

%
%
\begin{figure*}
\figurenum{5}
\centerline{\psfig{figure=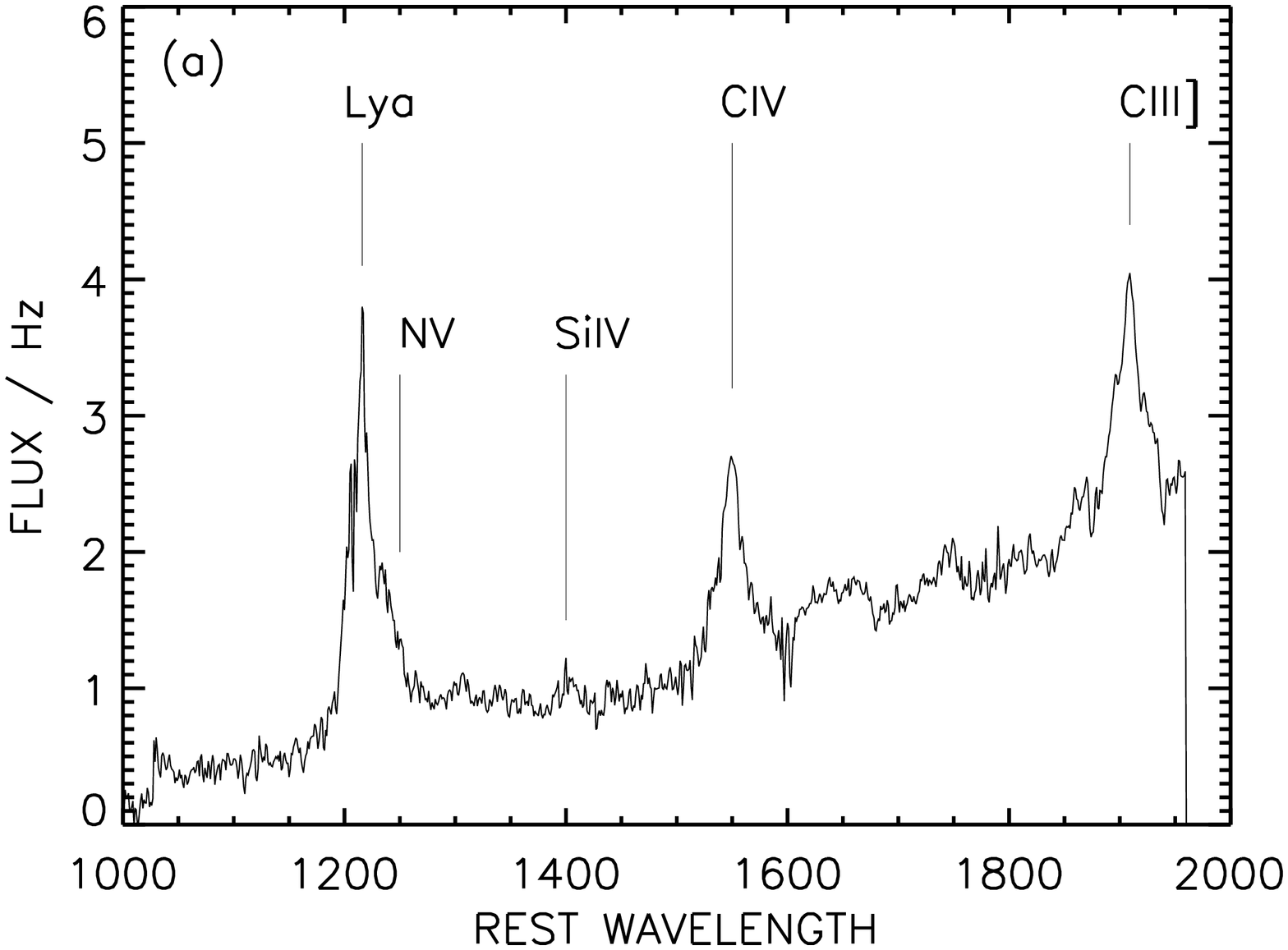,angle=90,width=3in}
\psfig{figure=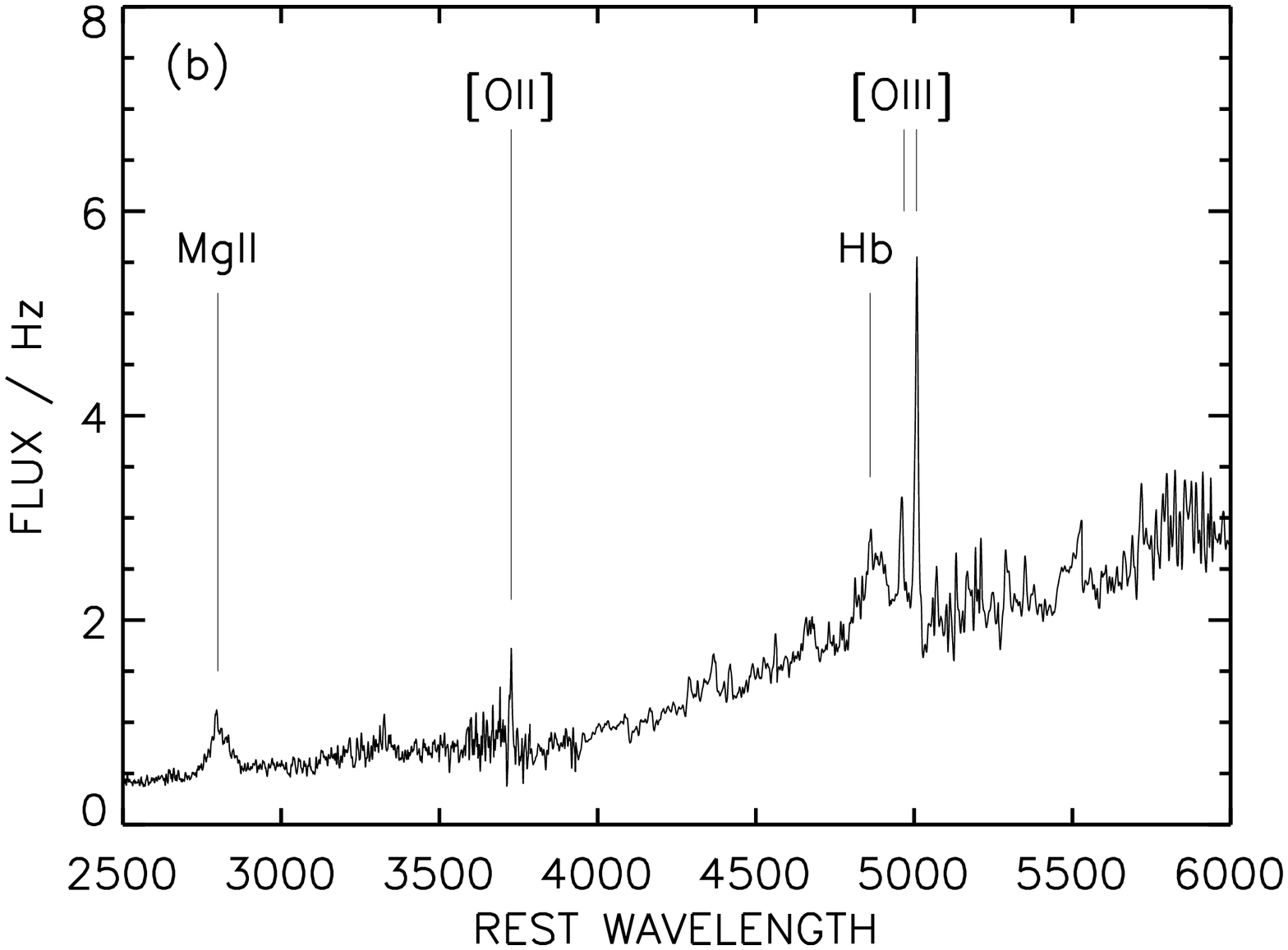,angle=90,width=3in}}
\centerline{\psfig{figure=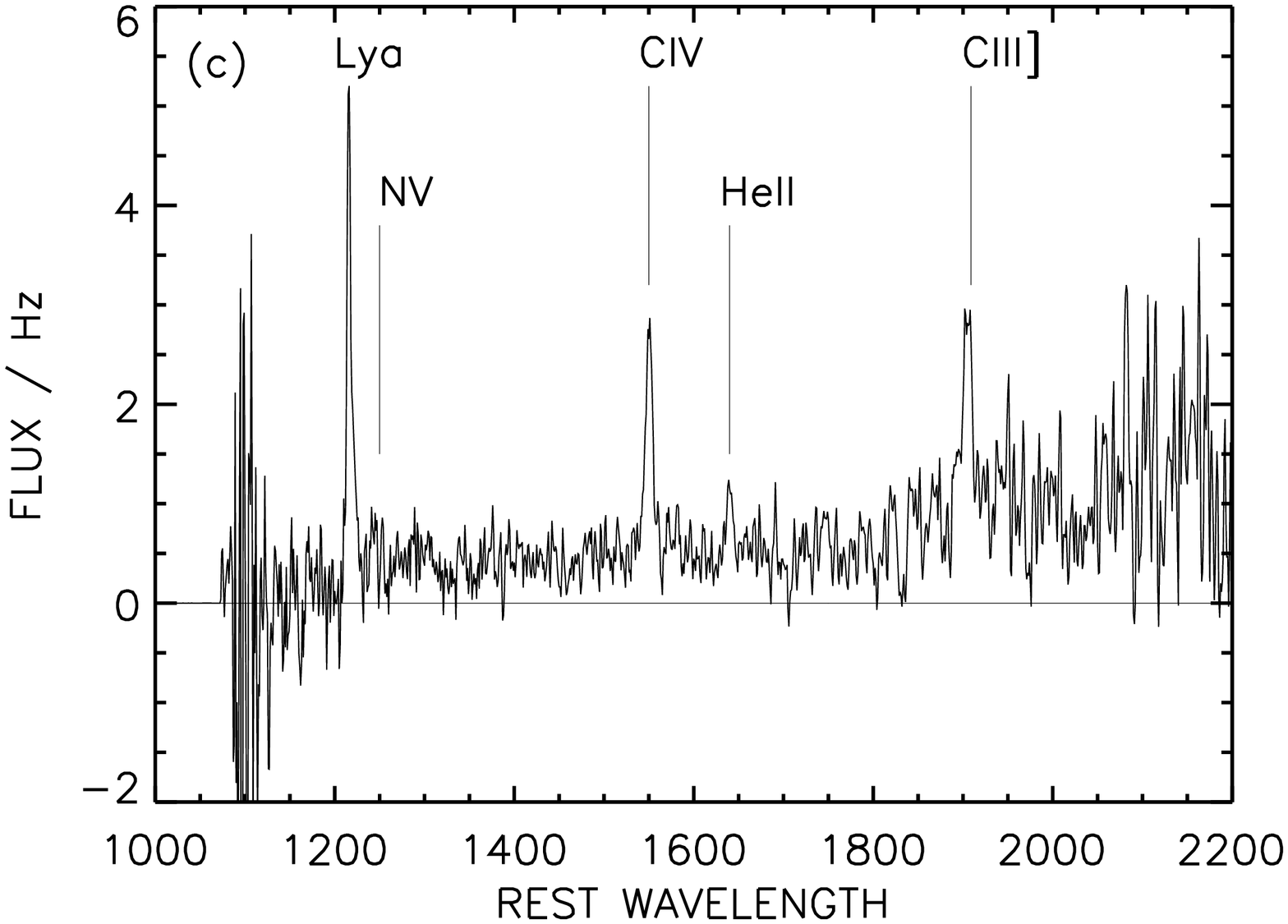,angle=90,width=3in}
\psfig{figure=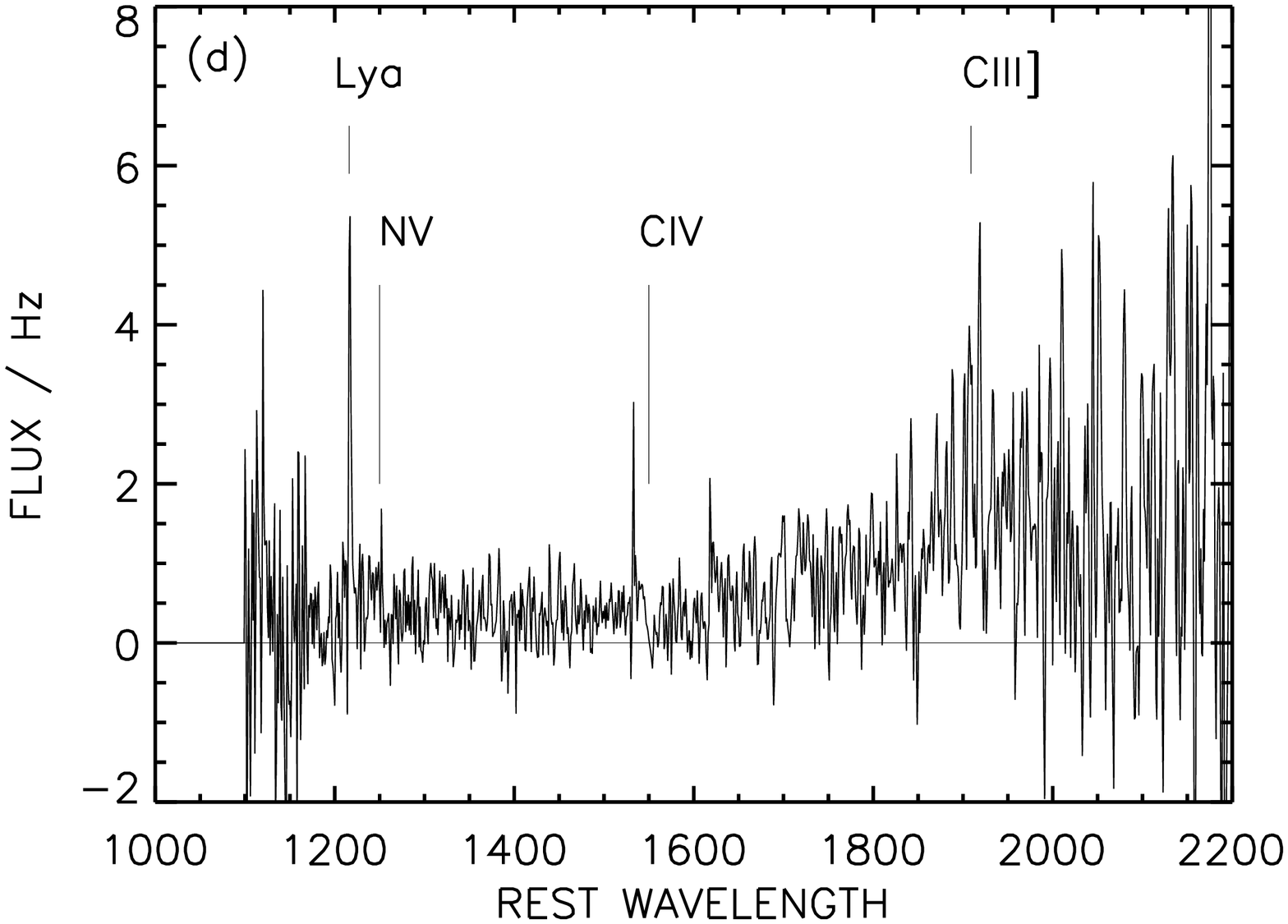,angle=90,width=3in}}
\centerline{\psfig{figure=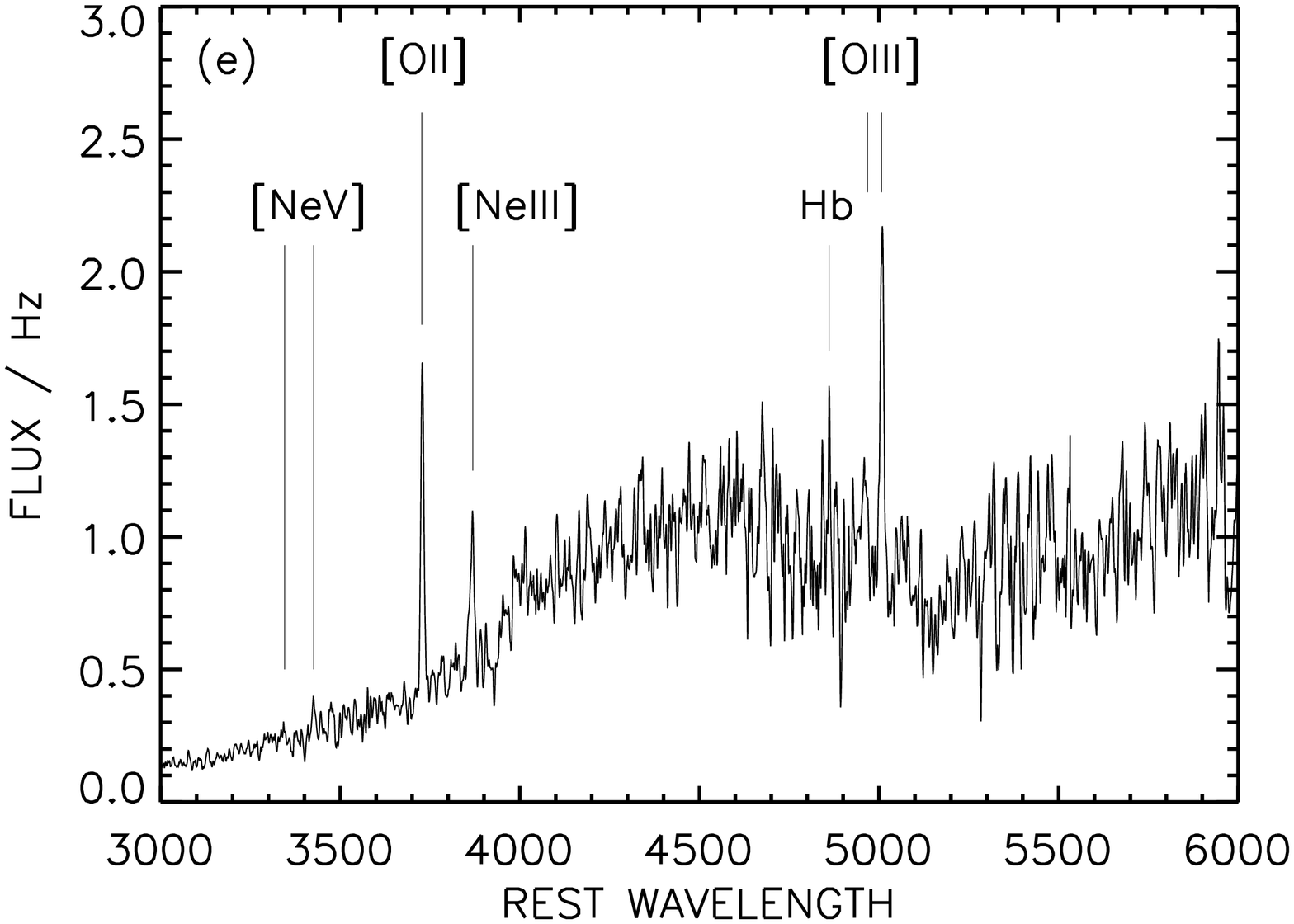,angle=90,width=3in}
\psfig{figure=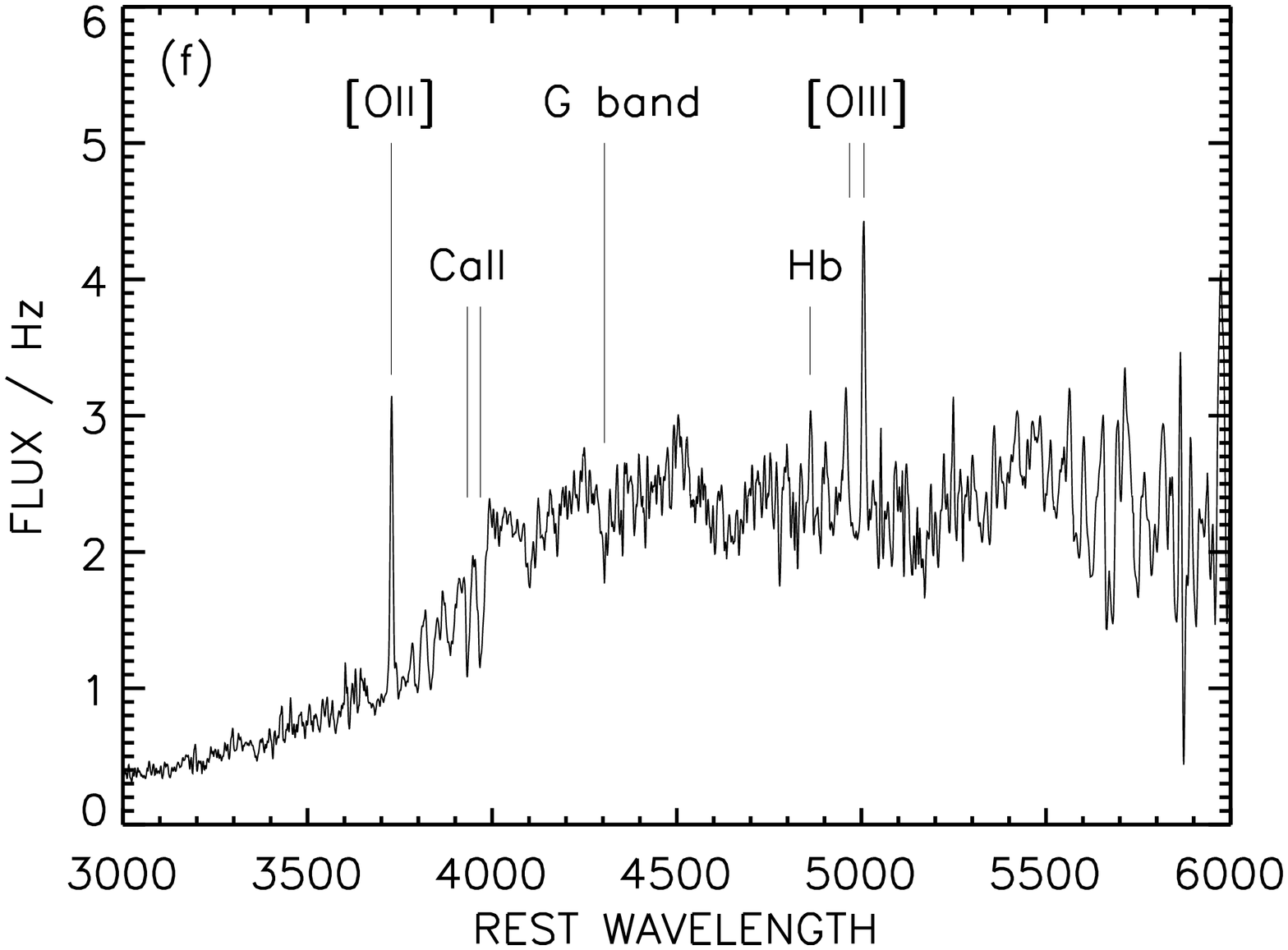,angle=90,width=3in}}
\centerline{\psfig{figure=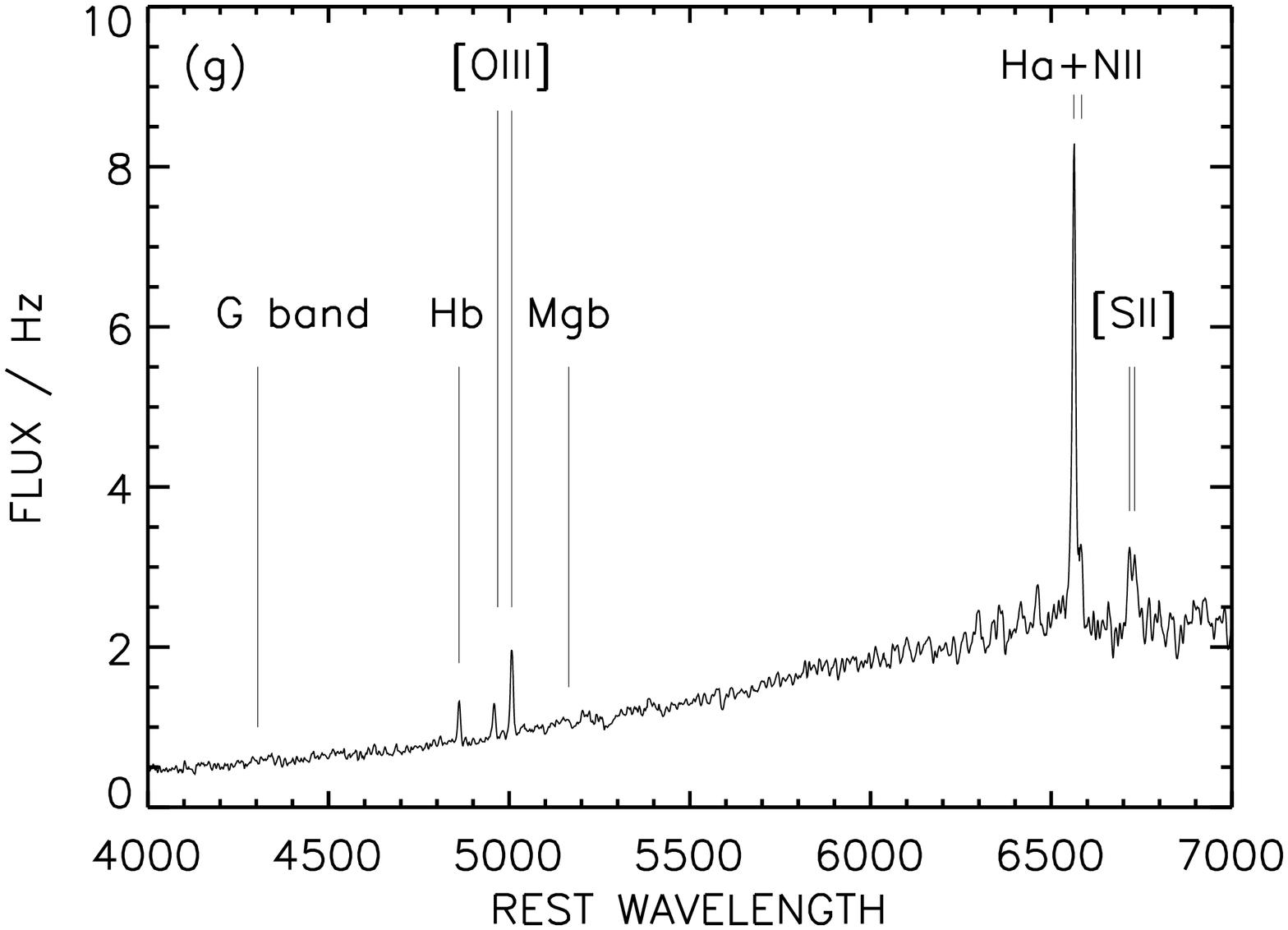,angle=90,width=3in}
\psfig{figure=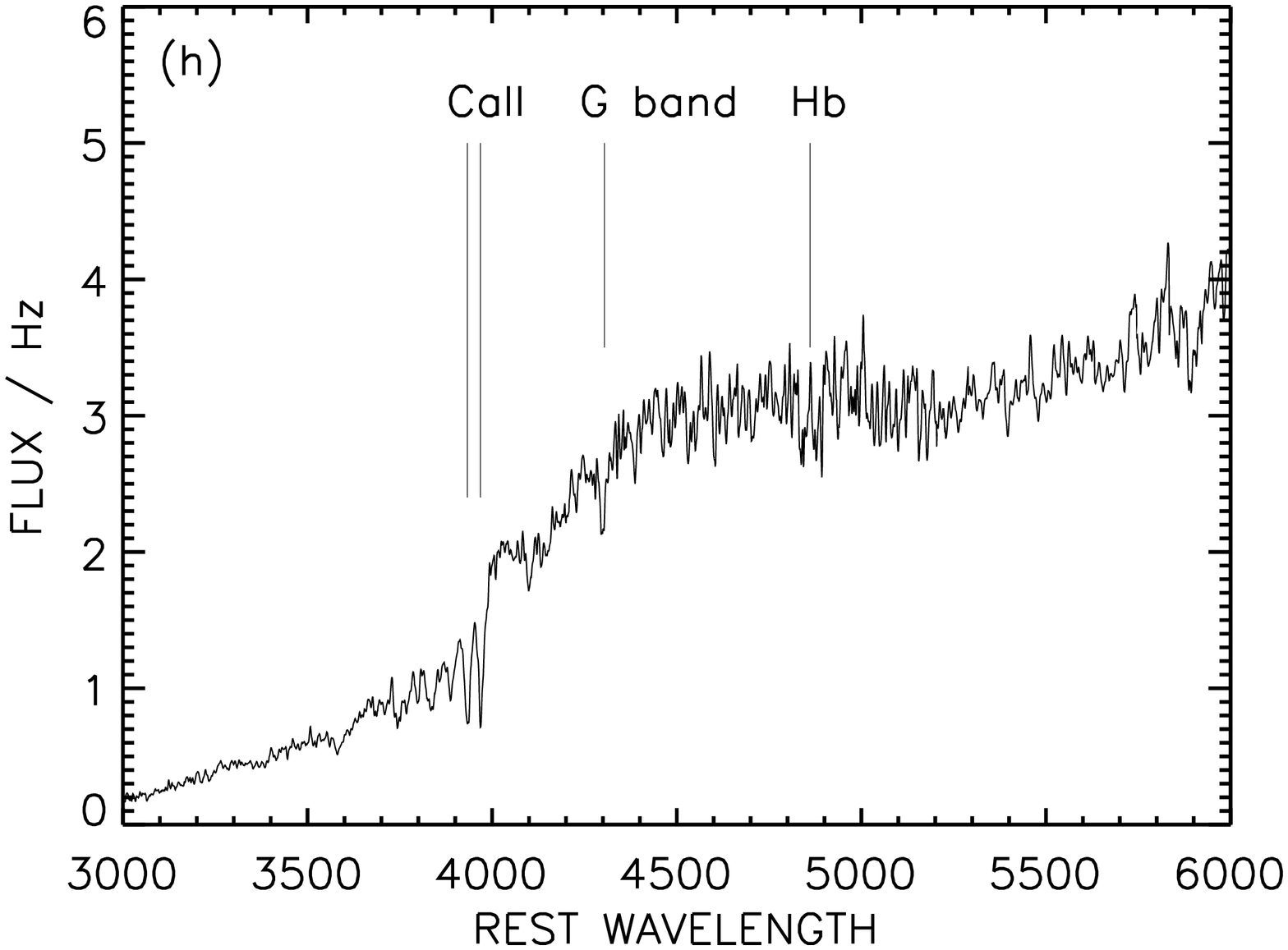,angle=90,width=3in}}
\vspace{7.5pt}
\figcaption{Unweighted averages of the normalized spectra
for eight spectral types:
(a)~broad emission lines in the ultraviolet, 
(b)~broad emission lines in the optical,
(c)~Ly$\alpha$ and strong narrow CIV emission,
(d)~Ly$\alpha$ emission and CIV absorption,
(e)~strong [NeIII], (f)~[OII] emission,
(g)~low redshift objects with H$\alpha$ emission,
and (h)~absorption. The composite spectra are marked
with the emission and absorption lines used to identify
the individual spectra of Fig.~6.
Table~\ref{tab3} lists
the number of galaxies per spectral type for objects
with spectra given in Fig.~6.
\label{fig5}
}
\end{figure*}

In Fig.~\ref{fig4}a (\ref{fig4}b) we plot $B-R$ color versus $2-8$~keV 
($0.5-2$~keV) flux for the CDF-N hard (soft) X-ray sample with $R<25$,
excluding any sources that are saturated.
Within the uncertainties the sources redden at faint X-ray fluxes,
with the effect being most pronounced in the faint hard X-ray 
sources. This can be explained by the host galaxy beginning to
contribute a larger fraction of the total optical 
light at these fluxes relative to that contributed by the AGN.

\section{Submillimeter Observations and Reduction}
\label{secsmm}

SCUBA jiggle map observations at 850~$\mu$m were taken
on the 15~m James Clerk Maxwell Telescope\footnote{The James Clerk Maxwell
Telescope is operated by the Joint Astronomy Centre on behalf
of the UK Particle Physics and Astronomy Research Council, the
Netherlands Organization for Scientific Research, and the
Canadian National Research Council.} during
observing runs in 2000 April and 2001 March and May. We
targeted areas in the CDF-N with large concentrations
of X-ray sources to maximize our submillimeter coverage of
the field at reasonable sensitivities. The data were obtained
in good weather conditions (median $\tau(850\mu$m)=0.282)
for a total integration time of 146.5~ks.
The maps were dithered to prevent any regions of the sky from
repeatedly falling on bad bolometers. The chop throw was
fixed at a position angle of 90~deg so that the negative beams
would appear $45''$ on either side east-west of the positive beam.
Regular ``skydips'' (\markcite{manual}Lightfoot et al.\ 1998)
were obtained to measure the zenith atmospheric opacities,
and the 225~GHz sky opacity was monitored at all times to check
for sky stability. Pointing checks were performed every
hour during the observations on the blazars 0923+392,
0954+685, 1144+402, 1216+487, 1219+285, 1308+326, or 1418+546.
The data were calibrated using jiggle maps of the primary
calibration source Mars or the secondary calibration sources
CRL618, OH231.8+4.2, or 16293-2422.
Submillimeter fluxes were measured using beam-weighted extraction
routines that include both the positive and negative portions of
the beam profile.

The data were reduced using
the dedicated SCUBA User Reduction Facility
(SURF; \markcite{surf}Jenness \& Lightfoot 1998).
Due to the variation in the density of bolometer samples across
the maps, there is a rapid increase in the noise levels at the
very edges, so the low exposure edges were clipped.
The SURF reduction routines arbitrarily normalize all the data
maps in a reduction sequence to the central pixel of the first
map; thus, the noise levels in a combined image are
determined relative to the quality of the central pixel in the
first map. In order to determine the absolute noise levels of
our maps, we first eliminated the $\gtrsim 3\sigma$ real sources
in each field by subtracting an appropriately normalized version
of the beam profile. We then iteratively adjusted the noise
normalization until the dispersion of the signal-to-noise
ratio measured at random positions became $\sim 1$.

The new data were combined in the reduction process with the
jiggle maps previously obtained by
\markcite{bcr00}Barger, Cowie, \& Richards (2000) which
targeted optically faint radio sources.
We also used the \markcite{hughes98}Hughes et al.\ (1998)
HDF-N proper ultradeep map, as rereduced by
Barger et al. We restrict our catalog to the 93~arcmin$^2$
area covered by our SCUBA jiggle maps at better than 5~mJy
sensitivity. The area sensitive to 2.5~mJy is 65~arcmin$^2$,
and the area sensitive to 1.25~mJy is 21~arcmin$^2$.

We first measured the submillimeter
fluxes at the positions of the 20~cm ($>5\sigma$) sources from
\markcite{richards00}Richards (2000) that fell on our
SCUBA maps. Any radio source that we detected above the $3\sigma$
level at 850~$\mu$m was included in a ``detection list''.
We next measured all $>5\sigma$
submillimeter sources that did not have a radio counterpart.
We identified two and added them to the list.
The $>5\sigma$ criterion for the non-radio submillimeter
sources ensures that the sources are real and that their positions
are relatively accurate, while the weaker $>3\sigma$ selection
criterion is appropriate for a pre-selected radio sample.

After compiling the detection list, we compared the coordinates
of our X-ray sources with the coordinates of the sources in the
detection list. When an X-ray source was found to lie within $3''$
of a source in the detection list, we identified the list source
as the counterpart to the X-ray source.
No {\it Chandra} X-ray source was within $10''$ of either of the
two $>5\sigma$ submillimeter sources without radio counterparts,
while all of the radio-detected
$>3\sigma$ submillimeter sources identified
as counterparts to X-ray sources were within $1.5''$ of the X-ray
source positions; thus, the identifications do not depend
critically on the choice of search radius. Once the
cross-identifications were made, the sources in the detection
list were removed from the SCUBA maps since the wide and complex
beam patterns of the bright SCUBA sources can produce spurious
detections at other positions.
Submillimeter fluxes and uncertainties were then measured at all
the unassigned X-ray positions using a recursive loop which
removed X-ray sources that were detected in the submillimeter
above the $3\sigma$ level in a descending
order of significance prior to remeasuring the fluxes. This
procedure again avoids multiple or spurious detections of a
single bright submillimeter source.

The submillimeter fluxes and uncertainties are given in
column~4 of Table~1. The interpretation of the submillimeter
data is presented in \markcite{barger01b}Barger et al.\ (2001b).

\section{Radio Observations}
\label{secradio}

We use the \markcite{richards00}Richards\ (2000) source catalog
of the very deep 20~cm VLA map centered on the HDF-N. The image
covers a $40'$ diameter region with an effective resolution of
$1.8''$ and a $5\sigma$ completeness limit of $40~\mu$Jy.
The absolute radio positions are known to $0.1''-0.2''$ r.m.s.
X-ray sources were identified with radio sources if there
were a $5\sigma$ radio source within $2''$ of the X-ray
position. The use of the radio data in black hole studies is 
discussed in \markcite{barger01c}Barger et al.\ (2001c) and H01.

\section{Spectroscopic Observations and Reduction}
\label{secspectra}

The region surrounding the HDF-N has been the subject of
a number of spectroscopic campaigns, and many of the
optical sources have known spectroscopic redshifts 
(e.g., \markcite{cohen00}Cohen et al.\ 2000;
\markcite{dawson01}Dawson et al.\ 2001). Additional redshifts 
have been obtained of the X-ray sample itself 
(H01; A. E. Hornschemeier et al., in preparation)
with the Hobby-Eberly Telescope and with the Low Resolution 
Imaging Spectrograph (LRIS; \markcite{oke95}Oke et al.\ 1995) on 
the Keck\footnote{The W.~M.~Keck Observatory is operated as a 
scientific partnership among the California
Institute of Technology, the University of California, and NASA,
and was made possible by the generous financial support of the
W.~M.~Keck Foundation.} 10~m telescopes. 
We have obtained our own spectra for as many of the X-ray sources 
as possible so that we can fully characterize the sources using a 
homogeneous dataset. In our final spectroscopic sample there 
are only 7 sources for which we do not have our own spectra,
all of which are taken from the
\markcite{cohen00}Cohen et al.\ (2000) compilation.

%
%
\begin{deluxetable}{cc}
\renewcommand\baselinestretch{1.0}
\tablenum{3}
\tablewidth{0pt}
\tablecaption{Number of X-ray Sources Per Spectral Type for Objects with
Spectra in Fig.~\ref{fig6}}
\small
\tablehead{Class & Number}
\startdata
Stars & 12 \cr
Broad-line emission (ultraviolet or optical) & 30  \cr
Ly$\alpha$ + CIV emission & 9 \cr
Ly$\alpha$ emission + CIV absorption & 2 \cr
[NeIII] emission & 22 \cr
[OII] or H$\alpha$ emission & 67 \cr
Absorption & 33 \cr
\enddata
\label{tab3}
\end{deluxetable}

Spectra were primarily obtained with LRIS in multi-slit mode.
For sources with $I\le 24.5$ counterparts within
a $2''$ radius, we positioned the slits at the optical centers.
For the remaining sources we positioned the slits at the
X-ray centroid positions. We used $1.4''$ wide slits throughout.
Initially we used the 300~lines~mm$^{-1}$
grating blazed at 5000~\AA, which gives a
wavelength resolution of $\sim 16$~\AA\ and a wavelength coverage of
$\sim 5000$~\AA. The wavelength range for each object depends on the
exact location of the slit in the mask but is generally between
$\sim 5000$ and 10000~\AA. After the blue side of
LRIS was implemented, we used the 400~lines~mm$^{-1}$
grating on the red side and the 600~lines~mm$^{-1}$
grism on the blue side, split by the 560 dichroic.
This gives a slightly higher resolution spectrum ($\sim 12$~\AA)
and nearly complete wavelength coverage from $\sim3500$ to 10000~\AA.
The observations were taken with exposure times of 1 to 1.5~hr
per slit mask. Each exposure was broken into three subsets
with the objects stepped along the slit by $2''$ in each
direction. The sky backgrounds were removed using the medians of
the images to avoid the difficult and time-consuming problems of
flat-fielding LRIS data. Fainter objects were observed a number
of times with the maximum exposure time for an individual source
around 6~hours. Details of the spectroscopic reduction
procedures can be found in \markcite{cowie96}Cowie et al.\ (1996).

Additional observations of X-ray sources with bright optical
counterparts were obtained with the multifiber
wide-field HYDRA spectrograph (\markcite{barden94}Barden et al.\ 1994)
on the WIYN\footnote{The WIYN Observatory is a joint facility
of the University of Wisconsin, Indiana University, Yale University,
and the National Optical Astronomy Observatory.} 3.5~m telescope
on UT 2001 February 19--20 and UT 2002 February 6. 
A 316~lines~mm$^{-1}$ grating was used to obtain a wavelength
range of $4900-10200$~\AA\ with a resolution of approximately 
$2.6$~\AA~pixel$^{-1}$.  
The red bench camera was used with the $3''$ `red'
HYDRA fibers to maximize the sensitivity at longer wavelengths.  
To obtain a wavelength solution for each fiber, CuAr comparison
lamps were observed in each HYDRA configuration.
Our HYDRA masks were designed to maximize the number of optically bright
sources in each configuration and to minimize the amount of overlap
between configurations. Fibers that were unable to be placed on a
source were assigned to a random sky location. We observed 5.3~hours 
on each of two HYDRA configurations in 2001 and 93~mins 
on a third configuration in 2002. 
To remove fiber-to-fiber variations, on-source observations were 
alternated with $\pm 7.5''$ offset `sky' exposures taken with the 
same exposure time. This effectively reduced our on-source
integration time to 2.6~hours for each configuration in 2001 and
to 53~mins for the configuration in 2002.

%
%
\begin{inlinefigure}
\psfig{figure=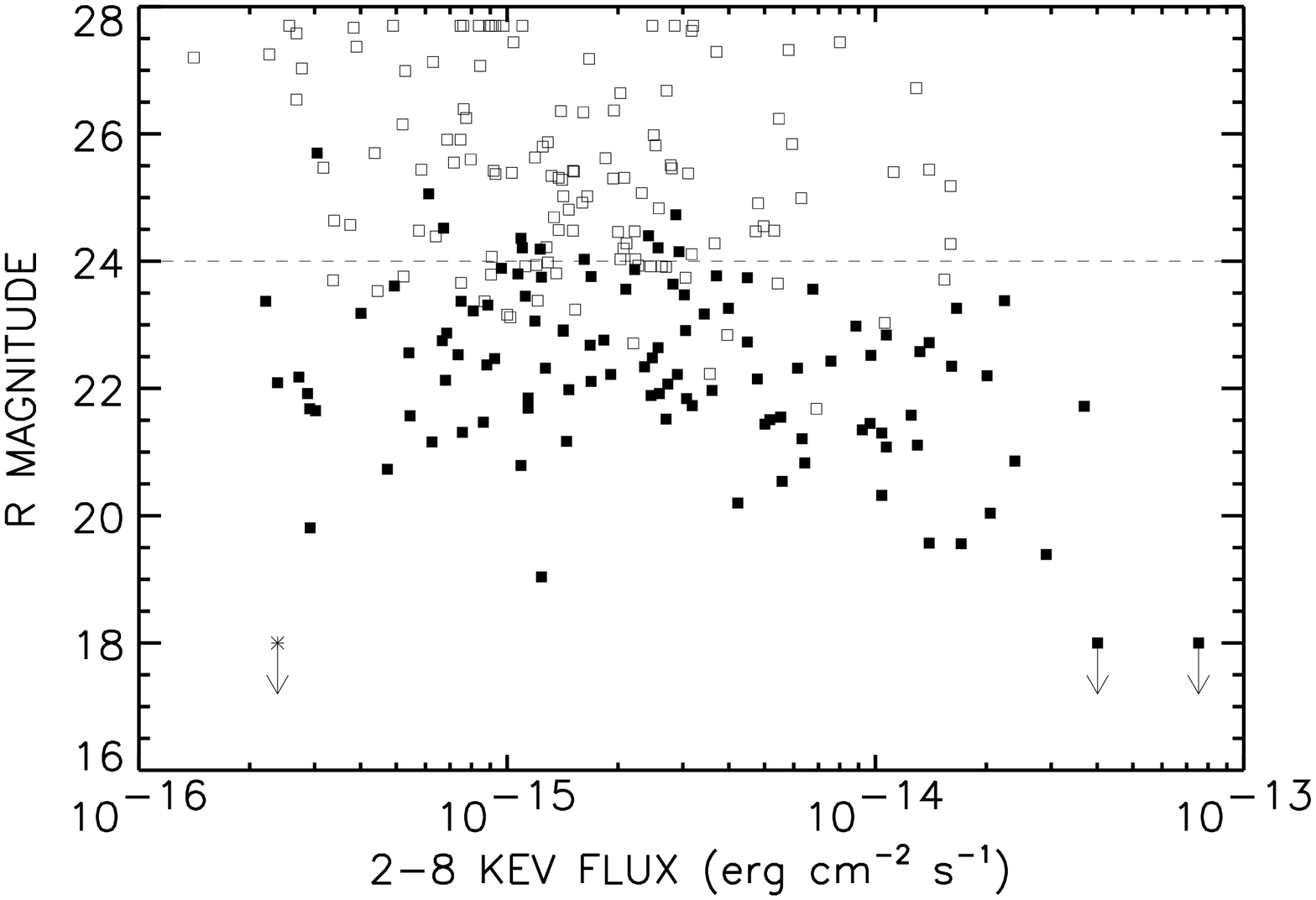,angle=90,width=3.5in}
\vspace{6pt}
\figurenum{7}
\caption{
$R$ magnitude versus $2-8$~keV flux for the hard X-ray sources
within a $10'$ radius of the approximate X-ray image center.
Solid (open) squares denote sources with (without) redshifts.
The dashed line shows the approximate $R$ magnitude to which
the sources can be spectroscopically identified with deep 
10~m telescope observations ($R=24$). The star is shown as an asterisk.
Objects saturated in $R$ are shown
at a nominal magnitude of $R=18$ with downward pointing arrows. 
Sources fainter than the $4\sigma$ limit of $R=27.7$
are plotted at this magnitude limit.
\label{fig7}
}
\addtolength{\baselineskip}{10pt}
\end{inlinefigure}

Reductions were performed with the standard IRAF\footnote{IRAF
is distributed by the National Optical Astronomy Observatories, which
are operated by the Association of Universities for Research in
Astronomy, Inc., under cooperative agreement with the National Science
Foundation.} task DOHYDRA. To optimize
sky subtraction, we performed a two-step process.  In
the first step, the DOHYDRA routine was used to create an average sky
spectrum using the fibers assigned to random sky locations. This
average sky spectrum was then removed from all of the remaining fibers.
In the offset images, this step effectively removed all of the sky
signal leaving behind only residuals caused by differences among the
fibers.  To remove these variations, we then subtracted the residuals
present in the sky-subtracted offsets from the sky-subtracted
on-source spectra.  We found this method to be very effective at
removing the residuals created by fiber-to-fiber varations with HYDRA.

\section{Spectroscopic Redshifts}
\label{secz}

Only spectra which could be confidently identified based on
multiple emission and/or absorption lines were included in the
sample. In Fig.~\ref{fig5} we show unweighted averages of our
normalized spectra to illustrate the lines used to
identify the different spectral types.
Given the positional uncertainties discussed in \S~\ref{secopt},
we have only cross-identified X-ray sources with spectroscopic
counterparts if the radial offsets are $\le 2''$ (see columns~13 
and 14 of Table~1). These redshifts are given in the lower left 
corner of each thumbnail image in Fig.~2 and are summarized in Table~1.
Objects where spectra were obtained but no identification could
be made are marked `obs' for observed in column~15 of Table~1.
In 13 cases we have spectroscopically identified an object
that lies outside our chosen cross-identification radius but 
still within $5''$ of the X-ray position. We use fainter 
text in Fig.~2 to give the redshifts for these nearby
sources for interest; however, apart from two special cases,
we do not give these redshifts in Table~1 nor do 
we use them in our subsequent analysis.
The two special cases are X-ray sources 218 and 290,
which appear to lie within the outer envelopes
of large extended galaxies with $R_{iso}<19$. 
Since there is very little probability that such bright galaxies 
would randomly fall so close to an X-ray source, we
include the redshifts for these two sources in Table~1
and use them in our subsequent analysis.

With this selection 182 of the 370 X-ray sources (49\%)
have spectroscopic identifications, including 12 stars.
The 175 identifiable spectra from the present work
(154 galaxies and 6 stars from Keck,
9 galaxies and 6 stars from WIYN)
are shown in Fig.~\ref{fig6}. (Redshift and B01 catalog number
are given at the base of each spectrum. The WIYN spectra are
denoted by the letter `W' and are not flux calibrated.)
The remaining 7 redshifts are taken from the compilation
of \markcite{cohen00}Cohen et al.\ (2000).
In the last column of Table~1 we denote with the letter
`B' sources with broad emission lines in their spectra (see
Figs.~\ref{fig5}a, b). There are 30 such sources in the sample.
Table~\ref{tab3} lists
the number of galaxies per spectral type (see Fig.~\ref{fig5})
for objects with spectra given in Fig.~\ref{fig6}.

Outside a $10'$ radius from the approximate {\it Chandra} image center
the X-ray point spread function degrades rapidly, so we do not include 
these sources in our subsequent figures or analysis.
Within $10'$, which corresponds to a 294~arcmin$^2$ area on 
the X-ray image, there are 330 {\it Chandra} sources, 
including 10 stars. {\it Hereafter, we refer to this as our
X-ray sample.} Of the 189 galaxy sources with $R\le 24$,
147 (78\%) have been spectroscopically identified.
The great majority of the identified extragalactic sources 
with $R\le 24$ lie below a redshift of one (104/147 or 71\%).
In Fig.~\ref{fig7} we illustrate for the hard X-ray sample how 
most sources with $R\le 24$ can be spectroscopically identified
while only a few can be identified at fainter optical magnitudes.

%
%
\begin{inlinefigure}
\psfig{figure=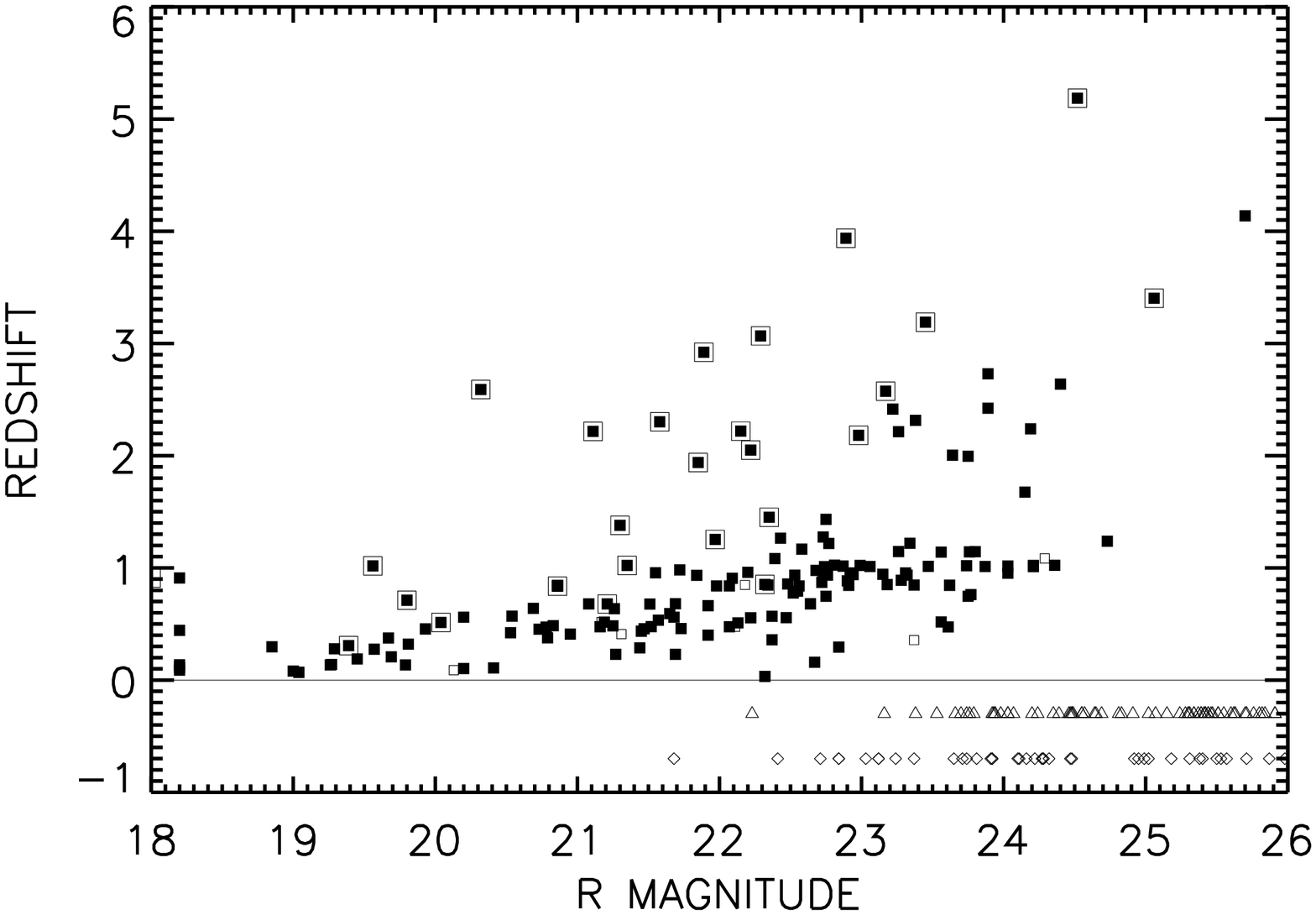,angle=90,width=3.5in}
\vspace{6pt}
\figurenum{8}
\caption{
Redshift versus $R$ magnitude for the
X-ray sources, excluding stars, within a $10'$ radius
of the approximate X-ray image center.
Objects saturated in $R$ are shown at a nominal magnitude of 18.2.
Objects with spectra
in Fig.~6 are denoted by solid squares, and the 7 with
redshifts from the literature are denoted by open squares.
Objects with broad lines are enclosed in a second larger symbol.
Objects without redshift identifications are shown at either $z=-0.3$
(open triangles for objects within a $6.5'$ radius of the
approximate X-ray image center) or $z=-0.7$ (open diamonds for objects
between radii of $6.5'$ and $10'$).
\label{fig8}
}
\addtolength{\baselineskip}{10pt}
\end{inlinefigure}

In Fig.~\ref{fig8} we show the redshift-magnitude relation for 
the galaxy population (squares; unidentified objects are plotted
below $z=0$ with different symbols). 
Broad-line sources (distinguished by a second larger
symbol) are systematically the most optically luminous of the
X-ray sources because of the AGN contribution to the light.
The gap between $z\sim1.5$ and 2 reflects the
difficulty of identifying sources with redshifts in this range,
where [OII]~3727~\AA\ has moved out of the optical window
and Ly$\alpha$~1216~\AA\ has not yet entered.

All of the sources with $z>1.6$ are either broad-line
AGN or have narrow Ly$\alpha$ and/or 
CIII]~1909~\AA\ emission. There are 14 broad-line AGN
versus 11 narrow-line emitters at these redshifts, 
but this may reflect selection bias in the spectroscopic 
identifications since the broad-line AGN are generally brighter 
and easier to identify. Most of the narrow-line sources
also have CIV~1550~\AA\ in emission, but two have 
CIV~1550~\AA\ in absorption.

%
%
\begin{inlinefigure}
\psfig{figure=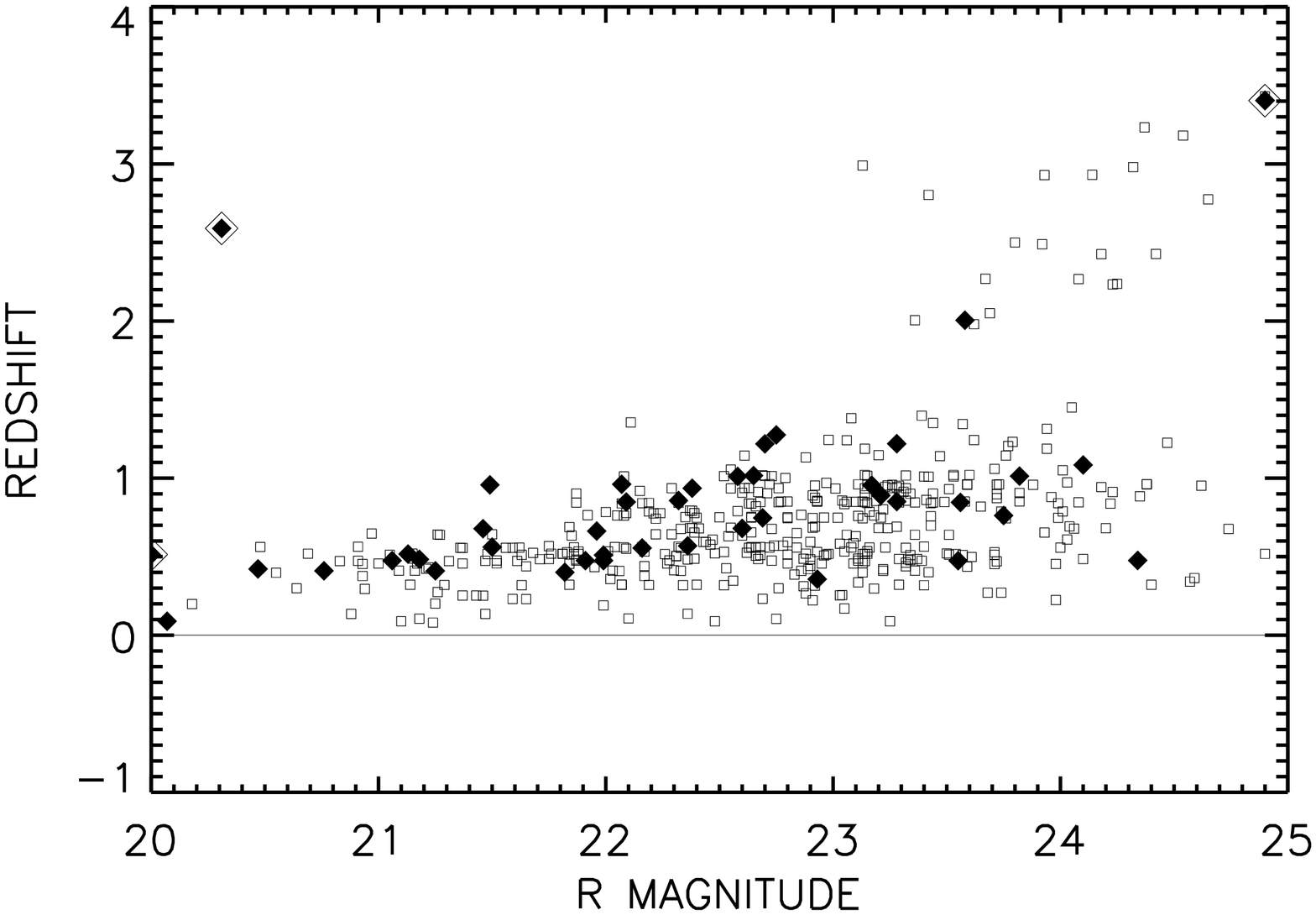,angle=90,width=3.5in}
\vspace{6pt}
\figurenum{9}
\caption{Redshift versus $R$ magnitude for
optically selected sources in a $6'\times 6'$ region
centered on the HDF-N (open squares). Sources with a counterpart
in the X-ray sample are denoted by solid diamonds, and
broad-line AGN are enclosed with a second larger symbol.
\label{fig9}
}
\addtolength{\baselineskip}{10pt}
\end{inlinefigure}

Intriguingly, while there are a substantial number of optically 
identified absorption-line sources with $z>1.6$ known in the 
HDF-N (e.g., see \markcite{brandt01b}Brandt et al.\ 2001b
and references therein) and flanking fields, none of these are 
present in the X-ray sample. We illustrate this in Fig.~\ref{fig9} 
where we show the redshift-magnitude diagram for an optically
selected sample in a $6'\times 6'$ region surrounding
the HDF-N. In this region 90\% of the 435 sources with
$R\le 23.5$ have been spectroscopically identified (open squares).
The sources with a counterpart in the X-ray sample are denoted
by solid diamonds.  All three of the high-redshift sources with 
X-ray emission have Ly$\alpha$ in emission (two of these are 
broad-line AGN), which suggests that only this type of object 
has X-ray counterparts at our current X-ray sensitivity.
Stacking analyses may show a weaker signal below the sensitivity
of the individual detections 
(\markcite{brandt01b}Brandt et al.\ 2001b;
\markcite{nandra02}Nandra et al.\ 2002).

Roughly a third of the low-redshift sources are
detected at 20~cm, and 3 out of the 11 high-redshift
narrow emission line sources are detected. However,
none of the high-redshift broad-line AGN are detected
at 20~cm.

%
%
\begin{inlinefigure}
\psfig{figure=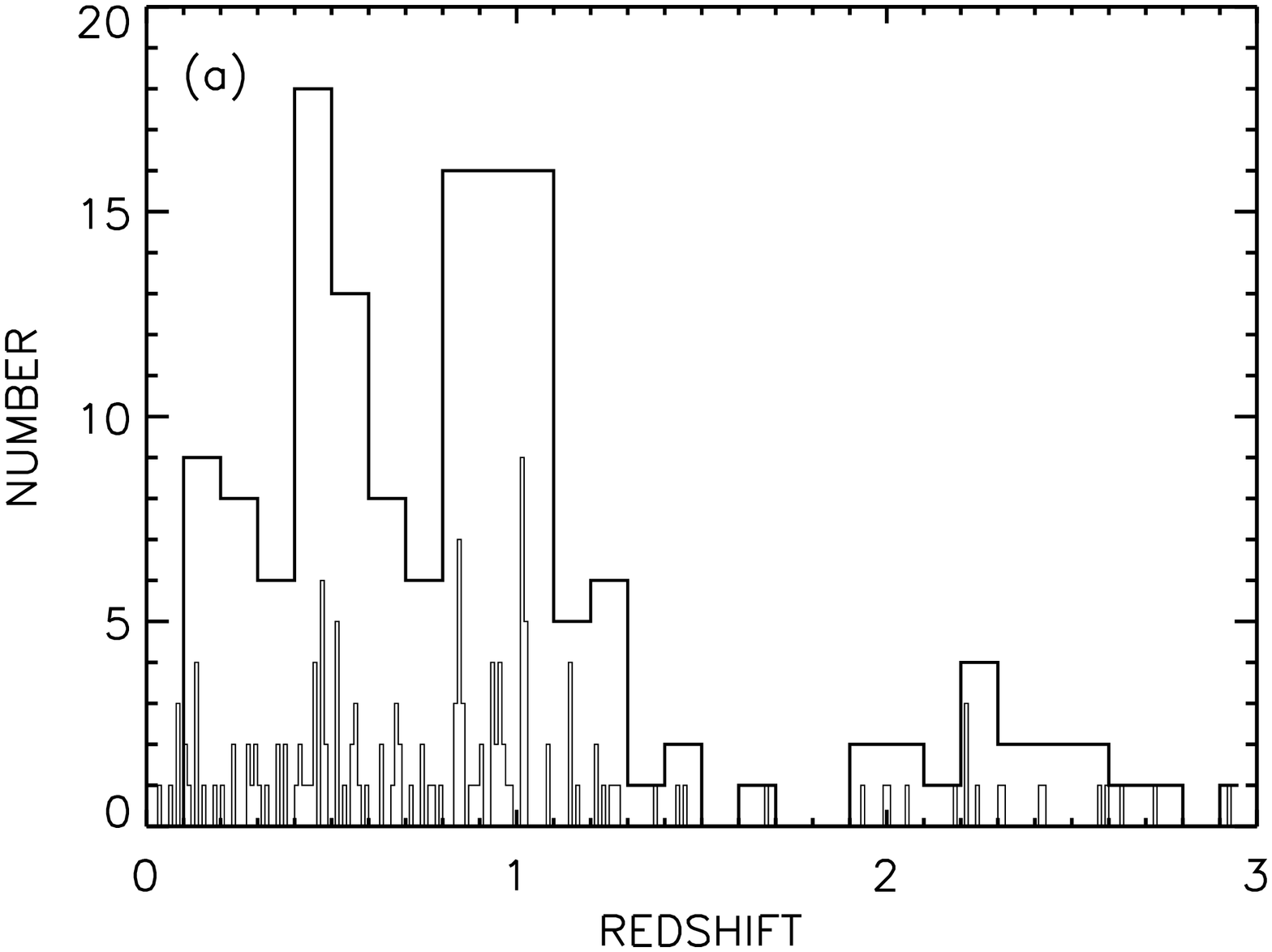,angle=90,width=3.5in}
\psfig{figure=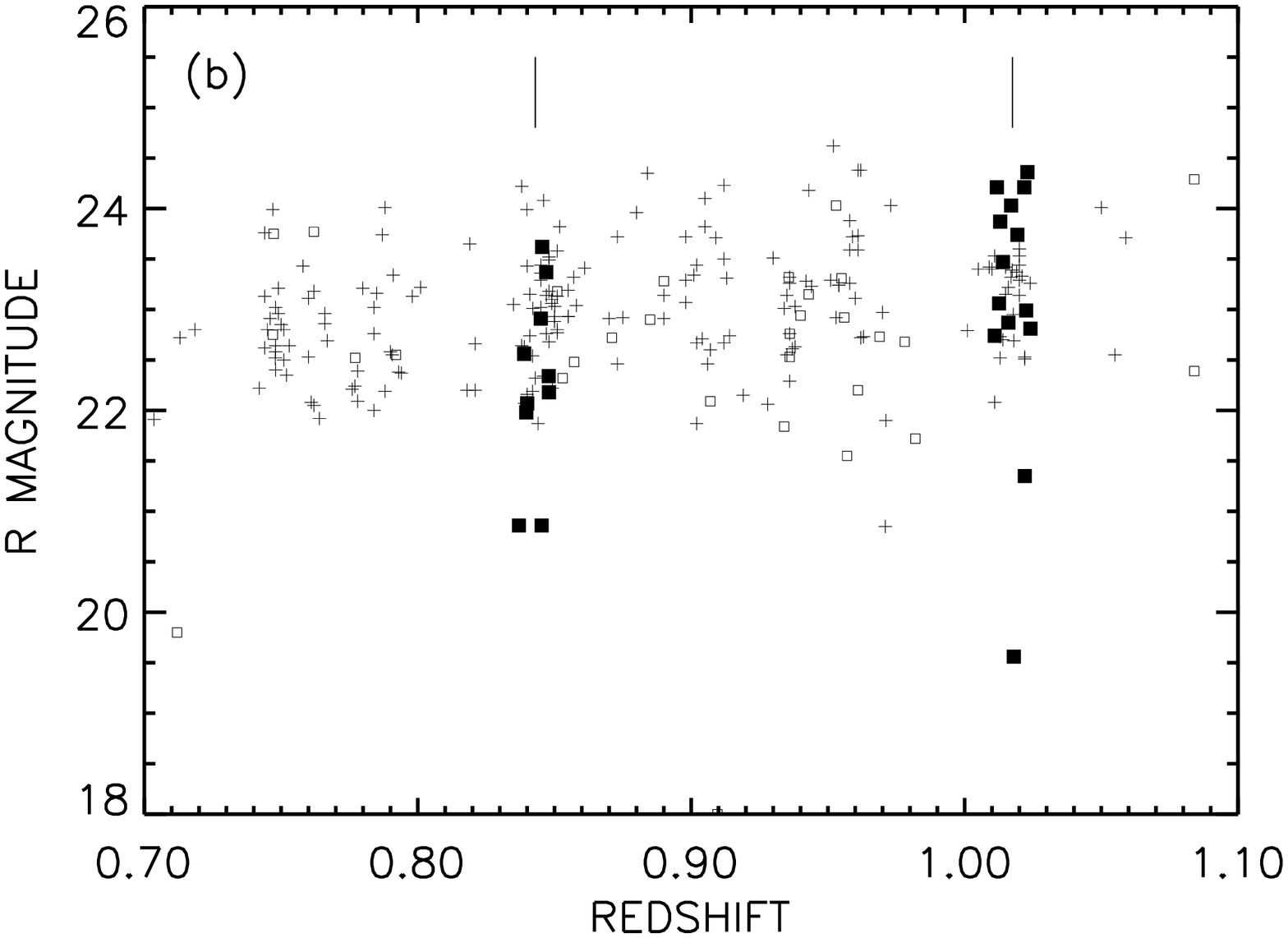,angle=90,width=3.5in}
\vspace{6pt}
\figurenum{10}
\caption{
(a)~Redshift distribution for two bin sizes ($\Delta z=0.1$ and
$\Delta z=0.01$) for sources within a $10'$ radius
of the approximate X-ray image center. (b)~$R$ magnitude
versus redshift blow-up
shows the two structures centered at $z=0.843$ and $z=1.0175$.
Sources within 1000~km~s$^{-1}$ of the center
positions are denoted by solid squares, other X-ray sources
are denoted by open squares, and field galaxies are denoted
by plus signs.
\label{fig10}
}
\addtolength{\baselineskip}{10pt}
\end{inlinefigure}

In Fig.~\ref{fig10}a we show the histograms of the redshift 
distribution of our X-ray sample for two bin sizes 
($\Delta z=0.1$ and $\Delta z=0.01$).
The lower resolution binning
suggests an excess of objects in two broad redshift
bins, which is also evident in the higher resolution binning. 
In Fig.~\ref{fig10}b we show an $R$ magnitude versus redshift 
blow-up of the two structures centered on $z=0.843$
and $z=1.0175$ that are seen in the higher resolution histogram.
Sources within 1000~km~s$^{-1}$ of the center 
positions are denoted by solid squares,
other X-ray sources are denoted by open squares, and field galaxies, 
for comparison, are denoted by plus signs. These are the only 
two structures which contain at least 10 X-ray sources with the 
specified velocity width. A smaller structure at $z=0.477$ 
contains 8 X-ray sources. 
Figure~\ref{fig10} suggests that there may be large
scale structure in the field (spatially the sources also look  
clustered), but the structures do not dominate
the number of X-ray sources in the sample so our redshift 
distribution should not be strongly affected by clustering.
However, the number of sources in these structures
(a total of 24 identified sources in the $z=0.843$
and $z=1.0175$ structures) is sufficiently large that
it could account for a part of the field-to-field
variation seen in the X-ray number counts 
(e.g., \markcite{cowie02}Cowie et al.\ 2002).
Similar redshift structures have also been
detected in the CDF-S exposure
(\markcite{hasinger02}Hasinger 2002). 

We note that
\markcite{cohen00}Cohen et al.\ (2000) also detected strong redshift
peaks at z=0.848 and z=1.016 in their optical survey of the
HDF-N region. \markcite{dawson02}Dawson et al. (2002) even claimed
the serendipitous discovery of ClG 1236+6215, a galaxy cluster with
redshift z=0.85; however, \markcite{bauer02}Bauer et al.\ (2002)
did not significantly detect diffuse X-ray emission associated
with this cluster.

\section{Colors and Photometric Redshifts}
\label{secphotoz}

Broad-band galaxy colors have been extensively used in recent
years to obtain photometric redshift estimates of galaxies (e.g.,
\markcite{bol00}Bolzonella, Pello, \& Miralles 2000). However,
the spectral energy distributions of X-ray sources
may be rather complex since they arise from both the host galaxy
and the AGN. Moreover, the spectral shape of the AGN relative
to that of the host galaxy may vary depending on the degree of
obscuration. In principle
it might be possible to model this with complex templates, but
here we simply try to determine empirically which of
the sources in our X-ray sample can have their redshifts estimated
from the color information using standard galaxy templates.
It is precisely for the cases where the AGN is obscured (and
hence a spectroscopic redshift hard to obtain) that the photometric
redshifts are of most interest, and it is in these cases that
the photometric redshift estimates may be expected to work best.

We begin by comparing color-color diagrams with
the tracks with redshift that would be followed by a
given galaxy type. In Fig.~\ref{fig11}a we show $B-R$ versus
$R-I$ for the 269 sources in our X-ray sample
with $I<25$, and in Fig.~\ref{fig11}b we show $B-I$ versus
$I-HK^\prime$ for the 208 sources
with $HK^\prime<21.5$.
(In both cases we have excluded sources saturated in the optical
images or contaminated by a nearby bright galaxy.)
In Fig.~\ref{fig11}a circles (squares) denote galaxies with
$z<0.65$ ($z>0.65$); in Fig.~\ref{fig11}b, circles (squares) denote
$z<1$ ($z>1$). The redshifts $z=0.65$ and $z=1$ correspond to
where the 4000~\AA\ break passes through the $R$ and $I$-bands,
respectively. We have distinguished sources with broad emission lines
in their observed spectra (which may be expected to be dominated by
the AGN light) by enclosing them in a second larger symbol.
Spectroscopically unidentified galaxies are shown as plus signs.
We overlay on the plots the
$z=0$ to $z=3$ tracks that would be followed by unevolving (i.e.,
invariant rest-frame spectra) E (top), Sb (middle), and Irr (bottom)
galaxies with spectral shapes taken from \markcite{cww}Coleman, Wu,
\& Weedman (1980).

From Fig.~\ref{fig11}a we see that nearly the whole X-ray sample
lies either in the region populated by the galaxy tracks or in
regions where a dominant AGN would migrate the colors.
The broad-line systems are all extremely blue
and lie in a small region of the $B-R$ versus $R-I$ plane. However,
they substantially overlap with the irregular galaxies, making
it hard to distinguish between low redshift irregulars and luminous
high-redshift AGN purely on the basis of color.
The redder galaxies fall in two well-defined regions.
The reason for the split is the passage of the
$4000$~\AA\ break through the $R$-band at this redshift, which
results in the colors, which were originally moving up the solid
arms of the unevolving galaxy tracks, crossing to the dashed arms
and then migrating down with redshift. There appear to be relatively
few sources with colors as red as an elliptical galaxy at $z>0.65$
but many sources with colors similar to Sb or Sa galaxies. There
are substantial numbers of spectroscopically unidentified galaxies
in both of the well-defined regions and also in the AGN/Irr region, 
but there are only a tiny number outside those areas.

%
%
\begin{inlinefigure}
\psfig{figure=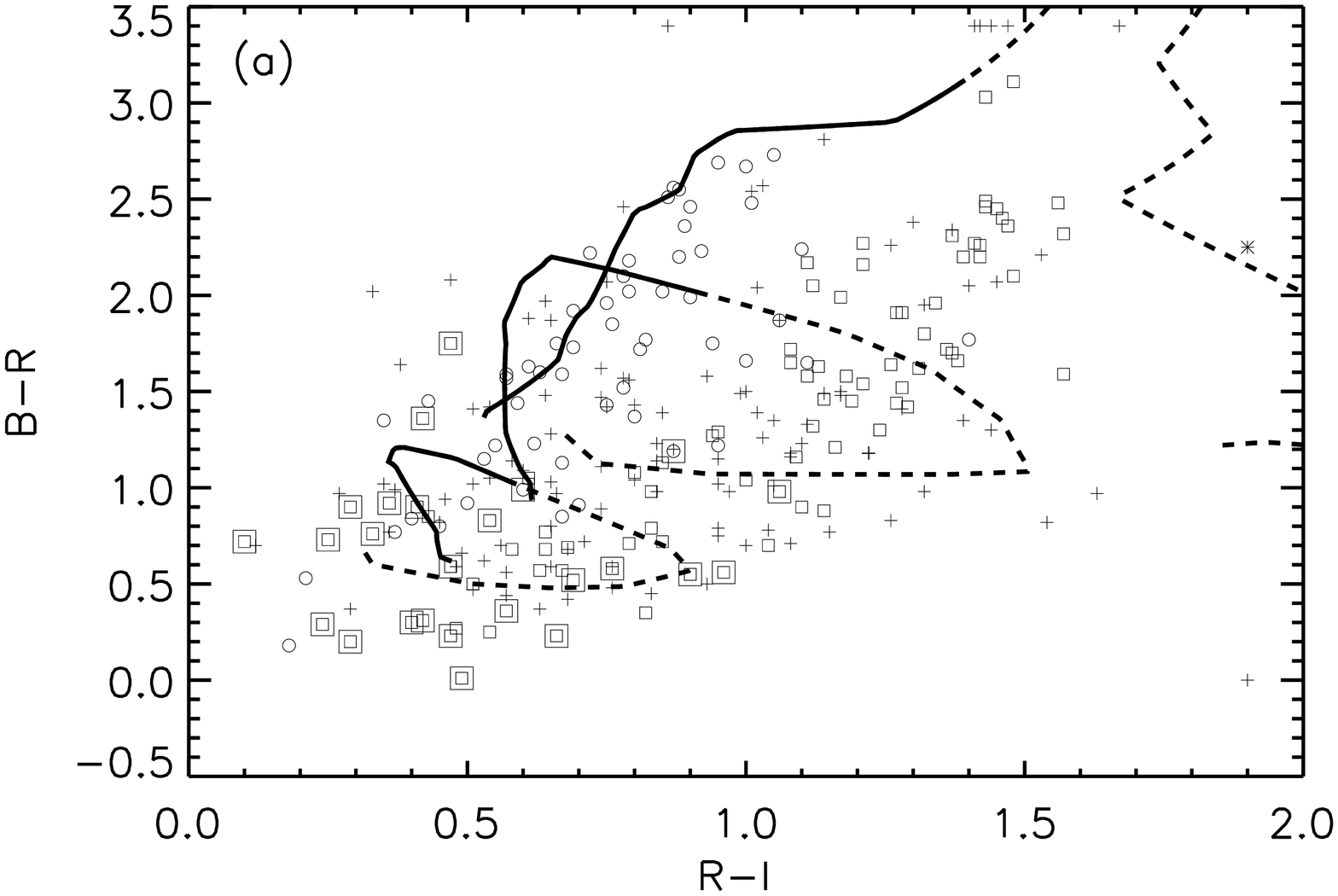,angle=90,width=3.5in}
\psfig{figure=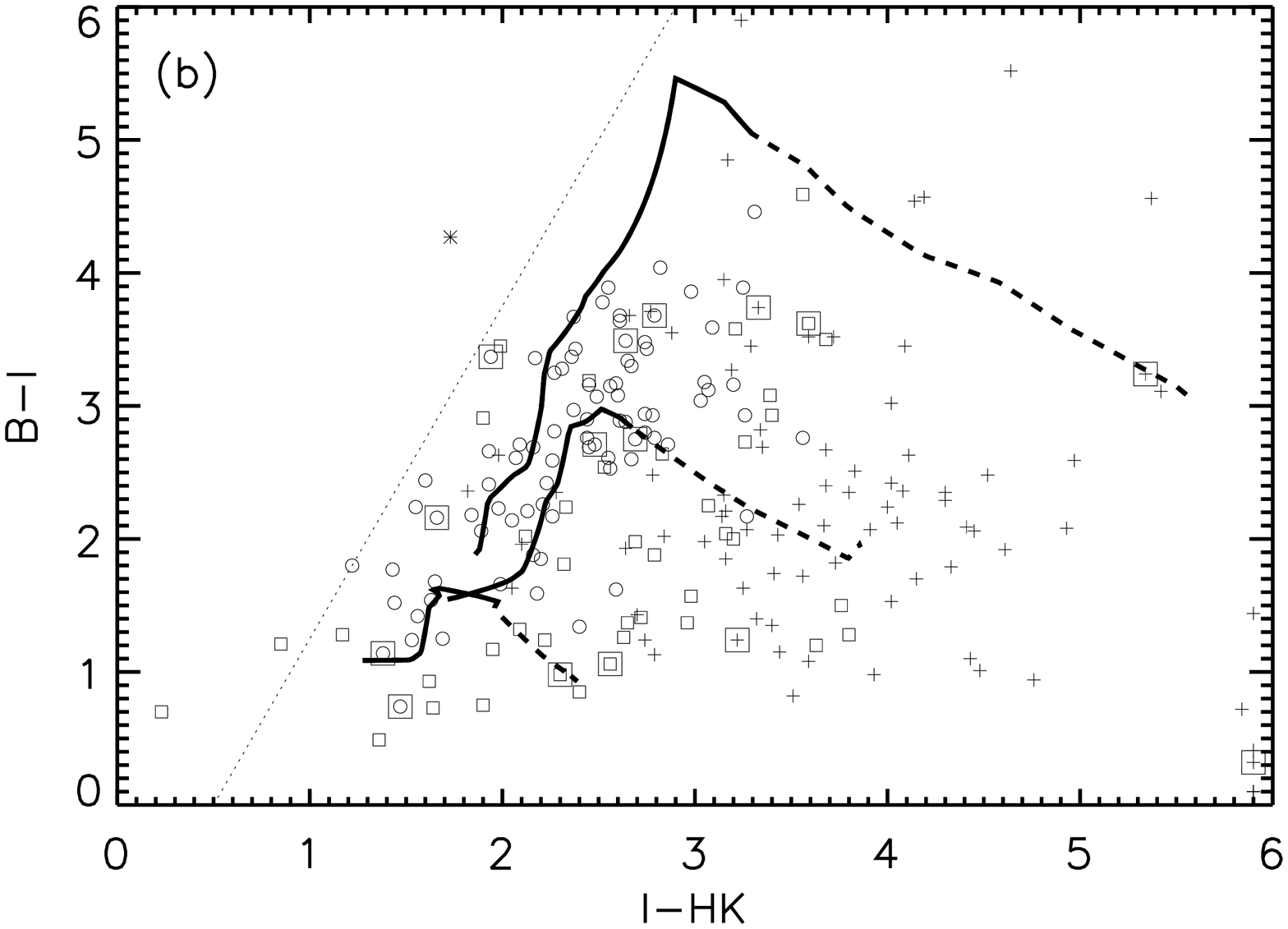,angle=90,width=3.5in}
\vspace{6pt}
\figurenum{11}
\caption{
(a)~$B-R$ versus $R-I$ for the $I<25$ X-ray sources within
a $10'$ radius of the approximate X-ray image center,
excluding the sources that are saturated or contaminated by
a bright neighbor. Plus signs
denote galaxies without redshift identifications. Circles (squares)
denote galaxies with $z<0.65$ ($z>0.65$). Broad-line sources are
enclosed in a second larger symbol.
Objects with $R-I>1.9$ ($B-R>3.4$) are plotted at $R-I=1.9$
($B-R=3.4$).
The overlaid curves show the tracks that would be followed by
unevolving E (top), Sb (middle), and
Irr (bottom) galaxies with spectral shapes from
Coleman, Wu, \& Weedman (1980). The redshift range of the tracks
is $z=0$ to $z=3$, with the $z<0.65$ ($z>0.65$) region
illustrated by a solid (dashed) line.
(b)~$B-I$ versus $I-HK^\prime$ for the $HK^\prime<21.5$ X-ray
sources within the $10'$ radius, excluding the sources that
are saturated or contaminated by a bright neighbor.
The symbols and the galaxy tracks are the same as in (a),
except here the circles (squares) denote galaxies with $z<1$ ($z>1$),
and the solid (dashed) tracks illustrate the region with $z<1$
($z>1$). Objects with $I-HK'>5.9$ ($B-I>5.9$) are plotted at
$I-HK'=5.9$ ($B-I=5.9$).
The dotted line shows the star-galaxy discriminator of
Gardner (1992), modified to the $HK^\prime$ system (Barger et al.\ 1999).
Stars are expected to lie to the left of this line, as is the case for
the one spectroscopically identified star which is not saturated
(asterisk).
\label{fig11}
}
\addtolength{\baselineskip}{10pt}
\end{inlinefigure}

The unidentified sources in Fig.~\ref{fig11}b seem to
preferentially lie at $z>1$, with only a small
number at lower redshifts. This is partially a selection effect
since the higher redshift galaxies have redder NIR colors and
hence are more likely to show up in an $HK'$ magnitude-limited sample.
As in Fig.~\ref{fig11}a, very few of the galaxies lie near the
elliptical track at high redshifts, with most of the sources having Sa/Sb
colors or bluer. There is a substantial spread in the $I-HK^\prime$
colors of the broad-line sources, which may reflect varying host
galaxy contributions in the NIR.
In Fig.~\ref{fig11}b we show the star-galaxy discriminator
(dotted line) used by \markcite{jg92}Gardner (1992) and modified
to the $HK^\prime$ system (\markcite{barger99}Barger et al.\ 1999).
Stars are expected to lie to the left of this line, and it is clear
from the figure that there are very few optically faint
stars in the X-ray sample (most have $I\lesssim 19$),
with only one (spectroscopically identified) star lying in this
region (asterisk).

We next used the Hyperz code of
\markcite{bol00}Bolzonella et al.\ (2000)
and the magnitudes (including $HK^\prime$, where available) of
Table~1 to estimate photometric redshifts. Seven updated
\markcite{bc93}Bruzual \& Charlot (1993) models corresponding to
present-day E, S0, Sa, Sb, Sc, Sd, and Irr galaxies were used, as
well as a single-burst model. These templates cover the entire
range of possible ages. The ratio of the photometric redshift estimate
from Hyperz to the observed spectroscopic redshift is shown versus
$B-I$ color in Fig.~\ref{fig12} for spectroscopically identified
sources which are not saturated in any of the optical bands or
contaminated by neighbors. Broad-line sources are enclosed in a
second larger symbol. As we saw from Fig.~\ref{fig11}, the present
analysis cannot distinguish the AGN from the Irr galaxies.
However, redder than $B-I=1.5$ (vertical dotted line), 
the photometric redshift estimates become
more robust, and most sources have photometric redshifts within
25\% of the spectroscopic redshift, as shown by the 
horizontal dashed lines.

%
%
\begin{inlinefigure}
\psfig{figure=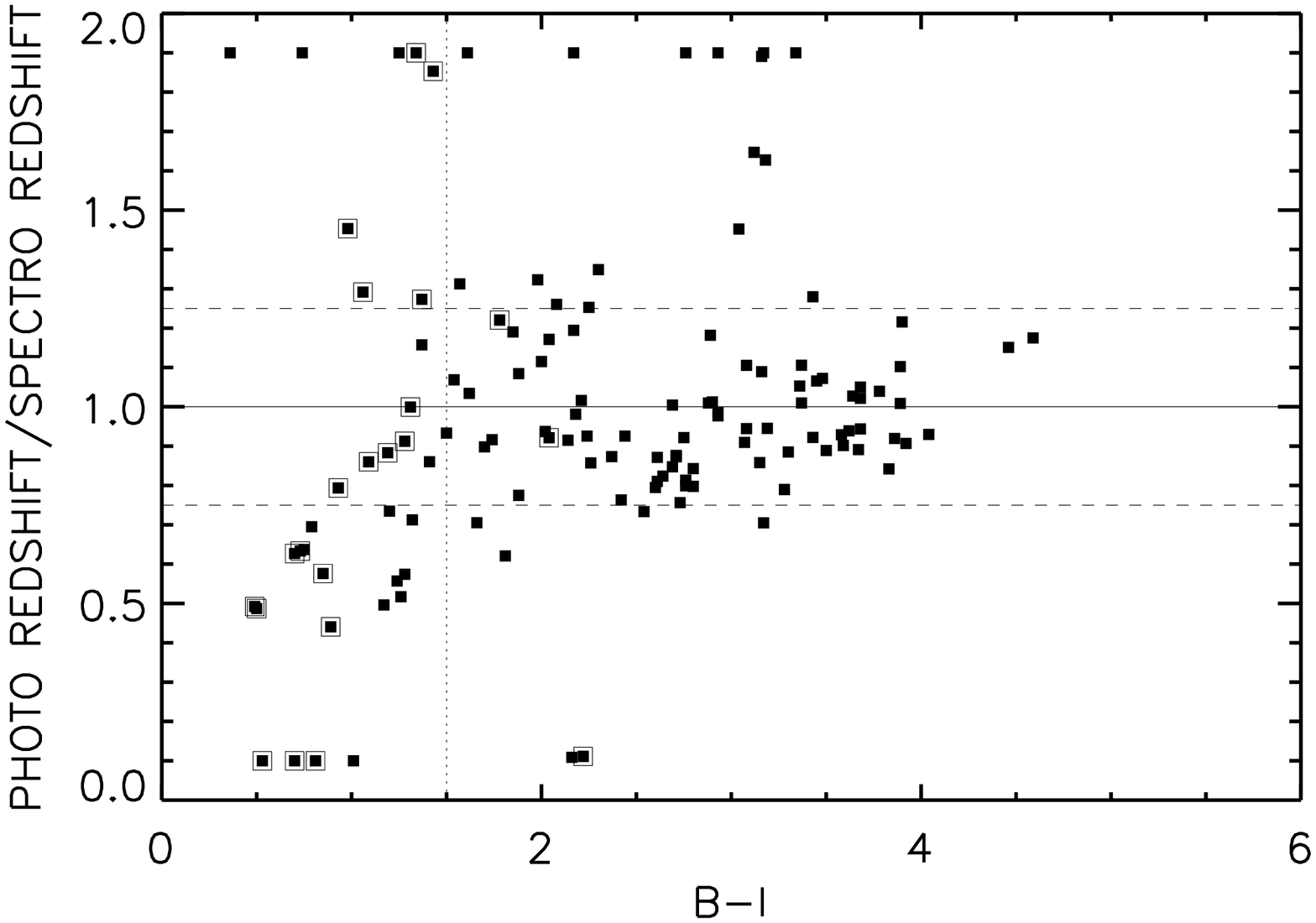,angle=90,width=3.5in}
\vspace{6pt}
\figurenum{12}
\caption{
Ratio of photometric (calculated with the Hyperz program and
the magnitudes in Table~1) to spectroscopic redshifts
versus $B-I$ color for the sources within a $10'$ radius of the
approximate X-ray image center, excluding the sources that are 
saturated or contaminated by a bright neighbor.
Broad-line sources are enclosed with a second larger
symbol. AGN contamination degrades the photometric redshift fits, but
redder than $B-I=1.5$ (vertical dotted line) the photometric redshifts
generally agree with the actual redshifts to within $25\%$
(horizontal dashed lines).
\label{fig12}
}
\addtolength{\baselineskip}{10pt}
\end{inlinefigure}

The use of compactness and apparent magnitude as prior constraints 
can improve the agreement between the photometric and spectroscopic 
redshifts for sources with blue colors since AGN are likely to be 
at high redshifts, compact, and bright, while the Irr galaxies are 
likely to be at low redshifts, extended, and faint. However, even 
with these constraints, the featurelessness of the blue spectra and 
the overlap of the apparent magnitudes means there is still a high 
degree of indeterminacy for the blue objects. 
Ultraviolet data would significantly improve the identification
of these types of objects.

In our subsequent analysis we restrict our use of the photometric 
redshifts to galaxies with $B-I>1.5$ and $I<25$ that are also not 
saturated or near a contaminating bright object.

%
%
\begin{inlinefigure}
\psfig{figure=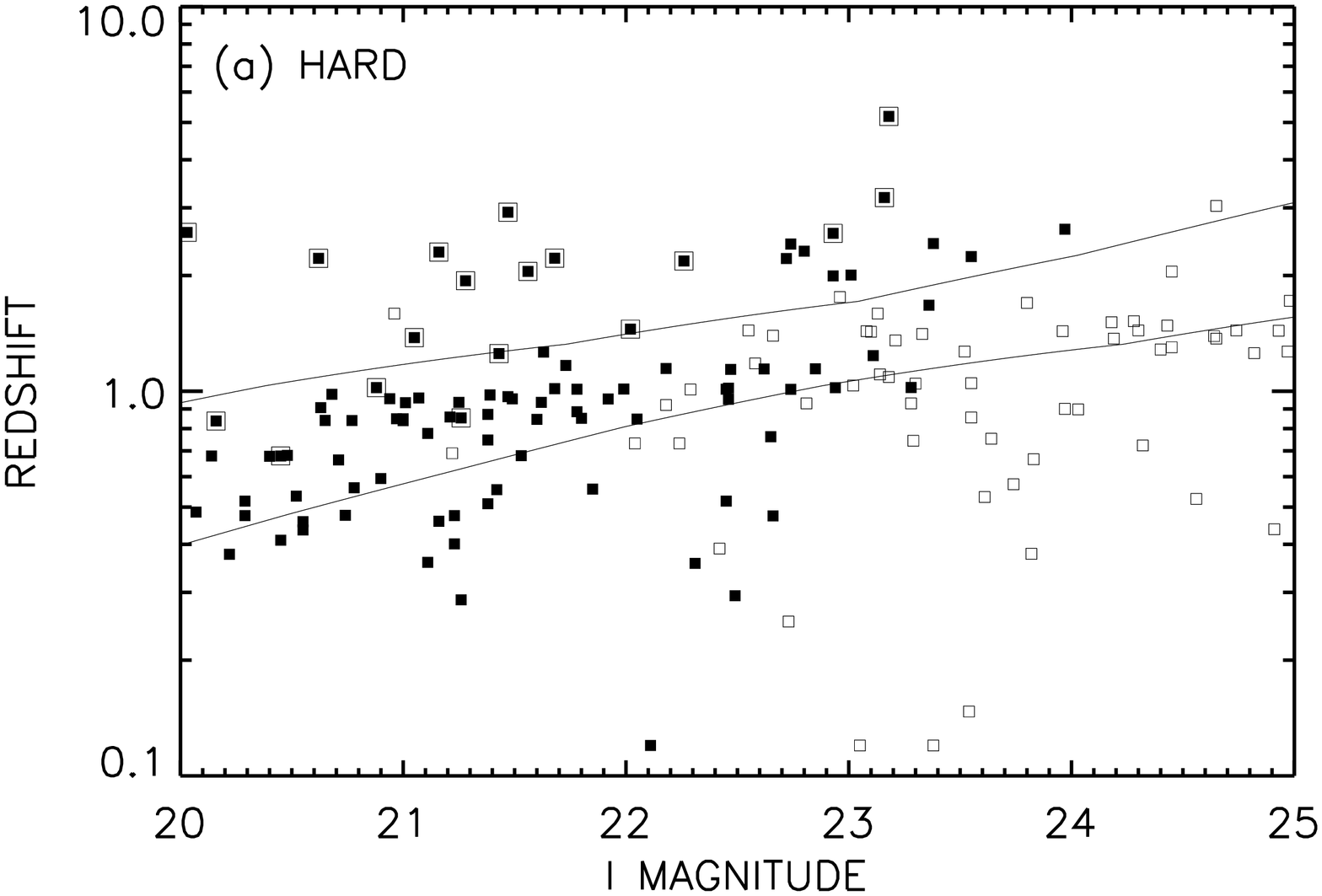,angle=90,width=3.5in}
\psfig{figure=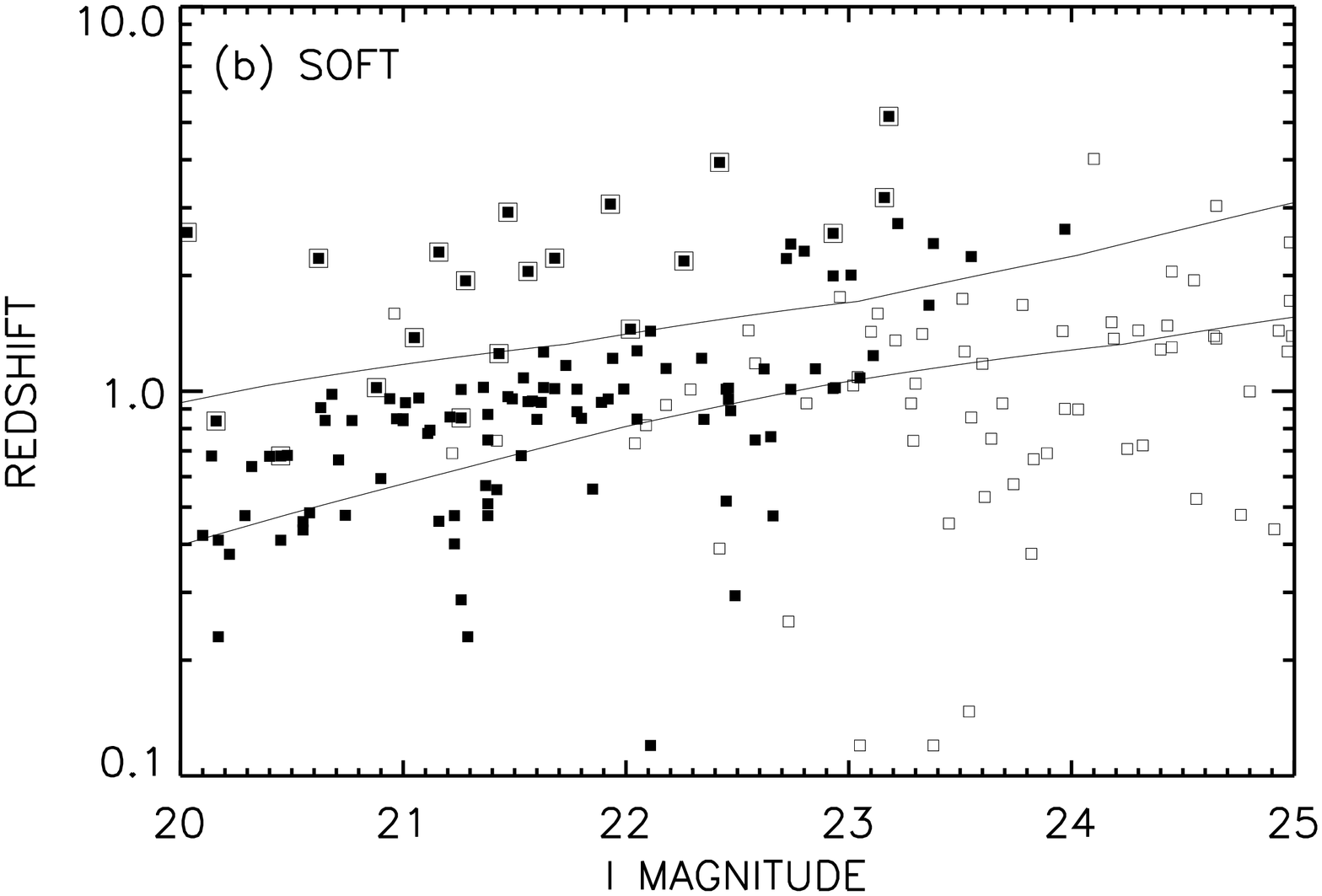,angle=90,width=3.5in}
\vspace{6pt}
\figurenum{13}
\caption{Spectroscopic redshift versus $I$ magnitude (solid squares)
for the (a)~$2-8$~keV and (b)~$0.5-2$~keV X-ray samples within a
$10'$ radius of the approximate X-ray image center.
Broad-line sources are enclosed in a second larger symbol.
The open symbols denote photometric redshifts for
unidentified or unobserved sources with $B-I>1.5$.
In (a)~22 objects bluer than
$B-I=1.5$ without spectroscopic redshifts are omitted from the
plot and in (b)~26.
The solid lines show the tracks of unevolving Sb galaxies with
absolute rest-frame magnitudes $M_I=-22$ (lower tracks) and
$M_I=-24.5$ (upper tracks).
\label{fig13}
}
\end{inlinefigure}

\section{Absolute Magnitudes of the Optical Counterparts}
\label{secabsmag}

In Figs.~\ref{fig13} we show redshift versus $I$ magnitude for
our $0.5-2$~keV and $2-8$~keV samples; we plot only objects
having optical magnitudes between $I=20$ and $I=25$. Sources with
spectroscopic (photometric) redshifts are denoted by solid (open)
symbols. Broad-line sources are enclosed in a second larger symbol.
The solid lines show the expected tracks for an unevolving
Sb galaxy with $M_I=-24.5$ (upper tracks) and $M_I=-22$ (lower tracks).
The $M_B^\ast=-20.4$ galaxy from \markcite{loveday92}Loveday et al.\ (1992)
roughly corresponds to an Sb galaxy with $M_I^\ast=-22.0$.

Of the 236 hard ($2-8$~keV) X-ray sources in our sample, 
177 lie in the $I=20$ to $I=25$
range, of which 155 (97 spectroscopic and 58 photometric) are shown
(22 $B-I<1.5$ objects are omitted). Of these, 29 are brighter
than $M_I=-24.5$ and 113 are brighter than $M_I^\ast=-22$.
Of the 288 soft X-ray sources in our sample, 
205 lie in the $I=20$ to $I=25$ range, of which 179
(115 spectroscopic and 64 photometric) are shown
(26 $B-I<1.5$ objects are omitted). Of these,
34 are brighter than $M_I=-24.5$ and 134 are brighter than
$M_I^\ast=-22$. Thus, most of the galaxies in both the hard and soft
samples are comparable to or more luminous than $M_I^\ast$.

\section{X-ray Properties}
\label{secrestrict}

In our analysis of the X-ray source properties, we consider
two restricted uniform flux-limited subsamples. {\it Hereafter, we 
refer to these as our `deep' and `bright' subsamples.}
For the deep subsample we consider a $6.5'$
radius high image quality and high exposure time
region around the approximate center of the X-ray image.
The minimum exposure time for a source in the $6.5'$ radius
region is 660~ks. The 90\% enclosed energy radius is less than
7 arcseconds for the $2-8$~keV band. Allowing for these variations,
we select sources detected above fluxes of
$10^{-16}$~erg~cm$^{-2}$~s$^{-1}$ ($0.5-2$~keV),
$5\times 10^{-16}$~erg~cm$^{-2}$~s$^{-1}$ ($2-8$~keV),
and $10^{-15}$~erg~cm$^{-2}$~s$^{-1}$ ($4-8$~keV).
For the bright subsample we consider the $10'$ radius region,
except now we select only sources
detected above fluxes of $10^{-15}$~erg~cm$^{-2}$~s$^{-1}$
($0.5-2$~keV), $5\times 10^{-15}$~erg~cm$^{-2}$~s$^{-1}$ 
($2-8$~keV), and $10^{-14}$~erg~cm$^{-2}$~s$^{-1}$ ($4-8$~keV).
The incompleteness and bias corrections are expected to be
small at the flux levels of both the deep and bright 
subsamples (\markcite{cowie02}Cowie et al.\ 2002).

\subsection{Redshift Distributions}
\label{seczdist}

We show the redshift distributions for the deep
(squares) and bright (inverted triangles)
subsamples versus X-ray flux in Figs.~\ref{fig14}.
Broad-line sources are enclosed in a second
larger symbol. The spectroscopically unidentified or unobserved
sources with photometric redshift estimates from \S~\ref{secphotoz}
are plotted as open symbols at
those redshifts, while sources without a photometric redshift
estimate are plotted below the $z=0$ line.
In Fig.~\ref{fig14}a we also include the {\it ASCA}
Large Sky Survey data
(circles) from \markcite{akiyama00}Akiyama et al.\ (2000) and the
SSA13 (diamonds) and A370 (triangles) data from
\markcite{barger01b}Barger et al.\ (2001a, b). In
Fig.~\ref{fig14}b we include the {\it ROSAT} Ultra Deep
Survey data (circles) from \markcite{lehmann01}Lehmann et al.\ (2001).

The percentage of the total light in the hard or soft deep subsample 
that comes 
from spectroscopically identified point sources in each redshift
interval is indicated. In the hard deep subsample at least 35\% 
arises below $z=1$ and at least 55\% arises below $z=2$. 
These percentages increase to 40\% and 73\% if we include the 
sources in the hard deep subsample with photometric redshifts. 
Similarly, in the soft deep subsample at least 31\% arises below 
$z=1$ and at least 62\% arises below $z=2$. These percentages 
increase to 36\% and 73\% if we include the sources in the soft deep 
subsample with photometric redshifts. Thus, the bulk of the total
$2-8$~keV and $0.5-2$~keV flux in the deep subsamples arises at 
recent times, which is broadly consistent with the results of 
other X-ray samples (\markcite{barger01c}Barger et al.\ 2001c; 
\markcite{hasinger02}Hasinger 2002).

%
%
\begin{inlinefigure}
\psfig{figure=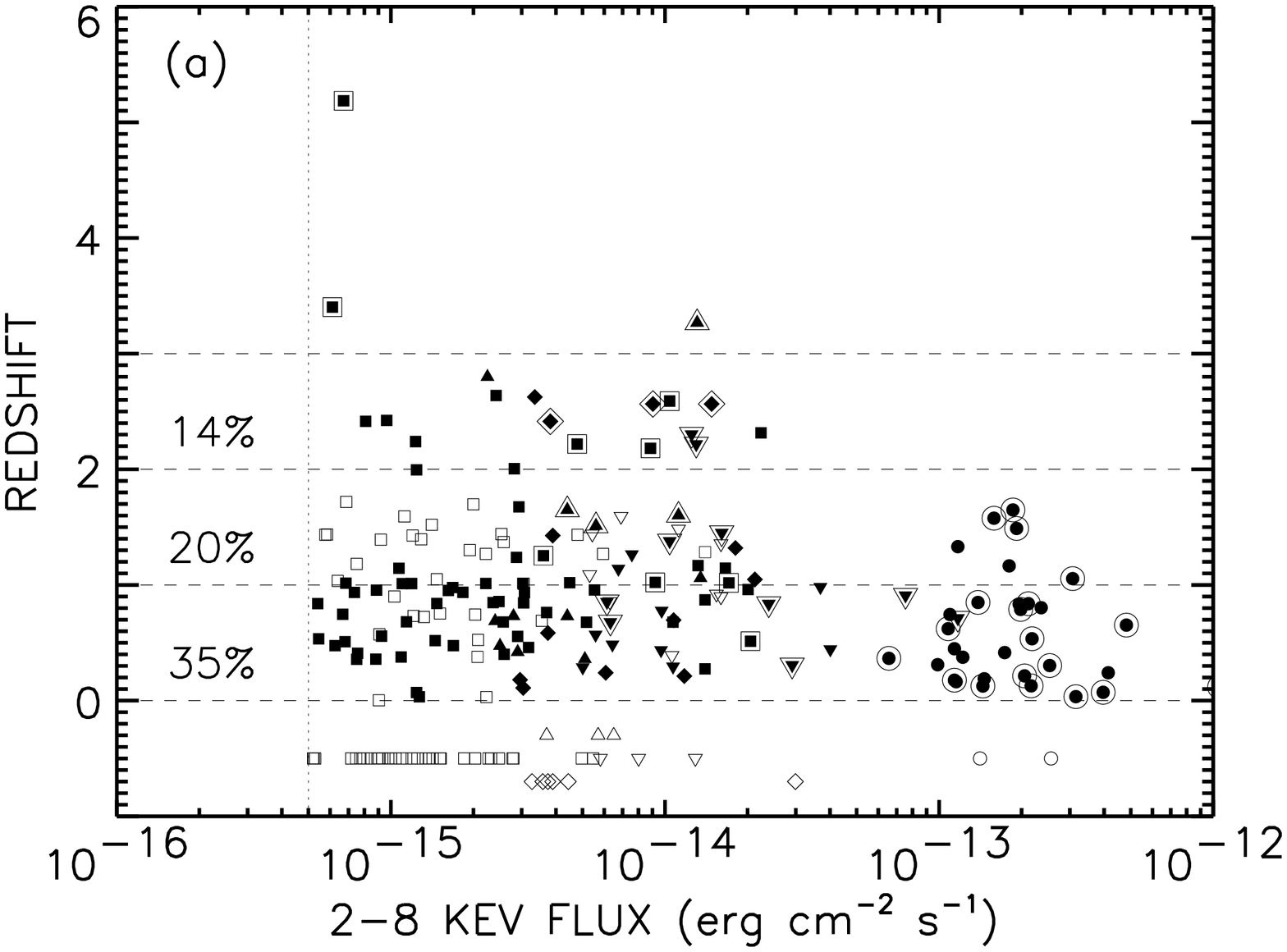,angle=90,width=3.5in}
\psfig{figure=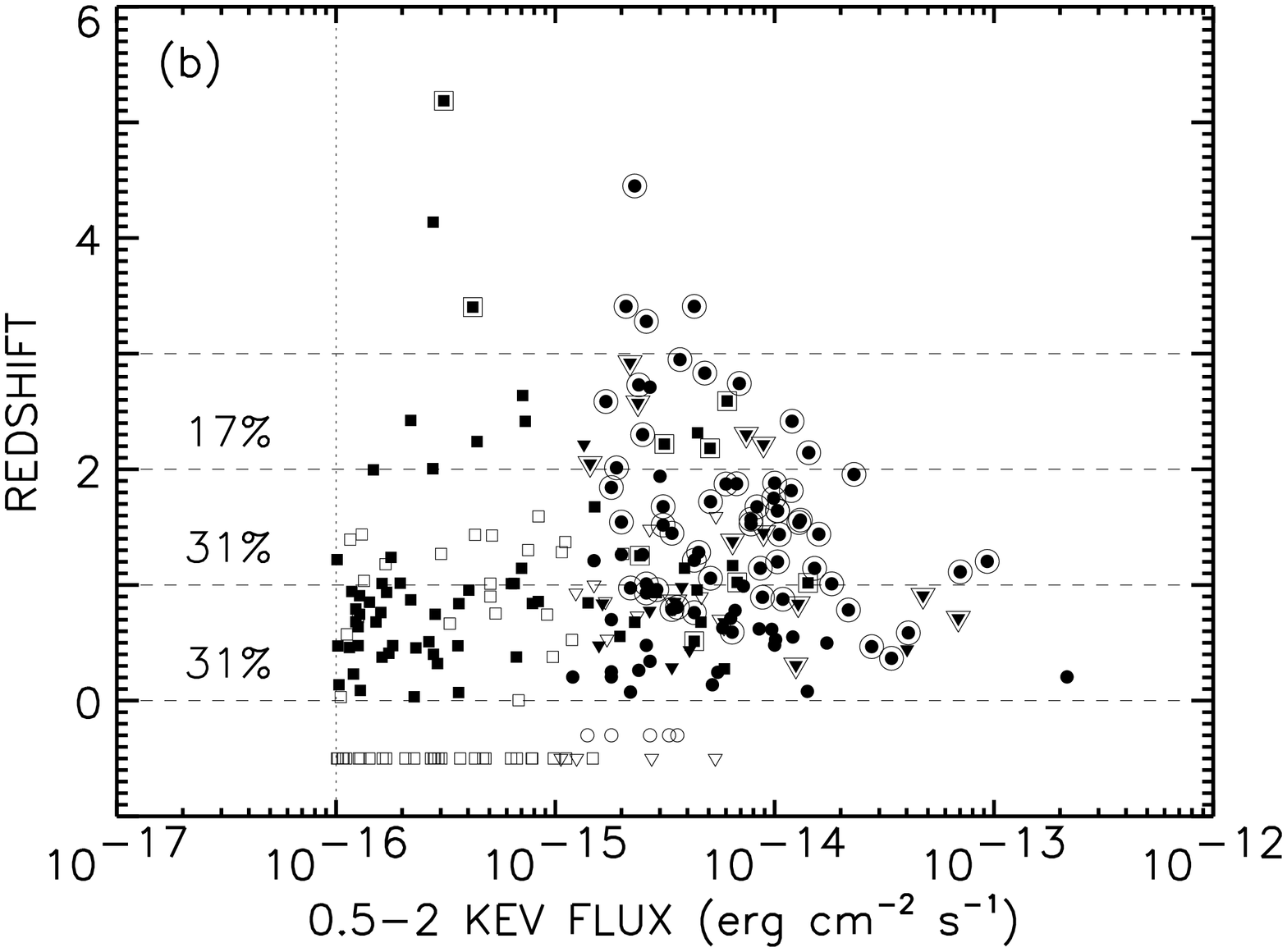,angle=90,width=3.5in}
\vspace{6pt}
\figurenum{14}
\caption{Redshift versus X-ray flux for the
deep (squares) plus bright (inverted triangles) (a)~hard and
(b)~soft X-ray subsamples (see \S~\ref{secrestrict}).
Broad-line sources are enclosed in a second larger symbol.
Spectroscopically unidentified or unobserved sources with
(without) photometric redshifts are plotted as open
symbols at those redshifts (below the $z=0$ line).
In (a) we also plot the Akiyama et al.\ (2000) data
(circles) and the Barger et al.\ (2001a,b)
SSA13 (diamonds) and A370 data (triangles).
In (b) we include the Lehmann et al.\ (2001) data (circles).
At the left is given the
percentage of the total light in the hard or soft deep
subsample that comes from spectroscopically identified point
sources in each redshift interval.
\label{fig14}
}
\addtolength{\baselineskip}{10pt}
\end{inlinefigure}

\subsection{Spectral Characteristics}
\label{secxspec}

The conversion from counts to flux, the correction
for opacity effects, and the $K$-correction
all depend on the shape of the source spectrum. For a 
power-law spectrum with photon index 
$\Gamma$ the counts~s$^{-1}$ in an energy band
$E_1$ to $E_2$ are given by
$N_{E_1-E_2}=\kappa~\int_{E_1}^{E_2}~A(E)~E^{-\Gamma}~dE$,
where $A(E)$ is the effective detector area at energy $E$
and $\kappa~E^{-\Gamma}$ has units of
photons~cm$^{-2}$~s$^{-1}$~keV$^{-1}$.
Once the normalization of the spectrum, $\kappa$, is determined
from the observed counts, the flux
is $f=\int_{E_1}^{E_2}~F(E)~dE$ where 
$F=\kappa~E^{1-\Gamma}$. The energy index is $\alpha=\Gamma-1$.

For their counts to flux conversion, B01 used an effective
$\Gamma$ for each source determined from the ratio
of the hard-band to soft-band counts. In this section we take a 
different approach and try to determine
the best choice of input spectrum by considering the 
$4-8$~keV sample, which is the least biased by any optical 
depth selection. 

We consider only sources within $6.5'$ of the approximate
X-ray image center that have $4-8$~keV 
fluxes above $2\times 10^{-15}$~erg~cm$^{-2}$~s$^{-1}$. All
the sources that satisfy these criteria are also detected 
in the $0.5-2$~keV and $2-8$~keV bands. Moreover, since the 
selected flux limit is well below the value 
$10^{-14}$~erg~cm$^{-2}$~s$^{-1}$ where most of the XRB 
contributions originate 
(e.g., \markcite{cowie02}Cowie et al.\ 2002), these sources 
should be representative of the `typical' source dominating 
the backgound. 

%
%
\begin{inlinefigure}
\psfig{figure=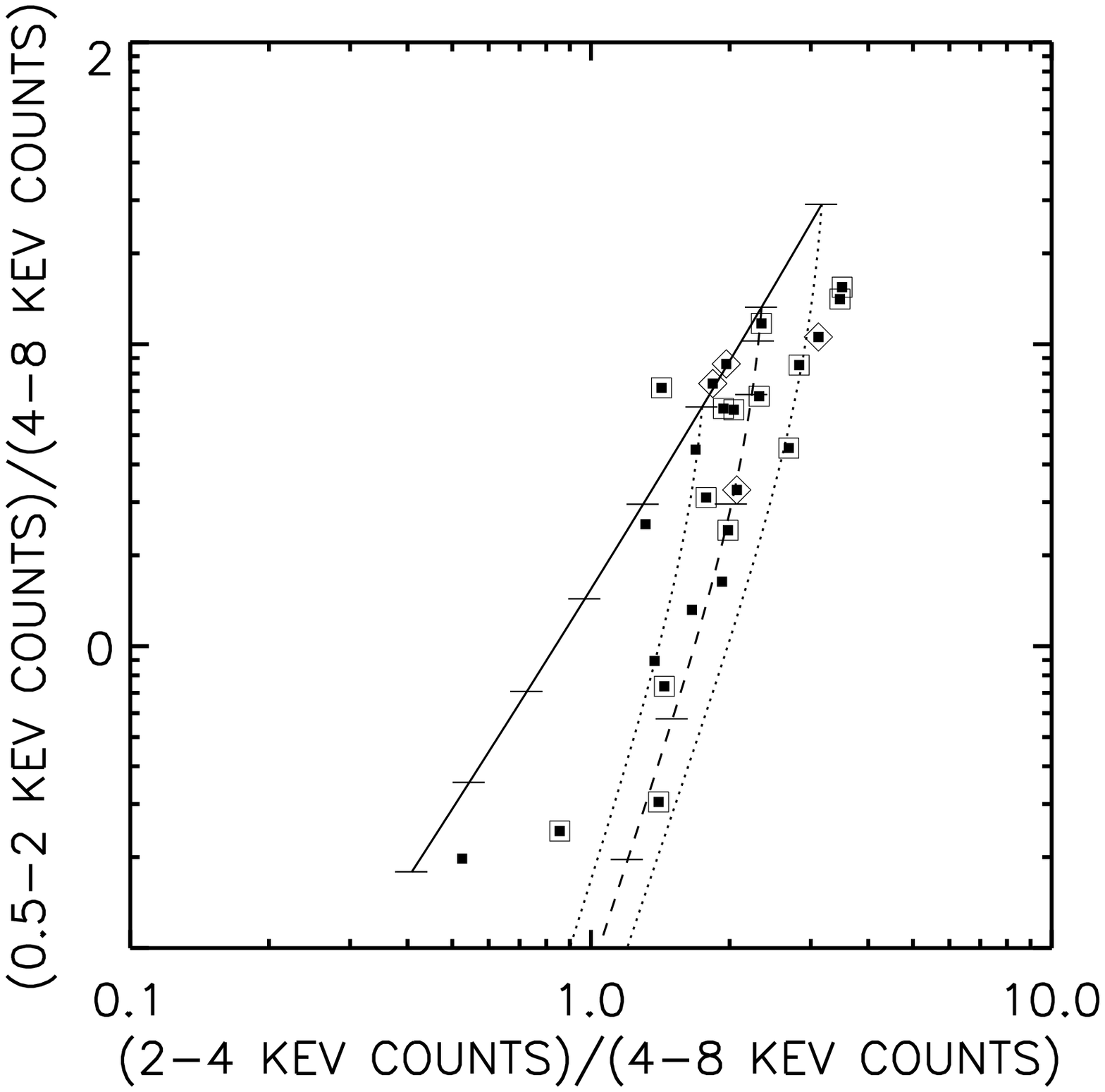,angle=90,width=4in}
\vspace{6pt}
\figurenum{15}
\caption{
Observed ($0.5-2$~keV counts)/($4-8$~keV counts)
versus ($2-4$~keV counts)/($4-8$~keV counts).
The solid squares show a $4-8$~keV sample within
$6.5'$ of the approximate X-ray image center with fluxes greater
than $2\times 10^{-15}$~erg~cm$^{-2}$~s$^{-1}$.
All of the sources in this sample are also detected in the
$0.5-2$~keV and $2-8$~keV bands. The large open squares (diamonds)
show sources with spectroscopic redshifts $z<1$ ($z>1$).
The solid line is the track for an unobscured power-law source with
varying photon index; the tick marks correspond to
$\Gamma=(-1, -0.5, 0, 0.5, 1, 1.5, 2, 2.5)$ with the values
increasing from bottom
to top. The dashed line shows the track of a power-law source
with an intrinsic $\Gamma=2$ versus rest-frame $N_H$; the
tick marks correspond to $\log N_H = (21, 21.5, 21.7, 22, 22.5, 22.7)$
with the values increasing from top to bottom.
The two dotted lines show the tracks of similar absorbed power-laws
with intrinsic spectral photon indices $\Gamma=1.5$ (left) and
$\Gamma=2.5$ (right).
\label{fig15}
}
\addtolength{\baselineskip}{10pt}
\end{inlinefigure}

In Fig.~\ref{fig15} we show 
($0.5-2$~keV counts)/($4-8$~keV counts) versus 
($2-4$~keV counts)/($4-8$~keV counts).
(The $2-4$~keV counts were determined from the difference 
of the $2-8$~keV and $4-8$~keV counts.) Sources with spectroscopic 
redshifts are denoted by a second larger symbol (squares for $z<1$ 
and diamonds for $z>1$). Two simple models are also shown.
The solid line is the track expected for a power-law source
with varying photon index $\Gamma$.
A single photon index cannot simultaneously match both the
hard and soft X-ray colors, and a strong break from harder
to softer spectra is required as the energy increases.
An only slightly more complicated model is one where we assume 
a fixed intrinsic photon index that is absorbed at low energies 
by photoelectric absorption from a varying column density of 
material in front of the source. The dashed line is the 
track expected for a power-law source with a fixed $\Gamma=2$ 
index and a varying opacity, computed using the cross-section
for solar abundances (\markcite{morrison83}Morrison \& McCammon 1983).
Over the energy range of interest for significant opacity, 
the cross-section, $\sigma(E)$, can be approximated by a 
single power-law

$$\sigma(E)=2.4\times 10^{-22}~E^{-2.6}~{\rm cm}^{-2}$$

\noindent
with $E$ in keV. The dotted lines show similar absorbed 
power-laws with intrinsic photon indices
$\Gamma =1.5$ (left line) and $\Gamma =2.5$ (right line).
As a consequence of the power-law energy dependence
of the cross-section, the effective hydrogen column density 
has a redshift dependence

$$N_{eff}=N_H/(1+z)^{2.6}$$

\noindent
It is clear from Fig.~\ref{fig15} that the X-ray colors of
the sources are reasonably well-described by this type of model
and that  most of the sources lie within the
$\Gamma=1.5$ to $\Gamma=2.5$ tracks.
The fact that the dashed line in Fig.~\ref{fig15} is
nearly vertical shows that absorption effects have little
consequence above 2~keV but produce significant flux
suppressions below 2~keV.

%
%
\begin{inlinefigure}
\psfig{figure=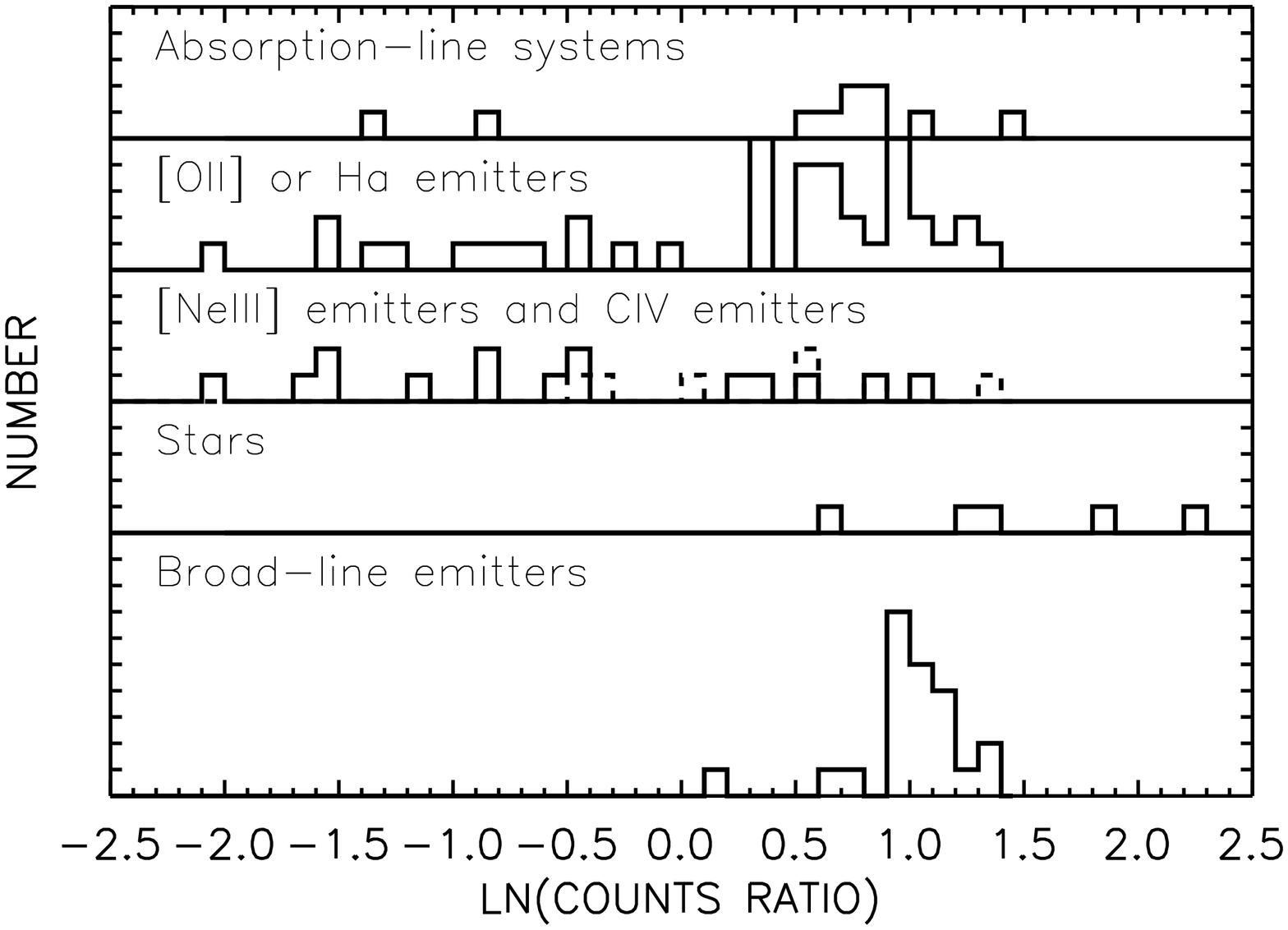,angle=90,width=3.5in}
\vspace{6pt}
\figurenum{16}
\caption{Histograms for the bright and deep subsamples
selected in the $2-8$~keV band
showing the distributions by optical spectral type
(see Table~\ref{tab3}) of the hardness of the sources, as
specified by the ($0.5-2$~keV)/($2-8$~keV) counts ratio.
The counts ratio for any source not detected in the soft band
was determined using the $0.5-2$~keV upper limit given in Table~3
of B01, so such sources could lie to the left of the current
positions. In the third panel from the bottom the solid
(dashed) line shows the distribution of sources with
[NeIII] emission (CIV emission).
\label{fig16}
}
\addtolength{\baselineskip}{10pt}
\end{inlinefigure}

In Fig.~\ref{fig16} we show for the bright and deep subsamples 
selected in the $2-8$~keV band the hardness distribution
(as specified by the ratio of the soft-band to hard-band counts)
with optical spectral type.
The broad-line sources have an extremely narrow hardness
range which would correspond to a spectral index of $\Gamma=1.7$
in the absence of optical depth effects. Nearly
all have $\Gamma<1.8$. The stars
are the softest X-ray sources, as is well-known. 
The remaining sources span a wide range of hardness, 
from very hard to comparable to that of the broad-line AGN. 
The line-of-sight column densities for these latter sources are 
only loosely related to the optical spectral characteristics;
this is likely due to different viewing angles for AGN
leading to different X-ray and optical obscurations.
Based on the distribution
of the counts ratios of the broad-line AGN in Fig.~\ref{fig16}, 
we adopt an intrinsic $\Gamma=1.8$ in the absence of
absorption in the subsequent discussion.

We now use the $\Gamma=1.8$ input spectrum to investigate 
the range of intrinsic column densities $N_H$ (i.e., the
column densities in the rest-frame of each source)
that produce the absorption.
In Fig.~\ref{fig17} we plot ($0.5-2$~keV counts)/($2-8$~keV counts)
versus redshift for the $0.5-2$~keV and $2-8$~keV deep subsamples.
Sources with spectroscopic (photometric) redshifts are denoted by 
solid (open) symbols. We overlay on the data fixed intrinsic 
$N_H$ curves generated assuming an intrinsic $\Gamma=1.8$ 
power-law spectrum and absorption by the photoelectric cross-section.
We do not show any curves after Compton thickness effects become important. 
At $z<1$ the sources in the sample appear to be well-described by the 
$\log N_H\leq23.5$ curves. At $z>1$ there are only a small number of 
spectroscopically unidentified sources which potentially could have 
column densities much in excess of $\log N_H=23.5$. Such sources 
may be very optically faint and thus preferentially lack redshifts.

%
%
\begin{inlinefigure}
\psfig{figure=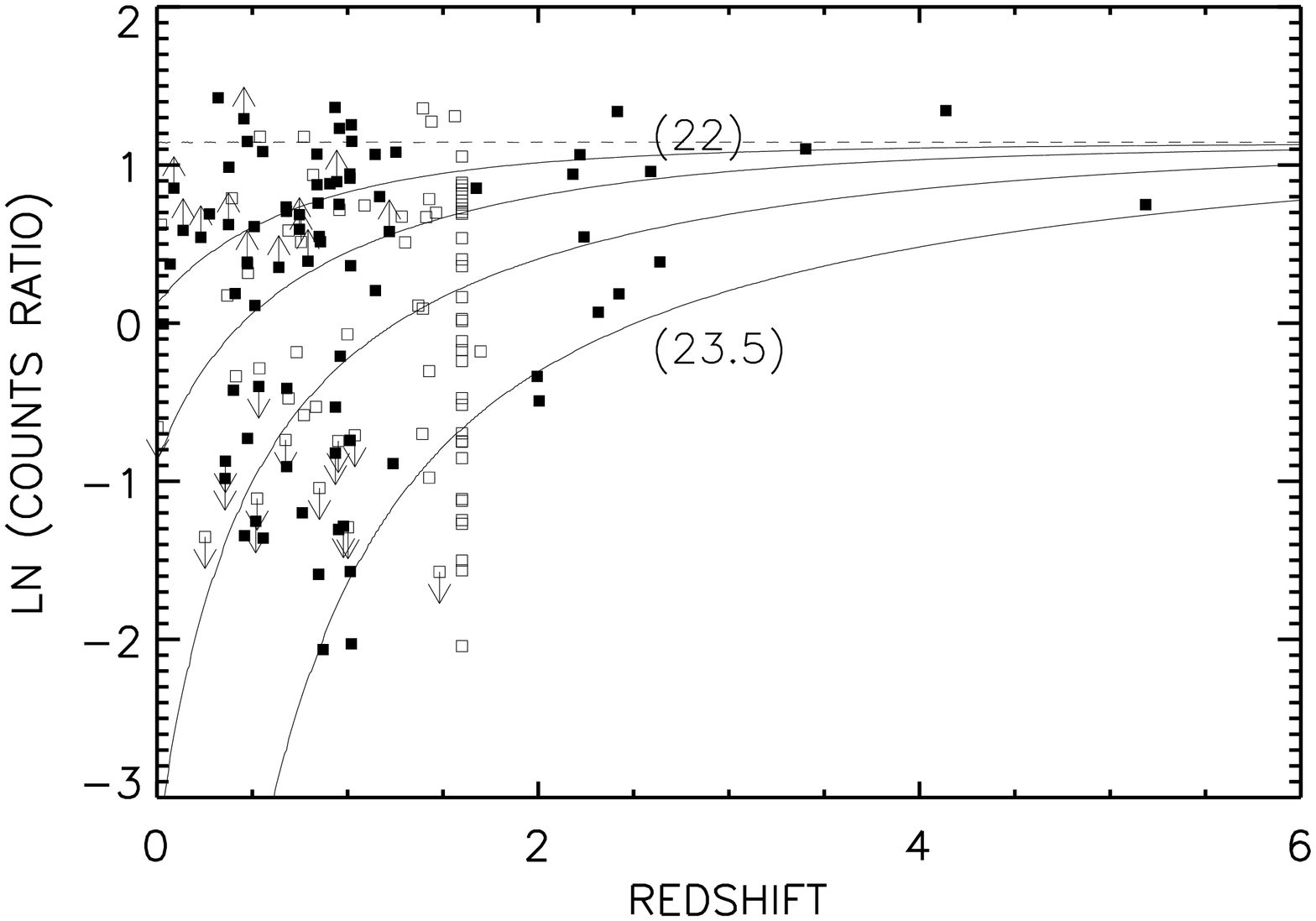,angle=90,width=3.5in}
\vspace{6pt}
\figurenum{17}
\caption{
The natural logarithm of the ($0.5-2$~keV)/($2-8$~keV) counts 
ratio versus redshift for the $0.5-2$~keV and $2-8$~keV deep 
subsamples.
Sources only detected in the soft (hard) band are shown 
with upward (downward) pointing arrows using the limits
from Table~3 of B01.
We overlay as solid curves the values expected for 
a source with an intrinsic $\Gamma=1.8$ spectrum 
(shown by the dashed line) and an intrinsic column 
density of gas and dust corresponding to 
$\log N_H = (22, 22.5, 23, 23.5)$ (solid lines). 
Sources with spectroscopic redshifts are denoted by solid symbols.
Unidentified or unobserved sources are denoted by open symbols at 
either the photometric redshift or at $z=1.5$, if the object is
too faint or too blue to have a photometric redshift.
\label{fig17}
}
\addtolength{\baselineskip}{10pt}
\end{inlinefigure}

In Fig.~\ref{fig18} we plot the deduced intrinsic $N_H$ values
for the individual sources in the $2-8$~keV selected bright and
deep subsamples with either spectroscopic or photometric redshifts.
The inferred $N_H$ values range from
about $10^{21}$~cm$^{-2}$ to $5\times 10^{23}$~cm$^{-2}$.
As expected, the broad-line AGN (denoted by a second larger symbol)
have comparatively low $N_H$ values.

Our description is simple, and we caution that when sources get
absorbed by material that is almost or fully Compton-thick,
other components could take over to determine
the spectral shape, including scattering of X-rays around the
absorber, multiple lines-of-sight through the absorber,
circumnuclear starburst emission, etc. These effects will
tend to give more flux at low energies than would be expected
from the simple model. Thus, column densities can be underestimated,
with the underestimation factor getting larger as the column density
gets larger. Consequently, there could still be some Compton-thick
sources in our sample, even if we do not recognize them as such.
This is an interesting issue, and in the next section we examine
whether the XRB {\it requires} contributions from a population
of Compton-thick sources.

%
%
\begin{inlinefigure}
\psfig{figure=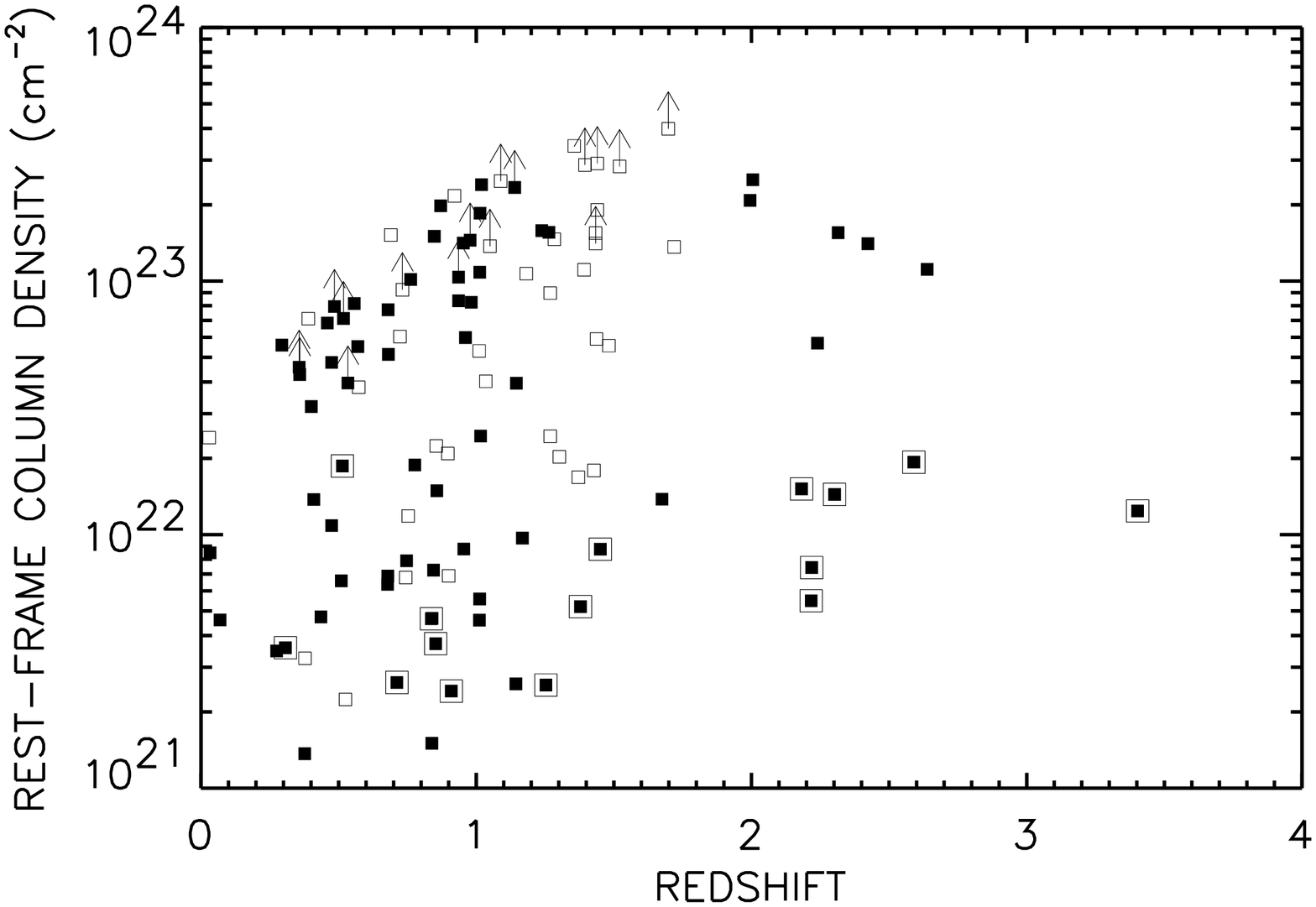,angle=90,width=3.5in}
\vspace{6pt}
\figurenum{18}
\caption{
Rest-frame column densities required to produce the observed
($2-8$~keV)/($0.5-2$~keV) counts ratios versus redshift
for the $2-8$~keV bright and deep subsamples with
spectroscopic (solid) or photometric (open) redshifts.
Sources not detected in the soft band are shown with upward
pointing arrows using the limits from Table~3 of B01.
The $z=5.18$ source is off-scale.
Sources with broad emission lines in their
spectra are indicated with a second larger symbol.
These broad-line sources are among the lowest column
density systems.
\label{fig18}
}
\addtolength{\baselineskip}{10pt}
\end{inlinefigure}

\subsection{High Energy XRB Contributions}
\label{secxrb}

An important question is how much of the very high energy 
($>8$~keV) XRB our current sample has resolved, and hence
how much light might still be needed from Compton-thick sources. 
In Fig.~\ref{fig19} we plot the total XRB contributions from the 
$2-8$~keV deep subsample (thick solid line), computed
for each source assuming an intrinsic $\Gamma=1.8$ power-law
together with an optical depth determined from
the ($0.5-2$~keV)/($2-8$~keV) counts ratio.
The dashed (\markcite{boldt87}Boldt 1987) and
dotted (\markcite{gruber84}Gruber et al.\ 1984;
\markcite{gruber92}Gruber 1992; 
\markcite{fabian92}Fabian \& Barcons 1992) lines
show the XRB measurements obtained by {\it HEAO-A}. 
The deep subsample produces about 87\% of this background 
at 3~keV and 57\% at 20~keV.

However, if the {\it Chandra} data were renormalized at 3~keV to 
the higher {\it ASCA} or {\it BeppoSAX} measurements (the latter,
taken from \markcite{vecchi99}Vecchi et al.\ 1999, are shown in 
Fig.~\ref{fig19} by the dot-dashed line), then the shape of the 
combined contributions from the {\it Chandra} sources
matches fairly well that of the observed XRB (thin solid line)
at both energies. Thus, whether the match to the 
high energy XRB requires either a substantial population of as-yet 
undetected Compton-thick sources or some change in the spectral 
shape of the current sources from the simple power-law dependence 
clearly depends critically on how the low energy and high energy
XRB measurements tie together (see also
\markcite{fabian02}Fabian, Wilman, \& Crawford 2002
for a theoretical discussion).

%
%
\begin{inlinefigure}
\psfig{figure=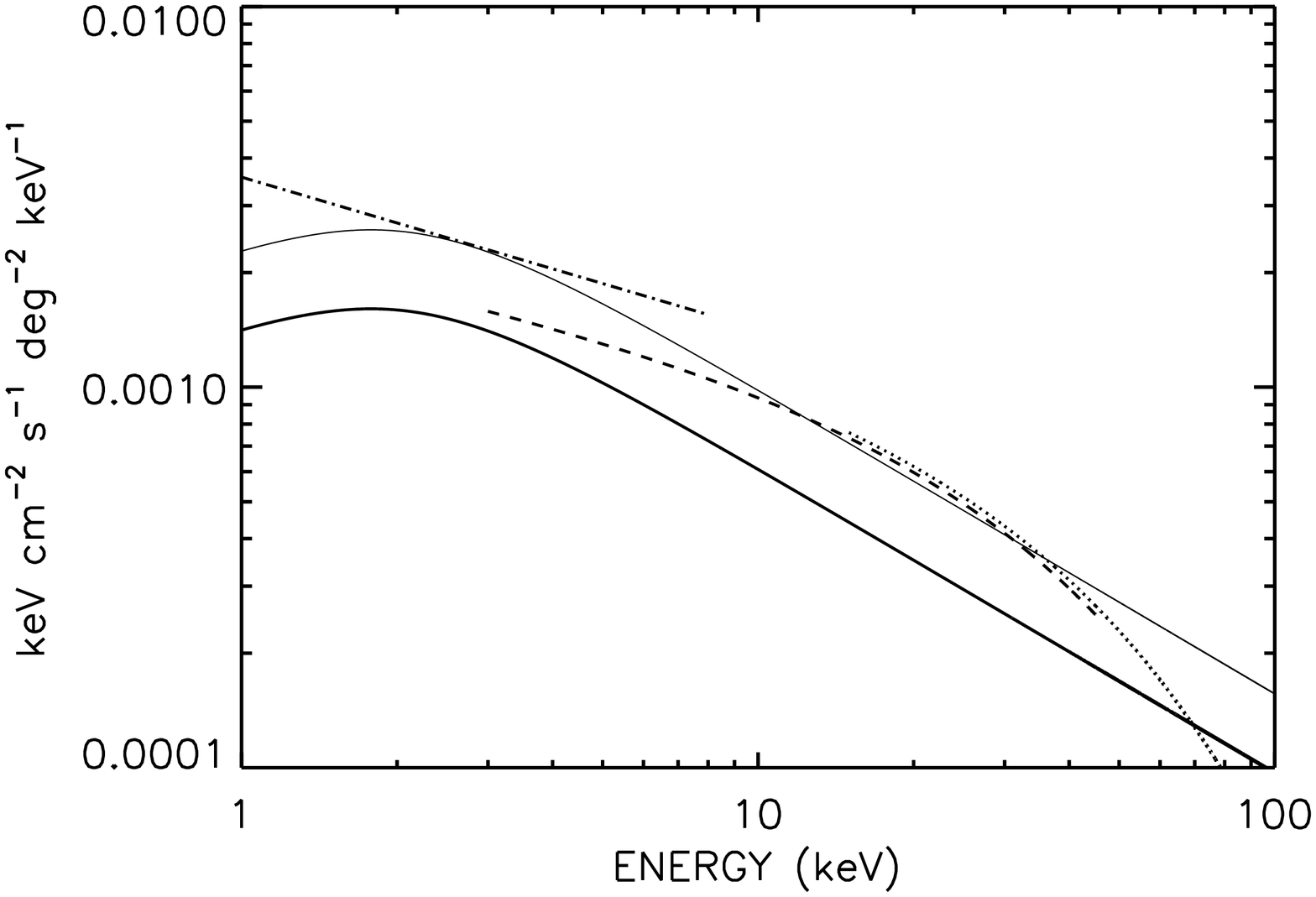,angle=90,width=3.5in}
\vspace{6pt}
\figurenum{19}
\caption{Total XRB contribution from the $2-8$~keV
deep subsample (thick solid line) based on an intrinsic
$\Gamma=1.8$ power-law dependence combined with an opacity 
determined from the ($0.5-2$~keV)/($2-8$~keV)
counts ratio for each source in the subsample.
The dashed (dotted) line shows the {\it HEAO-A} XRB 
measurement from Boldt 1987 (Gruber et al.\ 1984;
Gruber 1992; Fabian \& Barcons 1992). The
dot-dashed line is the {\it BeppoSAX} measurement
of Vecchi et al.\ (1999). Other measurements in the 
$1-10$~keV range lie between the {\it HEAO-A} and {\it BeppoSAX} 
measurements. The thin solid line is the total
XRB contribution from the $2-8$~keV deep subsample, 
renormalized to match the {\it BeppoSAX} measurement at 3~keV.
\label{fig19}
}
\addtolength{\baselineskip}{10pt}
\end{inlinefigure}

\subsection{X-ray Luminosities}
\label{secxlum}

Since the translation of observed {\it Chandra} counts to flux 
depends on the assumed spectral shape, the computation of 
the intrinsic rest-frame X-ray luminosities using the intrinsic 
$\Gamma=1.8$ spectrum and the inferred rest-frame $N_H$ values requires
a recomputation of the B01 fluxes, followed by a correction for 
the opacity (which reduces the observed flux) and a
$(1+z)^{-0.2}$ $K$-correction. It turns out that all of these effects 
are small in the $2-8$~keV band, and the final $2-8$~keV luminosities 
that we calculate only differ from those which would be computed 
directly with the B01 fluxes and a zero $K$-correction by an 
average factor of 0.83 with a weak redshift dependence. 
This factor is similar to the 0.85 constant factor used by 
\markcite{barger01a}Barger et al.\ (2001a) to empirically
correct their $2-8$~keV luminosities for these effects.

In Fig.~\ref{fig20} we show our rest-frame opacity-corrected and
$K$-corrected $2-8$~keV
luminosities versus redshift for the $2-8$~keV
bright and deep subsamples with spectroscopic identifications. 
Figure~\ref{fig20} shows very little evolution in the maximum
X-ray luminosities with decreasing redshifts until $z\lesssim 0.5$, 
at which point the 
volume becomes too small to probe effectively very 
high luminosity sources. If there were considerable evolution 
of the maximum luminosities, as is seen in the optical, then we 
would not have detected any high luminosity sources at low 
redshift.

There has been some discussion in the literature 
about rare finds of high-redshift type II quasars
(e.g., \markcite{norman02}Norman et al.\ 2002; 
\markcite{stern02a}Stern et al.\ 2002a). We note that
at $L_{2-8~keV}\gtrsim 10^{44}$~erg~s$^{-1}$ and $z>2$ we 
also find two such non-broad-line systems,
objects 184 and 287 from Table~1, both of which
are narrow Ly$\alpha$ emitters. At $z\approx1$ we also
see two such systems, objects 69 and 280 in Table~1, 
both of which have [NeIII] and [NeV]
emission; 280 also has narrow Balmer emission lines. However, 
we stress that high-redshift type II quasars are by no means 
responsible for producing the bulk of the XRB at these
energies (see \S~\ref{seczdist}).

%
%
\begin{inlinefigure}
\psfig{figure=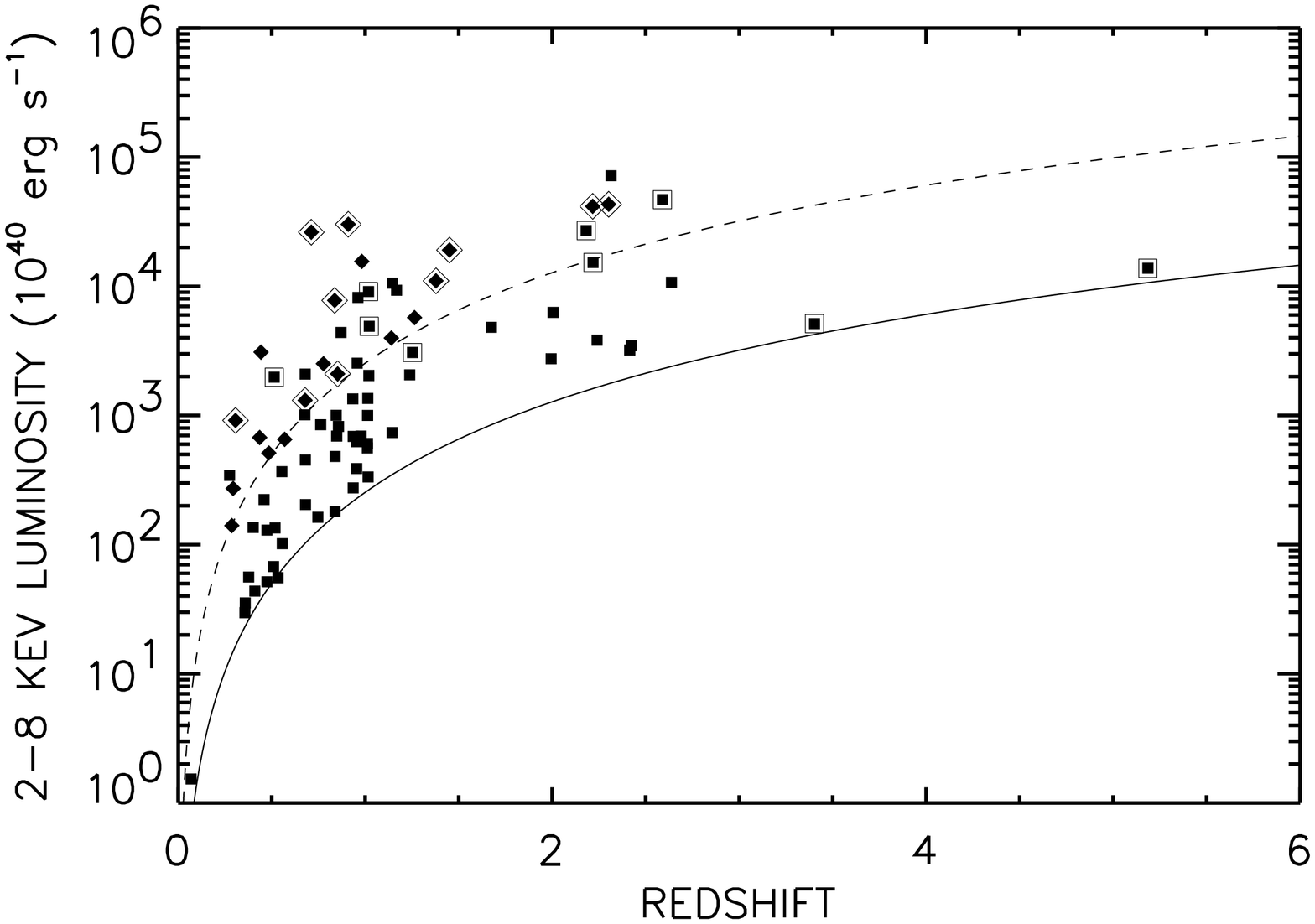,angle=90,width=3.5in}
\vspace{6pt}
\figurenum{20}
\caption{
Rest-frame opacity-corrected and $K$-corrected $2-8$~keV luminosity
versus redshift for the hard deep and bright subsamples with
spectroscopic identifications.
The solid squares (diamonds) denote sources in the deep
(bright) subsample.
Sources identified spectroscopically
as broad-line AGN are denoted by a second larger symbol.
The solid (dashed) curve shows the X-ray luminosity limit inferred
from the deep (bright) subsample flux limit of
$5\times 10^{-16}$~erg~cm$^{-2}$~s$^{-1}$
($5\times 10^{-15}$~erg~cm$^{-2}$~s$^{-1}$).
\label{fig20}
}
\addtolength{\baselineskip}{10pt}
\end{inlinefigure}

Because the optical depth effects are much larger at
low energies, the corrections described above depend sensitively 
on the choice of intrinsic spectrum; thus, we have not attempted 
to calculate intrinsic rest-frame $0.5-2$~keV luminosities.
Instead,
in Fig.~\ref{fig21} we show observed-frame $0.5-2$~keV luminosity
(based on the B01 fluxes with no opacity correction)
versus redshift for the soft bright and deep subsamples at $z<1$.
Sources with an intrinsic rest-frame 
$L_{2-8~keV}\gtrsim 10^{42}$~erg~s$^{-1}$ are denoted by
open symbols, and those with fainter hard X-ray luminosities are
denoted by solid symbols. X-ray luminosities of local starbursts 
never exceed $10^{42}$~erg~s$^{-1}$
(\markcite{zezas98}Zezas, Georgantopoulos, \& Ward 1998;
\markcite{moran99}Moran et al.\ 1999), so any source more luminous 
than this is very likely an AGN.

Below $10^{42}$~erg~s$^{-1}$ the sources could be more `typical' 
galaxies whose X-ray emission is dominated by a population of 
X-ray binaries, hot interstellar gas, or a low-luminosity AGN. 
We detect 18 low-luminosity sources in the 
$0.5-2$ deep subsample, which corresponds to a surface density 
of $490^{+140}_{-110}$~deg$^{-2}$. These sources contribute
a very small fraction (approximately 3\%) of the total
$0.5-2$~keV light in the deep subsample and are not
a significant contributor to the XRB.
A more thorough discussion of the optically bright, X-ray
faint source population to which normal galaxies belong
can be found in A. E. Hornschemeier et al., in preparation.

%
%
\begin{inlinefigure}
\psfig{figure=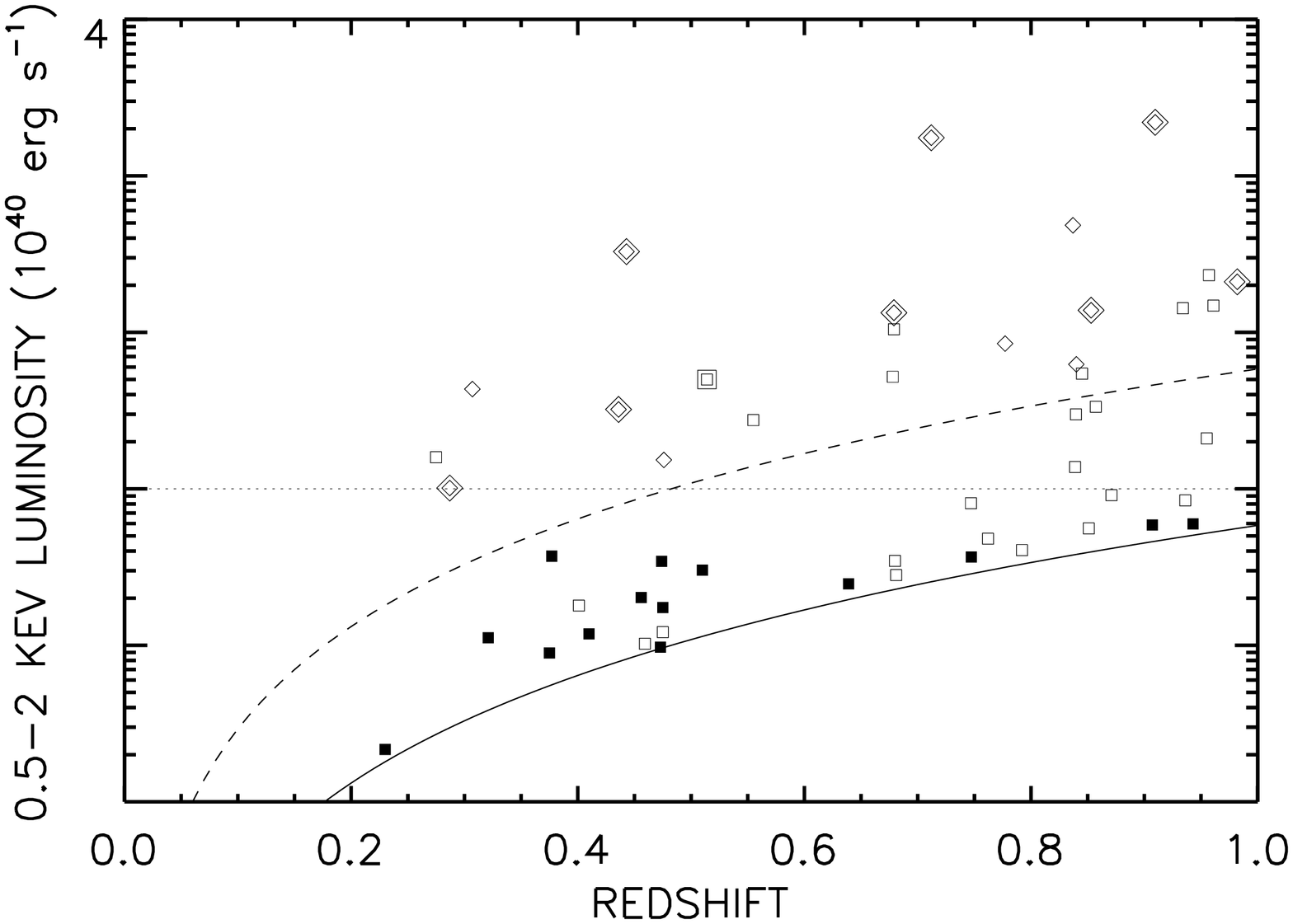,angle=90,width=3.5in}
\vspace{6pt}
\figurenum{21}
\caption{
Observed $0.5-2$~keV luminosity versus redshift for the soft
deep and bright subsamples with spectroscopic identifications.
The open squares (diamonds) denote sources in the deep
(bright) subsample with $L_{2-8~keV}\gtrsim 10^{42}$~erg~s$^{-1}$;
the solid symbols denote those with fainter hard X-ray
luminosities. Sources identified spectroscopically
as broad-line AGN are denoted by a second larger symbol.
The solid (dashed) curve shows the observed X-ray luminosity limit
inferred from the deep (bright) subsample flux limit of
$10^{-16}$~erg~cm$^{-2}$~s$^{-1}$
($10^{-15}$~erg~cm$^{-2}$~s$^{-1}$).
\label{fig21}
}
\addtolength{\baselineskip}{10pt}
\end{inlinefigure}

\subsection{Optical versus X-ray Luminosities}
\label{secovx}

We calculate $B$-band luminosities, $L_B$, for the X-ray sources
using the relation 

$$L_B=4\pi d_L^2 (1.5\times 10^{14}) f_{4500(1+z)} (1+z)^{-1}$$

\noindent
where $d_L$ is the luminosity distance in cm, $f_{4500(1+z)}$
is the rest-frame $B$-band flux in units of
erg~cm$^{-2}$~s$^{-1}$~Hz$^{-1}$ (estimated by interpolating from
the observed fluxes corresponding to the magnitudes given in
Table~1), and $1.5\times 10^{14}$~Hz is the width of the $B$-band filter.
In Fig.~\ref{fig22} we plot rest-frame $L_B$ versus 
$L_{2-8~keV}$ (from \S~\ref{secxlum}) for the $2-8$~keV bright and 
deep subsamples. The solid lines have slopes 
$\log(L_B/L_{2-8~keV})=100$, 10, 1, 0.1, and 0.01.
The open (solid) symbols denote sources with redshifts $z>1.2$
($z<1.2$). The second larger symbols denote spectroscopic 
broad-line AGN.

The $z>1.2$ sources approximately follow the line
$\log(L_B/L_{2-8~keV})=1$, which suggests that the central
AGN in these galaxies dominate their optical luminosities. 
At $L_{2-8~keV}\lesssim 10^{43}$~erg~s$^{-1}$ the
locus flattens above the $\log(L_B/L_{2-8~keV})=1$ line;
here the host galaxy light starts to dominate the total optical light. 
At the high X-ray luminosities most of the identified sources 
are broad-line AGN. Since the unobscured AGN light dominates the 
optical light in these systems, no matter what redshifts the 
sources are at we should have no trouble identifying them 
spectroscopically. However, at high redshifts and high X-ray 
luminosities, we may be unable to spectroscopically identify 
high-extinction AGN due to the obscuration of the optical light 
originating from the AGN.

%
%
\begin{inlinefigure}
\psfig{figure=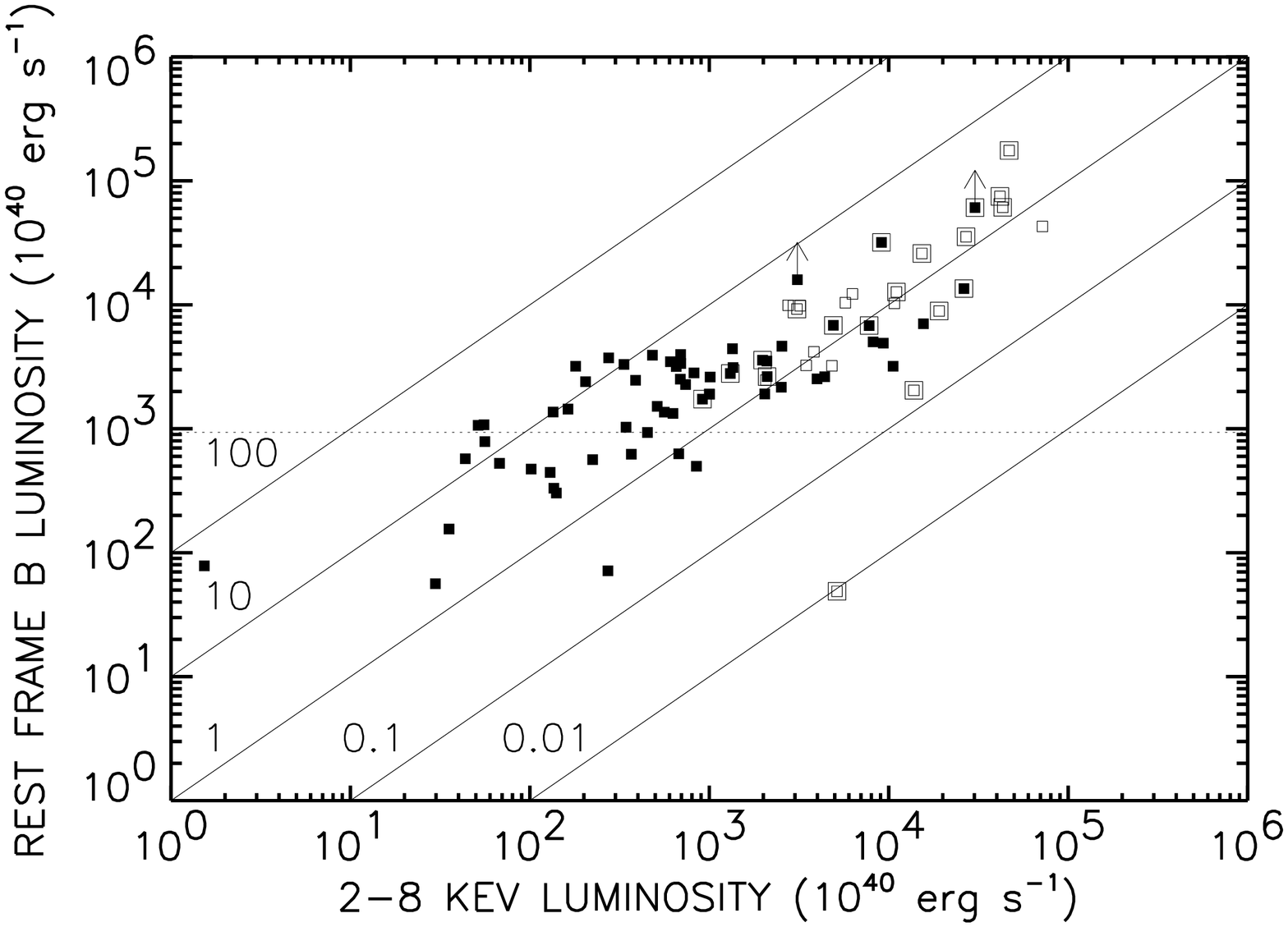,angle=90,width=3.5in}
\vspace{6pt}
\figurenum{22}
\caption{Rest-frame $B$-band luminosity versus rest-frame
opacity-corrected and $K$-corrected
$2-8$~keV luminosity for the spectroscopically identified
hard deep and bright subsamples.
The open (solid) symbols denote sources 
with redshifts $z>1.2$ ($z<1.2$).  
The second larger symbols denote spectroscopic broad-line AGN.
The two objects which are saturated in $R$ and $I$ are shown as 
lower limits with upward pointing arrows.
The solid lines have slopes $\log(L_B/L_{2-8~keV})=100$, 10, 1, 0.1, 
and 0.01. The dotted line is the luminosity of an 
$M_B^\ast=-20.4$ galaxy from Loveday et al.\ (1992).
\label{fig22}
}
\addtolength{\baselineskip}{10pt}
\end{inlinefigure}

\section{Summary}
\label{secsummary}

We presented a catalog of the optical, NIR, submillimeter,
and radio properties of the X-ray sources identified in the 
$\approx 1$~Ms {\it Chandra} exposure of the CDF-N.
We now have redshifts for 182 of the 370 X-ray sources, and
we presented the spectra for 175.
The redshift identifications are very complete (78\%) for the 
$R\le 24$ galaxy sources in a $10'$ radius around the approximate
X-ray image center. All of the X-ray sources with $z>1.6$ are 
either broad-line AGN or have narrow Ly$\alpha$ and/or 
CIII]~1909~\AA\ emission; none of the known $z>1.6$ absorption 
line systems in the field are detected in X-rays. 

We found spectroscopic evidence for large scale structure in 
the field which could account for a part of the 
field-to-field variation seen in the X-ray number counts.
However, since the observed structures do not dominate the number 
of X-ray sources in the sample, the redshift distribution 
should not be too strongly affected by clustering.

The broad-line sources are all extremely blue, making it hard 
to distinguish between low redshift irregulars and luminous 
high-redshift AGN purely on the basis of color. 
Very few of the redder X-ray sources, whether spectroscopically 
identified or not, lie outside the regions populated by 
unevolving spiral galaxy tracks. 

We estimated photometric redshifts for the sources using our
five broad-band colors (six when $HK'$ was available).
Above $B-I=1.5$ the photometric redshift estimates are quite
robust; for most sources with available spectroscopic redshifts 
the photometric redshifts are within 25\%. The majority of 
galaxies have absolute magnitudes comparable to or 
more luminous than $M_I^\ast=-22$.

We found that the spectral energy distributions of the X-ray
sources are well-described by a $\Gamma=1.8$ intrinsic spectrum
corrected for absorption. We determined the absorbing column 
densities for our sources with redshifts and found that they 
range from about $10^{21}$~cm$^{-2}$ to 
$5\times 10^{23}$~cm$^{-2}$.

We calculated intrinsic rest-frame hard and observed-frame soft 
X-ray luminosities. We found very little evolution in the maximum 
hard X-ray luminosities until $z\lesssim 0.5$, at which point 
the volume becomes too small to probe the very high
luminosity sources. At soft X-ray luminosities of
$L_{0.5-2~keV}<10^{42}$~erg~s$^{-1}$
the `typical' galaxy, whose X-ray emission may be dominated
by a population of X-ray binaries, hot interstellar gas,
or a low-luminosity AGN, becomes important.

We calculated $B$-band luminosities, $L_B$, for the
X-ray sources. Sources with $z>1.2$ approximately follow
the line $\log(L_B/L_{2-8~keV})=1$, which suggests that
the AGN dominate the optical luminosities. At
$L_{2-8~keV}\sim 10^{43}$~erg~s$^{-1}$ the
locus flattens above the line, which suggests this is where
the host galaxy light starts to dominate the optical light.

Redshift slices of the sources versus X-ray flux show that 
substantial contributions to the total X-ray flux in both bands 
(of order one-third using only the present spectroscopic 
sample) come from $z<1$. Thus, major accretion onto supermassive 
black holes has occurred since the Universe was half its 
present age.

We estimated that the {\it Chandra} sources that produce 
87\% of the {\it HEAO-A} XRB at 3~keV produce
57\% at 20~keV, provided that at high
energies the spectral shape of the sources continues to be 
well-described by a $\Gamma=1.8$ power-law. However, 
when the {\it Chandra} contributions are renormalized to the
{\it BeppoSAX} XRB at 3~keV, the shape matches fairly well the 
observed XRB at both energies. Thus,
whether a substantial population of as-yet
undetected Compton-thick sources or some change in the spectral
shape of the current sources from the simple power-law dependence
is required to resolve the high energy XRB depends critically 
on how the low energy and high energy XRB measurements tie together.

\acknowledgements
We thank the referee, David Helfand, for helpful comments
that improved the manuscript. We thank Andy Fabian, Scott Tremaine, 
and Omar Almaini for useful discussions.
We gratefully acknowledge support from the University of Wisconsin 
Research Committee with funds granted by the Wisconsin Alumni 
Research Foundation (AJB), NSF grants AST-0084847 (AJB, PI) and 
AST-0084816 (LLC), NASA grants DF1-2001X (LLC, PI),
NAS8-01128 (GPG, PI), and G02-3187A (WNB), 
NSF CAREER award AST-9983783 (WNB),
NASA GSRP grant NGT 5-50247 (AEH), and the 
Pennsylvania Space Grant Consortium (AEH).

%
%
\begin{inlinefigure}
\psfig{figure=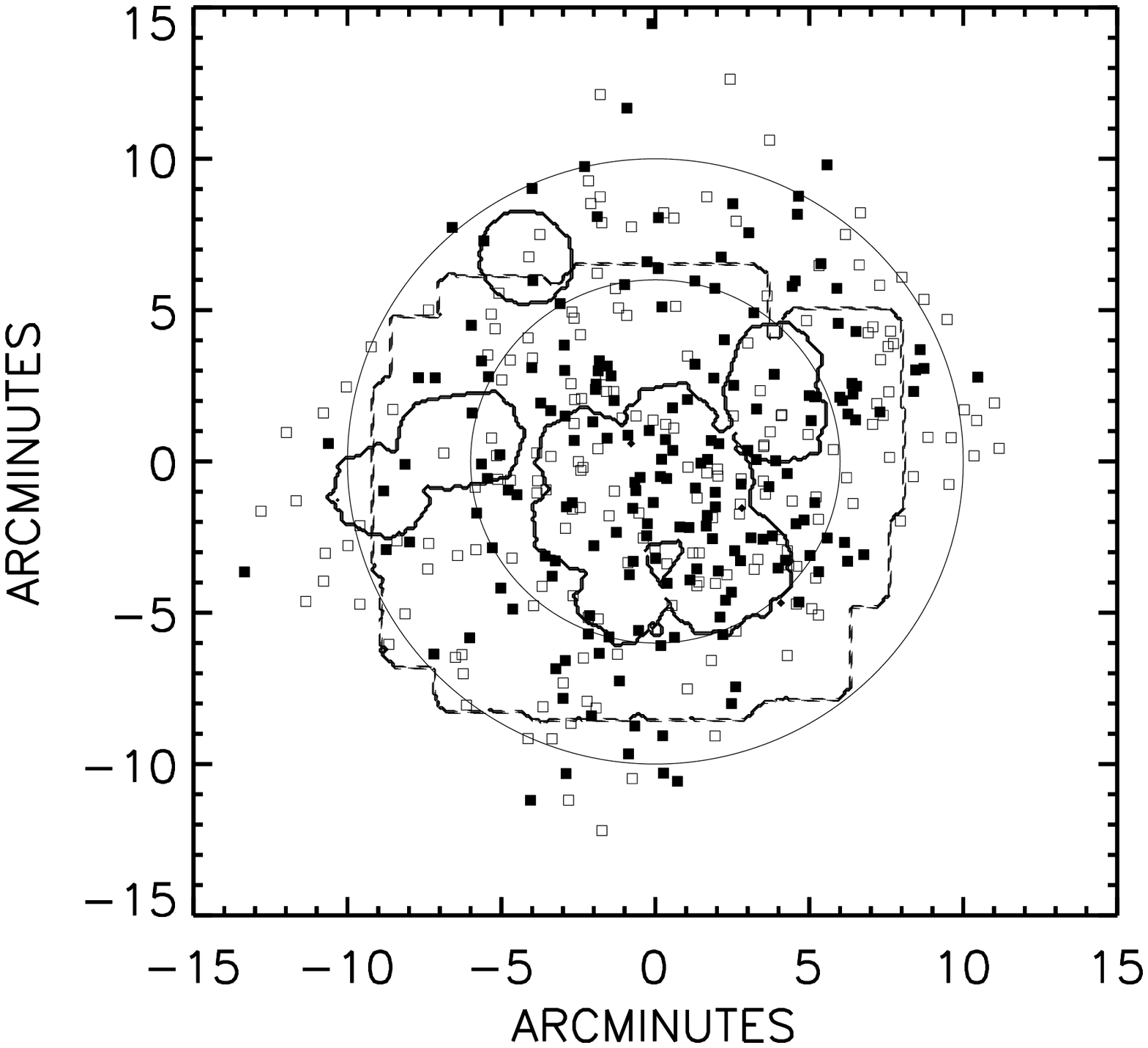,angle=90,width=8in}
\vspace{6pt}
\figurenum{1}
\caption{
Schematic diagram of the 370 X-ray point sources (squares)
in the B01 catalog of the CDF-N. Spectroscopically
identified X-ray sources are solid.
Radii of $6.5'$ and $10'$ from the approximate X-ray image center
are indicated by the large circles. The dashed (solid) contour
shows the area coverage of the $HK'$ (submillimeter) observations.
The optical data cover a larger field ($24'\times 24'$) than
that shown.
\label{fig1}
}
\end{inlinefigure}

%
%
\figurenum{2}
\figcaption[]{
$z'$-band thumbnail images of the 370 X-ray point sources
in the B01 catalog. The $1\sigma$ AB magnitude limit is 27.0.
The thumbnails are $18''$ on a side and are labeled with the 
B01 catalog number (column~1 of Table~1 in this paper) in the 
upper left corner. The expected position of each source is 
shown with a $2''$ radius circle if it is within $6.5'$ of the 
approximate X-ray image center and with a $3.6''$ radius circle 
otherwise. If an optically identified source is within $2''$ of 
the nominal X-ray position, then the spectroscopic redshift is 
shown in the lower left corner. If a bright optically identified 
source is within $5''$ of the nominal X-ray position, then the 
redshift of that source is printed with fainter text. Sources 
detected in the B01 $2-8$~keV catalog have an `H' printed in the 
lower right corner. Sources also detected at 20~cm or 850~$\mu$m 
have, respectively, an `R' or an `S' printed in the upper right
corner. North is up and East is to the left.
\label{fig2}
}

%
%
\figurenum{6}
\figcaption[]{
Spectra for 175 of the 182 X-ray sources
with spectroscopic redshifts versus rest-frame
wavelength. The hatched regions show the positions of
strong atmospheric bands and night sky emission lines.
In each case we give the redshift and the B01 number.
Sources with WIYN spectra are marked with a `W' and are
not flux calibrated.
\label{fig6}
}


\begin{references}

\reference{akiyama00}
Akiyama, M., et al.\ 2000, \apj, 532, 700

\reference{alexander01}
Alexander, D. M., Brandt, W. N., Hornschemeier, A. E., Garmire, G. P.,
Schneider, D. P., Bauer, F. E., \& Griffiths, R. E.\ 2001, \aj, 122, 2156

\reference{baldi02}
Baldi, A., Molendi, S., Comastri, A., Fiore, F., Matt, G. \&
Vignali, C.\ 2002, \apj, 564, 190

\reference{barden94}
Barden, S. C., Armandroff, T., Muller, G., Rudeen, A. C.,
Lewis, J., \& Groves, L. 1994,
Instrumentation in Astronomy VIII, ed. D.L. Crawford and E.R. Craine,
SPIE Vol.~2198, p.~87.

\reference{barger99}
Barger, A. J., Cowie, L. L., Trentham, N., Fulton, E., Hu, E. M.,
Songaila, A., \& Hall, D.\ 1999, \aj, 117, 102

\reference{bcr00}
Barger, A. J., Cowie, L. L., \& Richards, E. A.\ 2000, \aj, 119, 2092

\reference{barger01a}
Barger, A. J., Cowie, L. L., Mushotzky, R. F., \&
Richards, E. A.\ 2001a, \aj, 121, 662

\reference{barger01b}
Barger, A. J., Cowie, L. L., Steffen, A. T., Hornschemeier, A. E.,
Brandt, W. N., \& Garmire, G. P.\ 2001b, \apj, 560, L23

\reference{barger01c}
Barger, A. J., Cowie, L. L., Bautz, M. W., Brandt, W. N.,
Garmire, G. P., Hornschemeier, A. E., Ivison, R. J., \&
Owen, F. N.\ 2001c, \aj, 122, 2177

\reference{bauer02}
Bauer, F. E. et al.\ 2002, \apj, 123, 1163

\reference{boldt87}
Boldt, E.\ 1987, \physrep, 146, 215

\reference{bol00}
Bolzonella, M., Miralles, J.-M., \& Pell{\'o}, R.\ 2000, A\&A, 363, 467

\reference{brandt01a}
Brandt, W. N., et al.\ 2001a, \aj, 122, 1

\reference{brandt01b}
Brandt, W. N., Hornschemeier, A. E., Schneider, D. P., Alexander, D. M.,
Bauer, F. E., Garmire, G. P., \& Vignali, C.\ 2001b, \apj, 558, L5

\reference{brandt01c}
Brandt, W. N., et al.\ 2001c, \aj, 122, 2810 (B01)

\reference{bc93}
Bruzual, A. G. \& Charlot, S.\ 1993, \apj, 405, 538

\reference{campana01}
Campana, S., Moretti, A., Lazzati, D., \& Tagliaferri, G.\
2001, \apjl, 560, L19

\reference{cohen00}
Cohen, J. G., Hogg, D. W., Blandford, R., Cowie, L. L., Hu, E.,
Songaila, A., Shopbell, P., \& Richberg, K.\ 2000, \apj, 538, 29

\reference{cww}
Coleman, G. D., Wu, C-C.,
\& Weedman, D. W.\ 1980, \apjs, 43, 393

\reference{cowie96}
Cowie, L. L., Songaila, A., Hu, E. M., \& Cohen, J. G.\ 1996,
\aj, 112, 839

\reference{cowie01}
Cowie, L. L., et al.\ 2001, \apj, 551, L9

\reference{cowie02}
Cowie, L. L., Garmire, G. P., Bautz, M. W., Barger, A. J., 
Brandt, W. N., \& Hornschemeier, A. E.\ 2002, \apj, 566, L5

\reference{crawford01}
Crawford, C. S., Fabian, A. C., Gandhi, P., Wilman, R. J.,
\& Johnstone, R. M.\ 2001, \mnras, 324, 427

\reference{dawson01}
Dawson, S., Stern, D., Bunker, A., Spinrad, H., \& Dey, A.\ 2001,
\aj, 122, 598 

\reference{fabian92}
Fabian, A. C. \& Barcons, X.\ 1992, ARA\&A, 30, 429

\reference{fabian99}
Fabian, A. C. \& Iwasawa, K.\ 1999, \mnras, 303, L34

\reference{fabian02}
Fabian, A. C., Wilman, R. J., \& Crawford, C. S.\ 2002,
\mnras, 329, L18

\reference{fernandez99}
Fern{\'a}ndez-Soto, A., Lanzetta, K. M., \& Yahil, A.\ 1999, 
\apj, 523, 72

\reference{fiore01}
Fiore, F., et al.\ 2001, in Issues of Unifications of AGNs,
ed. R. Maiolino, A. Marconi, \& N. Nagar, in press (astro-ph/0109116)

\reference{jg92}
Gardner J. P.\ 1992, PhD Thesis, University of Hawaii.

\reference{gehrels86}
Gehrels, N.\ 1986, ApJ, 303, 336

\reference{giacconi02}
Giacconi, R., et al.\ 2002, \apjs, 139, 369

\reference{gruber84}
Gruber, D. E., Rothschild, R. E., Matteson, J. L.,
\& Kinzer, R. L.\ 1984, Max Planck Inst. Rep., 184, 129

\reference{gruber92}
Gruber, D. E.\ 1992, in 
The X-ray Background, ed. X. Barcons \& A. C. Fabian,
(Cambridge: Cambridge Univ. Press), p44

\reference{haiman99}
Haiman, Z. \& Loeb, A.\ 1999, \apj, 521, L9

\reference{hasinger98}
Hasinger, G., Burg, R., Giacconi, R., Hartner, G., Schmidt, M.,
Trumper, J., \& Zamorani, G.\ 1998, A\&A, 329, 482

\reference{hasinger01}
Hasinger, G., et al.\ 2001, A\&A, 365, L45

\reference{hasinger02}
Hasinger, G.\ 2002, in New Visions of the X-ray Universe in the
XMM-Newton and Chandra Era, ed. F. Jansen, in press
(astro-ph/0202430)

\reference{hawarden01}
Hawarden, T. G., Leggett, S. K., Letawsky, M. B., Ballantyne, D. R.,
\& Casali, M. M.\ 2001, \mnras, 325, 563

\reference{hodapp96}
Hodapp, K.-W.\ et al.\ 1996, NewA, 1, 177

\reference{horn01}
Hornschemeier, A. E., et al.\ 2001, \apj, 554, 742 (H01)

\reference{horn02}
Hornschemeier, A. E., Brandt, W. N., Alexander, D. M., Bauer, F. E.,
Garmire, G. P., Schneider, D. P., Bautz, M. W., \& Chartas, G.\ 2002,
\apj, 568, 82

\reference{hughes98}
Hughes, D.H. et al.\ 1998, \nat, 394, 241

\reference{surf}
Jenness, T. \& Lightfoot, J. F.\ 1998, Starlink User Note 216.3

\reference{kron80}
Kron, R.G. 1980, \apjs, 43, 305

\reference{lehmann}
Lehmann, I., et al.\ 2001, A\&A, 371, 833

\reference{manual}
Lightfoot, J. F., Jenness, T., Holland, W. S., \& Gear, W. K.\ 1998,
SCUBA System Note 1.2

\reference{loveday92}
Loveday, J., Peterson, B. A., Efstathiou, G., \& Maddox, S. J.\ 1992,
\apj, 390, 338

\reference{miya98}
Miyazaki, S., Sekiguchi, M., Imi, K., Okada, N., Nakata, F.,
\& Komiyama, Y.\ 1998, in Proc. SPIE Vol.~3355, Optical Astronomical 
Instrumentation, ed.\ S. D'Odorico, 363-374. 

\reference{moran99}
Moran, E. C., Lehnert, M. D., \& Helfand, D. J.\ 1999, \apj, 526, 649

\reference{morrison83}
Morrison, R. \& McCammon, D.\ 1983, \apj, 270, 119

\reference{mushotzky00}
Mushotzky, R. F., Cowie, L. L., Barger, A. J., \& Arnaud, K. A.\ 2000,
\nat, 404, 459

\reference{nandra02}
Nandra, K., Mushotzky, R. F., Arnaud, K. A., Steidel, C. C.,
Gardner, J. P., Teplitz, H. I., \& Windhorst, R. A.,
\apj, in press (astro-ph/0205215)

\reference{norman02}
Norman, C., et al.\ 2002, \apj, 571, 218

\reference{oke95}
Oke, J.B., et al.\ 1995, \pasp, 107, 375

\reference{richards00}
Richards, E. A.\ 2000, \apj, 533, 611

\reference{rosati02}
Rosati, P., et al.\ 2002, \apj, 566, 667

\reference{schmidt98}
Schmidt, M., et al.\ 1998, A\&A, 329, 495

\reference{stern02a}
Stern, D., et al.\ 2002a, \apj, 568, 71

\reference{stern02b}
Stern, D., et al.\ 2002b, \aj, 123, 2223

\reference{stocke91}
Stocke, J. T., Morris, S. L., Gioia, I. M., Maccacaro, T.,
Schild, R., Wolter, A., Fleming, T. A., \& Henry, J. P.\ 1991,
\apjs, 76, 813

\reference{tozzi01}
Tozzi, P., et al.\ 2001, \apj, 562, 42

\reference{vecchi99}
Vecchi, A., Molendi, S., Guainazzi, M., Fiore, F., \& Parmar, A.N.\ 1999,
A\&A, 349, L73

\reference{zezas98}
Zezas, A., Georgantopoulos, I., \& Ward, M. J.\ 1998, \mnras, 301, 915

\end{references}
\end{document}